\documentclass[american,reprint, floatfix,amsmath,amssymb, superscriptaddress,nofootinbib,prx,aps,longbibliography]{revtex4-1}

\usepackage{geometry}
 \usepackage[hyperlink]{}

\usepackage{hyperref}
\hypersetup{
     colorlinks   = true,
     linkcolor    = blue,
     urlcolor     = blue,
     citecolor    = blue
}

\usepackage{graphicx}
\usepackage{bm}
\usepackage{dcolumn}
\usepackage{amssymb}
\usepackage{amsmath}
\usepackage{slashed}
\usepackage{amsfonts}
\usepackage{epsfig}
\usepackage{stackrel}
\usepackage{paralist}
\usepackage[rgb]{xcolor}
\usepackage{pdfcomment}
\usepackage{epstopdf}

\usepackage{units}
\usepackage{amsthm}
\usepackage{amsmath}
\usepackage{amssymb}

\usepackage{centernot}

\usepackage{multirow}

\usepackage[caption=false]{subfig}

\global\long\def\ket#1{\left|#1\right\rangle }%
\global\long\def\bra#1{\left\langle #1\right|}%
\global\long\def\braket#1#2{\left\langle #1\right|\left.#2\right\rangle }%

\begin{document}
\title{Intrinsic  sign problem in fermionic and bosonic chiral topological matter}
\author{Omri Golan}
\email{omri.golan@weizmann.ac.il}

\affiliation{Department of Condensed Matter Physics, Weizmann Institute of Science,
Rehovot 76100, Israel}
\author{Adam Smith}
\affiliation{Department of Physics, TFK, Technische Universit{\"a}t M{\"u}nchen, James-Franck-Stra{\ss}le
1, D-85748 Garching, Germany}
\author{Zohar Ringel}
\affiliation{Racah Institute of Physics, The Hebrew University of Jerusalem, Jerusalem
9190401, Israel}
\begin{abstract}
The infamous \textit{sign problem} leads to an exponential complexity
in Monte Carlo simulations of generic many-body quantum systems. Nevertheless,
many phases of matter are known to admit a sign-problem-free representative,
allowing efficient simulations on classical computers. Motivated by
long standing open problems in many-body  physics, as well as fundamental
questions in quantum complexity, the possibility of \textit{intrinsic
sign problems}, where a phase of matter admits no sign-problem-free
representative, was recently raised but remains largely unexplored.
Here, we establish the existence of an intrinsic sign problem in
a broad class of gapped, chiral, topological phases of matter. Within
this class, we exclude the possibility of stoquastic Hamiltonians
for bosons (or 'qudits'), and of sign-problem-free determinantal Monte
Carlo algorithms for fermions. The intrinsically sign-problematic
class of phases we identify is defined in terms of topological invariants
with clear observable signatures: the chiral central charge, and the
topological spins of anyons. We obtain analogous results for phases
that are \textit{spontaneously} chiral, and present evidence for an
extension of  our results that applies to both chiral and non-chiral
topological matter. 
\end{abstract}
\maketitle

\section{Introduction}

Monte Carlo simulations are arguably the most powerful tools for
numerically evaluating thermal averages of classical many-body systems,
by randomly sampling the phase-space  according to the Boltzmann
probability distribution \citep{doi:10.1080/01621459.1949.10483310}.
 Though the phase-space of an $N$-body system scales exponentially
with $N$, a Monte Carlo approximation with a fixed desired error is usually
obtained in polynomial time \citep{troyer2005computational,barahona1982computational}.
In \textit{Quantum} Monte Carlo (QMC), one attempts to perform Monte Carlo
computations of thermal averages in quantum many-body systems, by
following the heuristic idea that quantum systems in $d$ dimensions
are equivalent to classical systems in $d+1$ dimensions \citep{Assaad,li2019sign}.

The difficulty with any such quantum to classical mapping, henceforth
referred to as a \textit{method}, is the infamous \textit{sign problem},
where the mapping can produce complex, rather than non-negative, Boltzmann
weights $p$, which do not correspond to a probability distribution.
  Faced with a sign problem, one can try to change the method used
and obtain $p\geq0$, thus \textit{curing the sign problem} \citep{marvian2018computational,klassen2019hardness}.
Alternatively, one can perform QMC using the weights $\left|p\right|$,
which is often done but generically leads to an exponential computational
complexity in evaluating physical observables, limiting ones ability
to simulate large systems at low temperatures \citep{troyer2005computational}.

Conceptually, the sign problem can be understood as an obstruction
to mapping quantum systems to classical systems, and accordingly,
from a number of complexity theoretic perspectives, a generic curing
algorithm in polynomial time is not believed to exist  \citep{troyer2005computational,bravyi2006complexity,hastings2016quantum,marvian2018computational,hangleiter2019easing,klassen2019hardness}.
In many-body physics, however, one is mostly interested
in universal phenomena, i.e phases of matter and the transitions between
them, and therefore representative Hamiltonians which are sign-free
often suffice \citep{kaul2013bridging}. In fact, QMC simulations
continue to produce unparalleled results, in all branches of many-body
quantum physics, precisely because new sign-free models are constantly
being discovered \citep{RevModPhys.67.279,PhysRevLett.83.3116,PhysRevX.3.031010,PhysRevD.88.021701,kaul2013bridging,GazitE6987,berg2019monte,li2019sign}.

Designing sign-free models requires \textit{design principles} (or
``de-sign'' principles) \citep{kaul2013bridging,wang2015split}
- easily verifiable properties that, if satisfied by a Hamiltonian
and method, lead to a sign-free representation of the corresponding
partition function.   An important example is
the condition $\bra iH\ket j\leq0$ where $i\neq j$ label a local
basis, which implies non-negative weights $p$ in a wide range of
methods  \citep{kaul2013bridging,hangleiter2019easing}.
Hamiltonians satisfying this condition in a given basis are known
as \textit{stoquastic} \citep{bravyi2006complexity}, and have
proven very useful in both application and theory of QMC in bosonic
(or spin, or 'qudit') systems \citep{kaul2013bridging,troyer2005computational,bravyi2006complexity,hastings2016quantum,marvian2018computational,hangleiter2019easing,klassen2019hardness}.

Fermionic Hamiltonians are not expected to be stoquastic in any local
basis \citep{troyer2005computational,li2019sign}, and alternative
methods, collectively known as determinantal quantum Monte-Carlo (DQMC),
are therefore used \citep{PhysRevD.24.2278,Assaad,Santos_2003,li2019sign,berg2019monte}.
 The search for design principles that apply to DQMC, and applications
thereof, has naturally played the dominant role in tackling the sign
problem in fermionic systems, and has seen a lot of progress in recent
years \citep{chandrasekharan2013fermion,wang2015split,li2016majorana,wei2016majorana,wei2017semigroup,berg2019monte,li2019sign}.
Nevertheless, long standing open problems in quantum many-body physics
continue to defy solution, and remain inaccessible for QMC. These
include the nature of high temperature superconductivity and the associated
repulsive Hubbard model  \citep{Santos_2003,PhysRevB.80.075116,PhysRevX.5.041041,kantian2019understanding},
dense nuclear matter and the associated lattice QCD at finite baryon
density \citep{Hands_2000,PhysRevD.66.074507,10.1093/ptep/ptx018},
and the enigmatic fractional quantum Hall state at filling $5/2$
and its associated Coulomb Hamiltonian \citep{banerjee2018observation,PhysRevB.98.045112,PhysRevLett.121.026801,PhysRevB.97.121406,*PhysRevB.98.167401,PhysRevB.99.085309,hu2020microscopic},
all of which are fermionic.

One may wonder if there is a fundamental reason that no design principle
applying to the above open problems has so far been found, despite
intense research efforts. More generally, 
\begin{quote}
\textit{Are there phases of matter which do not admit a sign-free representative?
Are there physical properties that cannot be exhibited by sign-free
models?}
\end{quote}
We refer to such phases of matter, where the sign problem simply cannot
be cured, as having an \textit{intrinsic sign problem} \citep{hastings2016quantum}.
From a practical perspective, intrinsic sign problems may prove useful
in directing research efforts and computational resources. From a
fundamental perspective, intrinsic sign problems identify certain
phases of matter as inherently quantum - their physical properties
cannot be reproduced by a partition function with positive Boltzmann
weights.

To the best of our knowledge, the first intrinsic sign problem was
discovered by Hastings \citep{hastings2016quantum}, who proved that
no stoquastic, commuting projector, Hamiltonians exist for the 'doubled
semion' phase \citep{PhysRevB.71.045110}, which is bosonic and topologically
ordered. In a parallel work \citep{Paper1}, we generalize this result
considerably - excluding the possibility of stoquastic Hamiltonians
in a broad class of bosonic non-chiral topological phases of matter.
Additionally, Reference \citep{ringel2017quantized} demonstrated,
based on the algebraic structure of edge excitations, that no translationally
invariant stoquastic Hamiltonians exist for bosonic chiral topological
phases.

\medskip{}

In this paper, we establish a new criterion for intrinsic sign problems
in chiral topological matter, and take the first step in analyzing
intrinsic sign problems in fermionic systems. First, based on the
well established 'momentum polarization' method for characterizing
chiral topological matter \citep{PhysRevB.88.195412,PhysRevLett.110.236801,PhysRevB.90.045123,PhysRevB.90.115133,PhysRevB.92.165127},
we obtain a variant of the result of Ref.\citep{ringel2017quantized}
- excluding the possibility of stoquastic Hamiltonians in a broad
class of bosonic chiral topological phases. We then develop a formalism
with which we obtain analogous results for systems comprised of both
bosons \textit{and} fermions - excluding the possibility of sign-free
DQMC simulations. 

All of the above mentioned topological phases are gapped, 2+1 dimensional,
and described at low energy by a topological field theory \citep{doi:10.1142/S0129055X90000107,kitaev2006anyons,freed2016reflection}.
The class of such phases in which we find an intrinsic sign problem
is defined in terms of robust data characterizing them: the chiral
central charge $c$, a rational number, as well as the set $\left\{ \theta_{a}\right\} $
of topological spins of anyons, a subset of roots of unity. Namely,
we find that 

\begin{quote}
\textit{An intrinsic sign problem exists if $e^{2\pi ic/24}$ is not
the topological spin of some anyon, i.e $e^{2\pi ic/24}\notin\left\{ \theta_{a}\right\} $.
}
\end{quote}
The above criterion applies to 'most' chiral topological phases, see
Table \ref{tab:1} for examples.  In particular, we identify an intrinsic
sign problem in $96.7\%$ of the first one-thousand fermionic Laughlin
phases, in all chiral triplet superconductors, and in the three non-abelian candidate phases for the quantum Hall state at filling 5/2. We also find
intrinsic sign problems in $91.6\%$ of the first one-thousand $SU\left(2\right)_{k}$
Chern-Simons theories. Since, for $k\neq1,2,4$,  these allow for
universal quantum computation by manipulation of anyons \citep{Freedman:2002aa,nayak2008non},
our results support the strong belief that quantum computation cannot
be simulated with classical resources, in polynomial time \citep{Arute:2019aa}.
This conclusion is strengthened by examining the Fibonacci anyon model,
which is known to be universal for quantum computation \citep{nayak2008non},
and is found to be intrinsically sign-problematic. 

We stress that both $c$ and $\left\{ \theta_{a}\right\} $ have clear
observable signatures in both the bulk and boundary of chiral topological
matter, some of which have been experimentally observed. The chiral
central charge controls the boundary thermal Hall conductance  \citep{kane1997quantized,read2000paired,cappelli2002thermal,kapustin2019thermal},
 which was recently measured in quantum Hall and spin systems \citep{jezouin2013quantum,banerjee2017observed,banerjee2018observation,Kasahara:2018aa}.
In the bulk it is predicted to contribute to the Hall (or odd) viscosity
at finite wave-vector, as well as in curved background \citep{abanov2014electromagnetic,klevtsov2015geometric,bradlyn2015topological,PhysRevB.100.104512}.
The chiral central charge also affects the angular momentum at conical
defects \citep{can2016emergent}, as was recently observed in an optical
realization of integer quantum Hall states \citep{schine2016synthetic,schine2018measuring}.
The topological spins determine the exchange statistics of anyons, predicted to appear in interferometry experiments \citep{nayak2008non},
though experimental observation remains elusive \citep{PhysRevLett.122.246801}. A measurement of anyonic statistics via current correlations \citep{PhysRevLett.116.156802} was recently reported 
 in the Laughlin 1/3 quantum Hall state \citep{Bartolomei173}.

\medskip{}

The paper is organized as follows. In Sec.\ref{sec:Signs-from-geometric}
we collect relevant facts regarding chiral topological matter,
arriving at the 'momentum polarization' Eq.\eqref{eq:12-3-1}. Section \ref{sec:No-stoquastic-Hamiltonians}
then obtains Result \hyperref[Result 1]{1} - an intrinsic sign problem
in bosonic chiral topological matter. In Sec.\ref{sec:Spontaneous-chirality}
we perform a similar analysis for the case where chirality (or time
reversal symmetry breaking), appears spontaneously rather than explicitly,
arriving at Result \hyperref[Result 2]{2}. We then turn to fermionic
systems. In Sec.\ref{sec:Determinantal-quantum-Monte} we develop
a formalism which unifies and generalizes the currently used DQMC
algorithms, and the corresponding design principles. In Sec.\ref{sec:No-sign-free-DQMC},
we obtain within this formalism Result \hyperref[Result 1F]{1F} and Result
\hyperref[Result 2F]{2F}, the fermionic analogs of Results \hyperref[Result 1]{1}
and \hyperref[Result 2]{2}. Section \ref{sec:Generalization-and-extension}
describes a conjectured extension of our results that applies beyond
chiral phases, and unifies them with the intrinsic sign problems found
in our parallel work \citep{Paper1}. In Sec.\ref{sec:Discussion-and-outlook}
we discuss our results and provide an outlook for future work.

\begin{table*}[t]

\caption{Examples of intrinsic sign problems based on the criterion $e^{2\pi ic/24}\protect\notin\left\{ \theta_{a}\right\} $, in terms of the chiral central charge $c$ and the topological spins $\theta_{a}=e^{2\pi ih_{a}}$. The number of spins $h_a$ is equal to the dimension of the ground state subspace on the torus. We mark bosonic/fermionic phases by (B/F). The quantum Hall Laughlin phases corresond to $U(1)_q$ Chern-Simons theories. The $\ell$-wave superconductor is chiral, e.g $p+ip$ for $\ell=1$, and spinless. Data for the spinfull case is identical to that of the Chern insulator, with $-\ell$ odd (even) in place of $\nu$, for triplet (singlet) pairing. The modulo 8 ambiguity in the central charge of the Fibonacci anyon model corresponds to the stacking of a given realization  with copies of the $E_{8}$ $K$-matrix phase. Data for the three quantum Hall Pfaffian phases is given at the minimal filling 1/2. The physical filling 5/2 is obtained by stacking with a $\nu=2$ Chern insulator, and an intrinsic sign problem appears in this case as well. \label{tab:1} }

\begin{ruledtabular} 
\renewcommand*{\arraystretch}{1.3}

\begin{tabular}{lllll} \textbf{Phase of matter} & \textbf{Parameterization} & $\boldsymbol{c}$  & $\boldsymbol{\left\{h_{a}\right\}}$ & \textbf{Intrinsic sign problem?}

\tabularnewline Laughlin (B) \citep{hu2020microscopic} & Filling $1/q,\;(q\in2\mathbb{N})$ & $1$ & $\left\{ a^{2}/2q\right\} _{a=0}^{q-1}$ & In $98.5\%$ of first $10^3$

\tabularnewline Laughlin (F) \citep{hu2020microscopic} & Filling $1/q,\;(q\in2\mathbb{N}-1)$ & $1$ & $\left\{ \left(a+1/2\right)^{2}/2q\right\} _{a=0}^{q-1}$ & In $96.7\%$ of first $10^3$

\tabularnewline Chern insulator (F) {[}App.\ref{subsec:Beyond-the-assumption}{]} & Chern number $\nu\in\mathbb{Z}$ & $\nu$ & $\left\{ \nu/8\right\} $ & For $\nu\notin 12\mathbb{Z}$

\tabularnewline $\ell$-wave superconductor (F) \citep{PhysRevB.100.104512} & Pairing channel $\ell\in2\mathbb{Z}-1$ & $-\ell/2$ & $\left\{ -\ell/16\right\} $ & Yes

\tabularnewline Kitaev spin liquid (B) \citep{kitaev2006anyons} & Chern number $\nu\in2\mathbb{Z}-1$ & $\nu/2$ & $\left\{ 0,1/2,\nu/16\right\} $ & Yes

\tabularnewline $SU\left(2\right)_{k}$ Chern-Simons (B) \citep{bonderson2007non} & Level $k\in\mathbb{N}$ & $3k/\left(k+2\right)$ & $\left\{ a\left(a+2\right)/4\left(k+2\right)\right\} _{a=0}^{k}$ & In  $91.6\%$ of first $10^3$

\tabularnewline $E_{8}$ $K$-matrix (B) \citep{PhysRevB.94.155113} & Stack of $n\in\mathbb{N}$ copies& $8n$ & $\left\{ 0\right\} $ & For $n\notin 3\mathbb{N}$

\tabularnewline Fibonacci anyon model (B) \citep{bonderson2007non} &  & $14/5$ (mod $8$)& $\left\{ 0,2/5\right\} $ & Yes

\tabularnewline Pfaffian (F) \citep{hsin2020effective} &  & $3/2$ & $\left\{   0,1/2,1/4,3/4,1/8,5/8     \right\}$ & Yes

\tabularnewline PH-Pfaffian (F) \citep{hsin2020effective} &  & $1/2$ & $\left\{   0,0,1/2,1/2,1/4,3/4     \right\}$ & Yes

\tabularnewline Anti-Pfaffian (F) \citep{hsin2020effective} &  & $-1/2$ & $\left\{   0,1/2,1/4,3/4,3/8,7/8     \right\}$ & Yes

\tabularnewline \end{tabular}

 \end{ruledtabular} \end{table*}

\section{Chiral topological matter, and signs from geometric manipulations
\label{sec:Signs-from-geometric}}

In this section we review the necessary details regarding chiral topological
phases of matter, and, following Refs.\citep{PhysRevB.88.195412,PhysRevLett.110.236801,PhysRevB.90.045123,PhysRevB.90.115133,PhysRevB.92.165127},
obtain the 'momentum polarization' Eq.\eqref{eq:12-3-1}, which is
the main tool we use in the remainder of this paper. Later on we will
show that gapped bosonic and fermionic Hamiltonians which are sign-free
(due to the an appropriate design principle) cannot obey the aforementioned Eq.\eqref{eq:12-3-1},
unless the chiral central charge and topological spins obey a tight
constraint.

\subsection{Chiral topological matter}

A gapped Hamiltonian is said to be in a topological phase of matter
if it cannot be deformed to a trivial reference Hamiltonian, without
closing the gap. If a symmetry is enforced, only symmetric deformations
are considered, and it is additionally required that the symmetry
is not spontaneously broken \citep{wang2017symmetric,zeng2019quantum}.
For Hamiltonians defined on a lattice, as considered in this paper,
a natural trivial Hamiltonian is given by the atomic limit of decoupled
lattice sites, where the symmetry acts independently on each site.

Topological phases with a unique ground state on the 2-dimensional
torus exist only with a prescribed symmetry group,
and are termed symmetry protected topological phases (SPTs) \citep{chen2011symmetry,Kapustin_2015,Kapustin_2017}.
When such phases are placed on the cylinder, they support anomalous
boundary degrees of freedom which cannot be realized on isolated 1-dimensional
spatial manifolds, as well as corresponding quantized bulk response
coefficients. Notable examples are the integer quantum Hall states,
topological insulators, and topological superconductors \citep{qi2011topological}.

Topological phases with a degenerate  ground state subspace  on
the torus are termed topologically ordered, or symmetry enriched if
a symmetry is enforced \citep{doi:10.1142/S0217979290000139,PhysRevB.82.155138}.
Beyond the phenomena exhibited by SPTs, these support localized quasiparticle
excitations with anyonic statistics and fractional charge under the
symmetry group. Notable examples are fractional quantum Hall states
\citep{nayak2008non,PhysRevLett.110.067208}, quantum spin liquids
\citep{Savary_2016}, and fractional topological insulators \citep{PhysRevLett.103.196803,doi:10.1146/annurev-conmatphys-031115-011559}.

\begin{figure}[t]
\begin{centering}
\includegraphics[width=1\columnwidth]{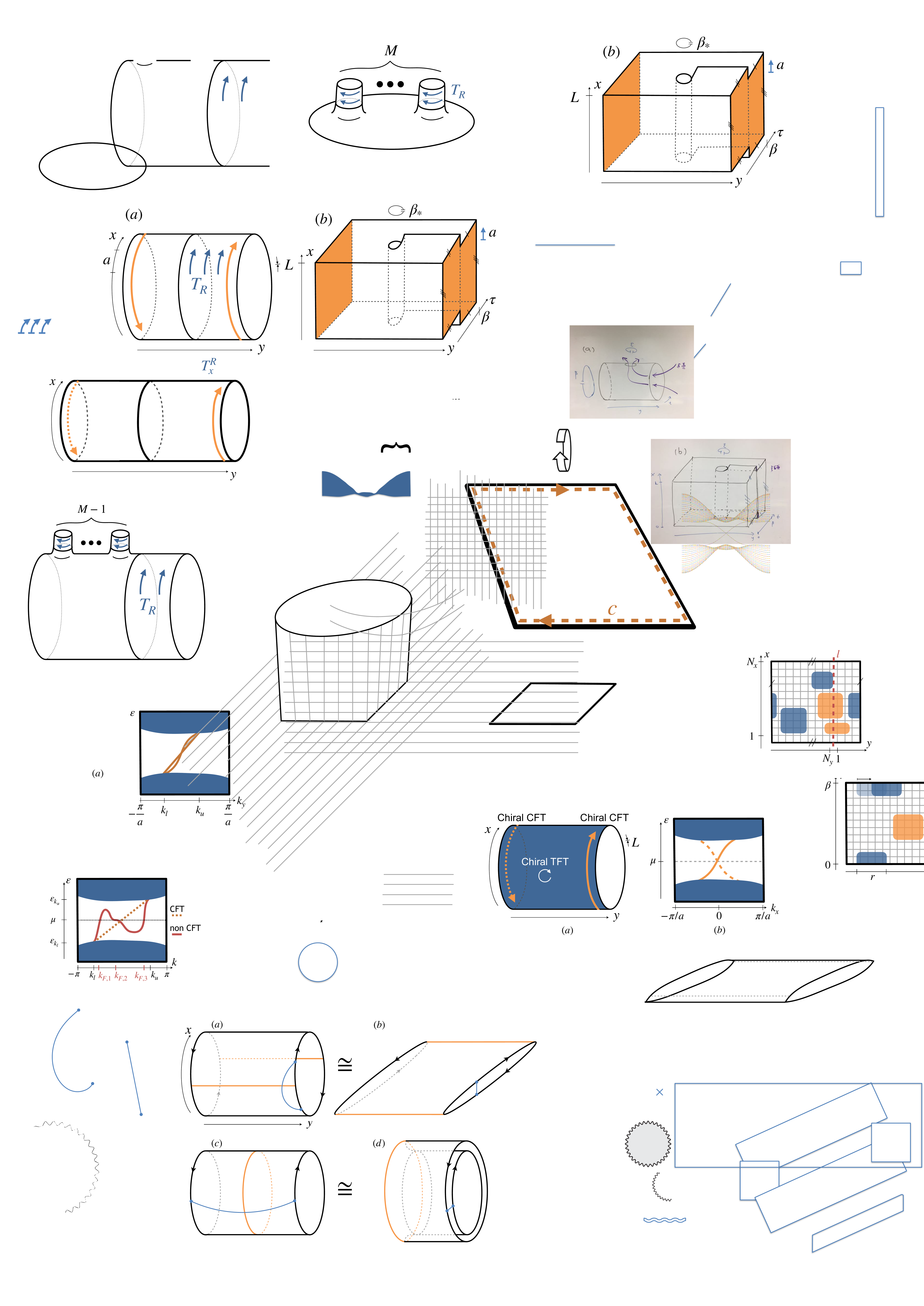}
\par\end{centering}
\caption{Chiral topological phases of matter on the cylinder. (a) The low energy
description of a chiral topological phase is comprised of two, counter
propagating, chiral conformal field theories (CFTs) on the boundary,
and a chiral topological field theory (TFT) in the bulk.   (b)
Examples: schematic single-particle spectrum of a Chern insulator
and of the Majorana fermions describing the Kitaev spin liquid. Assuming
discrete translational symmetry with spacing $a$ in the $x$ direction,
one can plot the single-particle eigen-energies $\varepsilon$ on
the cylinder as a function of (quasi) momentum $k_{x}$.  This reveals
an integer number of chiral dispersion branches whose eigen-states
are supported on one of the two boundary components. In the Chern
insulator (Kitaev spin liquid) these correspond to the Weyl (Majorana-Weyl)
fermion CFT, with $c=\pm1$ ($c=\pm1/2$) per branch. The velocity,
$v=\left|\partial\varepsilon/\partial k_{x}\right|$ at the chemical
potential $\mu$, is a non-universal parameter. \label{fig:Chiral-topological-phases}
}
\end{figure}

In this work we consider \textit{chiral} topological phases, where
the boundary degrees of freedom that appear on the cylinder propagate
unidirectionally. At energies small compared with the bulk gap, the
boundary can be described by a chiral conformal field theory (CFT)
\citep{ginsparg1988applied,di1996conformal}, while the bulk reduces
to a chiral topological field theory (TFT) \citep{kitaev2006anyons,freed2016reflection},
see Fig.\ref{fig:Chiral-topological-phases}(a). We consider both
bosonic and fermionic phases. These may be protected or enriched by
an on-site symmetry, but we will not make use of this symmetry in
our analysis - only the chirality of the phase will be used. 

A notable example for chiral topological phases is given by Chern
insulators \citep{PhysRevLett.61.2015,qi2008topological,ryu2010topological}:
SPTs protected by the $U\left(1\right)$ fermion number symmetry,
which admit free-fermion Hamiltonians. The single particle spectrum
of a Chern insulator on the cylinder is depicted in Fig.\ref{fig:Chiral-topological-phases}(b).
Another notable example is the topologically ordered Kitaev spin liquid
\citep{kitaev2006anyons,takagi2019concept}, which can be described
by Majorana fermions with a single particle spectrum similar to Fig.\ref{fig:Chiral-topological-phases}(b),
coupled to a $\mathbb{Z}_{2}$ gauge field.

 Note that the velocity $v$ of the boundary CFT is a non-universal
parameter which generically changes as the microscopic Hamiltonian
is deformed.  Furthermore, different chiral branches may have
different velocities.

The chirality of the boundary CFT and bulk TFT is manifested by their
non-vanishing chiral central charge $c$, which is rational and
\textit{universal} - it is a topological invariant with respect to
continuous deformations of the Hamiltonian which preserve the bulk
energy gap, and therefore constant throughout a topological phase
\citep{Witten_1989,kitaev2006anyons,gromov2015framing,bradlyn2015topological}.
On the boundary $c$ is defined with respect to an orientation of
the cylinder, so the two boundary components have opposite chiral
central charges. 

\subsection{Boundary finite size corrections\label{subsec:Boundary-finite-size}}

The non-vanishing of $c$ implies a number of geometric, or 'gravitational',
physical phenomena \citep{ginsparg1988applied,di1996conformal,abanov2014electromagnetic,klevtsov2015geometric,bradlyn2015topological,gromov2016boundary,PhysRevB.98.064503,PhysRevB.100.104512}.
In particular, the boundary supports a non-vanishing energy current
$J_{E}$, which receives a correction 
\begin{align}
 & J_{E}\left(T\right)=J_{E}\left(0\right)+2\pi T^{2}\frac{c}{24},\label{eq:1-2}
\end{align}
at a temperature $T>0$, and in the thermodynamic
limit $L=\infty$, where $L$ is the circumference of the cylinder.
Note that we set $K_{\text{B}}=1$ and $\hbar=1$ throughout. Within
CFT, this correction is universal since it is independent of $v$.
Taking the two counter propagating boundary components of the cylinder
into account, and placing these at slightly different temperatures,
leads to a thermal Hall conductance $K_{H}=c\pi T/6$ \citep{kane1997quantized,read2000paired,cappelli2002thermal},
a prediction that recently led to the first measurements of $c$ 
\citep{jezouin2013quantum,banerjee2017observed,banerjee2018observation,Kasahara:2018aa}.

In analogy with Eq.\eqref{eq:1-2}, the boundary of a chiral topological
phase also supports a non-vanishing ground state (at $T=0$) momentum
density $p$, which receives a universal correction on a cylinder
with finite circumference $L<\infty$, 
\begin{align}
 & p\left(L\right)=p\left(\infty\right)+\frac{2\pi}{L^{2}}\left(h_{0}-\frac{c}{24}\right).\label{eq:2-2}
\end{align}

Equation \eqref{eq:2-2} is the main property of chiral topological
matter that we use below, so we discuss it in detail. First, the rational
number $h_{0}$ is a \textit{chiral} conformal weight, which is an
additional piece of data characterizing the boundary CFT. Like the
chiral central charge, the two boundary components of the cylinder
have opposite $h_{0}$s.  From the bulk TFT perspective, $h_{0}$
corresponds to the topological spin of an anyon quasi-particle, defined
by the phase $\theta_{0}=e^{2\pi ih_{0}}$ accumulated as the anyon
undergoes a $2\pi$ rotation \citep{kitaev2006anyons}. The set $\left\{ \theta_{a}\right\} _{a=1}^{N}$
of topological spins of anyons is associated with the $N$-dimensional
ground state subspace on the torus, and the unique $\theta_{0}=e^{2\pi ih_{0}}$
defined by \eqref{eq:2-2} corresponds to the generically unique ground
state on the cylinder, with a finite-size energy separation $\sim1/L$
from the low lying excited states, see Appendix  \ref{subsec:Further-details-regarding}. 

As the equation $\theta_{0}=e^{2\pi ih_{0}}$ suggests, only $h_{0}\mod1$
is universal for a topological phase. The integer part of $h_{0}$
can change as the Hamiltonian is deformed on the cylinder, while maintaining
the bulk gap, and even as a function of $L$ for a fixed Hamiltonian.
Additionally, the choice of $\theta_{0}$ from the set $\left\{ \theta_{a}\right\} $
is non-universal, and can change due to bulk gap preserving deformations,
or as a function of $L$. Both types of discontinuous jumps in $h_{0}$
may be accompanied by an accidental degeneracy of the ground state
on the cylinder. Therefore, the universal and $L$-independent statement
regarding $h_{0}$ is that, apart from accidental degeneracies, $e^{2\pi ih_{0}}=\theta_{0}\in\left\{ \theta_{a}\right\} $
- a fact that will be important in our analysis. 

The non-trivial behavior of $h_{0}$ described above appears when
the boundary corresponds to a non-conformal deformation of a CFT,
by e.g a chemical potential. As demonstrated analytically and numerically
in Appendix \ref{subsec:Beyond-the-assumption}, such behavior appears
already in the simple context of Chern insulators with non-zero Fermi
momenta, as would be the case in Fig.\ref{fig:Chiral-topological-phases}(b)
if the chemical potential $\mu$ is either raised or lowered. 

\subsection{Momentum polarization\label{subsec:Momentum-polarization}}

In this section we describe a procedure for the extraction of $h_{0}-c/24$
in Eq.\eqref{eq:2-2}, given a lattice Hamiltonian on the cylinder.
Since the two boundary components carry opposite momentum densities,
the ground state on the cylinder does not carry a total momentum,
only a 'momentum polarization'. It is therefore clear that some sort
of one-sided translation will be required.

\begin{figure}[t]
\begin{centering}
\includegraphics[width=1\columnwidth]{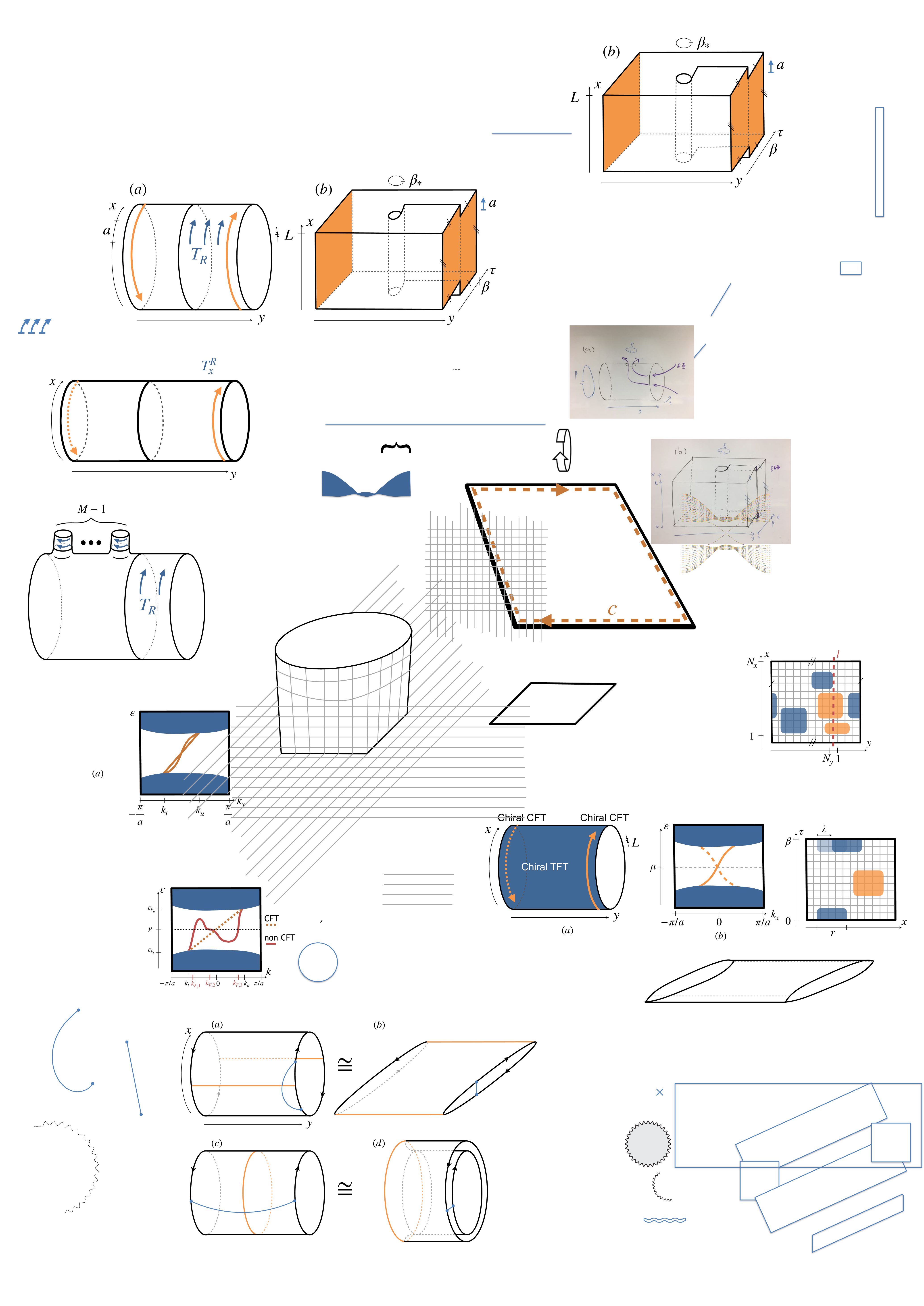}
\par\end{centering}
\caption{Momentum polarization. (a) Hamiltonian, or spatial, point of view.
The operator $T_{R}$ translates the right half of the cylinder by
one unit cell, a distance $a$, in the $x$ direction. It acts as
the identity on the left boundary component, and as a translation
on the right boundary component. The object $\tilde{Z}/Z$ is the
thermal expectation value of $T_{R}$. (b) Field theory, or space-time,
point of view. The object $\tilde{Z}$ is the partition function on
a space-time carrying a screw dislocation. The space-time region occupied
by the boundary components of the spatial cylinder is colored in orange.
The screw dislocation can be described as an additional boundary component,
on which $T_{R}$ acts as a translation, with a high effective temperature
$1/\beta_{*}$. \label{fig:Defect} }
\end{figure}
 Following Ref.\citep{PhysRevB.88.195412}, we define $\tilde{Z}:=\text{Tr}\left(T_{R}e^{-\beta H}\right)$,
which is related to the usual partition function $Z=\text{Tr}\left(e^{-\beta H}\right)$
($\beta=1/T$), by the insertion of the operator $T_{R}$, which translates
the right half of the cylinder by one unit cell in the periodic $x$
direction, see Fig.\ref{fig:Defect}(a). The object $\tilde{Z}$ satisfies
\begin{align}
\tilde{Z} & =Z\exp\left[\alpha N_{x}+\frac{2\pi i}{N_{x}}\left(h_{0}-\frac{c}{24}\right)+o\left(N_{x}^{-1}\right)\right],\label{eq:12-3-1}
\end{align}
where $N_{x}$ is the number of sites in the $x$ direction, $\alpha\in\mathbb{C}$
is non-universal and has a negative real part, and $o\left(N_{x}^{-1}\right)$
indicates corrections that decay faster than $N_{x}^{-1}$ as $N_{x}\rightarrow\infty$.
 Equation \eqref{eq:12-3-1} is valid at temperatures
low compared to the finite-size energy differences on the boundary,
$\beta^{-1}=o\left(N_{x}^{-1}\right)$, see Appendix \ref{subsec:Further-details-regarding}.

Equation \eqref{eq:12-3-1} follows analytically from the low energy
description of chiral topological matter in terms of chiral TFT and
CFT \citep{PhysRevB.88.195412}, and was numerically scrutinized
in a large number of examples in Refs.\citep{PhysRevB.88.195412,PhysRevLett.110.236801,PhysRevB.90.045123,PhysRevB.90.115133,PhysRevB.92.165127},
as well as in Appendix \ref{subsec:Beyond-the-assumption}. Nevertheless,
we are not aware of a rigorous proof of Eq.\eqref{eq:12-3-1} for
gapped lattice Hamiltonians. Therefore, in stating our results we
will use the assumption 'the Hamiltonian $H$ is in a chiral topological
phase of matter', the content of which is that $H$ admits a low energy
description in terms of a chiral TFT with chiral central charge $c$
and topological spins $\left\{ \theta_{a}\right\} $, and in particular,
Eq.\eqref{eq:12-3-1} holds for any bulk-gap preserving deformation
of $H$, with $e^{2\pi ih_{0}}\in\left\{ \theta_{a}\right\} $, apart
from accidental degeneracies on the cylinder.
In the remainder of this section we further discuss the content of
Eq.\eqref{eq:12-3-1} and its expected range of validity, in light
of the Hamiltonian and space-time interpretations of $\tilde{Z}$. 

From a Hamiltonian perspective, $\tilde{Z}/Z$ is the thermal expectation
value of $T_{R}$, evaluated at a temperature $\beta^{-1}$ low enough
to isolate the ground state. The exponential decay expressed in Eq.\eqref{eq:12-3-1}
appears because $T_{R}$ is not a symmetry of $H$, and $-\text{Re}\left(\alpha\right)$
can be understood as the energy density of the line defect where $T_{R}$
is discontinuous, see Fig.\ref{fig:Defect}(a). In fact, we expect Eq.\eqref{eq:12-3-1} to hold
irrespective of whether the \textit{uniform} translation is a symmetry
of $H$, or of the underlying 'lattice' on which $H$ is defined,
which may be any polygonalization of the cylinder (see Ref.\citep{PhysRevLett.115.036802}  for a similar scenario). The only expected requirement is that  the low energy description of $H$ is homogeneous. Furthermore, if Eq.\eqref{eq:12-3-1} only holds after a disorder
averaging of $\tilde{Z}/Z$, our results and derivations in the following sections remain unchanged.

There is also a simple space-time interpretation of $\tilde{Z}$,
which will be useful in the context of DQMC. The usual partition function
$Z=\text{Tr}\left(e^{-\beta H}\right)$ has a functional integral
representation in terms of bosonic fields $\phi$ (fermionic fields
$\psi$) defined on space, the cylinder $C$ in our case, and the
imaginary time circle $S_{\beta}^{1}=\mathbb{R}/\beta\mathbb{Z}$,
with periodic (anti-periodic) boundary conditions \citep{altland2010condensed}.
In $\tilde{Z}=\text{Tr}\left(T_{R}e^{-\beta H}\right)$, the insertion
of $T_{R}$ produces a twisting of the boundary conditions of $\phi,\psi$
in the time direction, such that $\tilde{Z}$ is the partition function
on a space-time carrying a screw dislocation, see Fig.\ref{fig:Defect}(b).

The above interpretation of $\tilde{Z}$, supplemented by Eq.\eqref{eq:2-2},
allows for an intuitive explanation of Eq.\eqref{eq:12-3-1}, which
loosely follows its analytic derivation \citep{PhysRevB.88.195412}.
As seen in Fig.\ref{fig:Defect}(b), the line where $T_{R}$ is discontinuous
can be interpreted as an additional boundary component at a high effective
temperature, $\beta_{*}\ll L/v$. Since the effective temperature
is much larger than the finite size energy-differences $2\pi v/L$
on the boundary CFT, the screw dislocation contributes no finite size
corrections to $\tilde{Z}$. This leaves only the contribution of
the boundary component on the right side of the cylinder, where $T_{R}$
produces the phase $e^{iaLp\left(L\right)}$, assuming $\beta_{*}\ll L/v\ll\beta$.
Equation \eqref{eq:2-2} then leads to the universal finite size correction
$\left(2\pi i/N_{x}\right)\left(h_{0}-c/24\right)$.

\section{Excluding stoquastic Hamiltonians for chiral topological matter\label{sec:No-stoquastic-Hamiltonians}}

In this section we consider bosonic (or 'qudit', or spin) systems,
and a single design principle  - existence of a local basis in which
the many-body Hamiltonian is stoquastic. A sketch of the derivation
of Result \hyperref[Result 1]{1} is that the momentum polarization $\tilde{Z}$
is positive for Hamiltonians $H'$ which are stoquastic in an on-site
and homogenous basis, and this implies that $\theta_{0}=e^{2\pi ic/24}$
for any Hamiltonian $H$ obtained from $H'$ by conjugation with a
local unitary. 

\subsection{Setup\label{subsec:Setup}}

The many body Hilbert space is given by $\mathcal{H}=\otimes_{\mathbf{x}\in X}\mathcal{H}_{\mathbf{x}}$,
where the tensor product runs over the sites $\mathbf{x}=\left(x,y\right)$
of a 2-dimensional lattice $X$, and $\mathcal{H}_{\mathbf{x}}$ are
on-site 'qudit' Hilbert spaces of finite dimension $\mathsf{d}\in\mathbb{N}$.
With finite-size QMC simulations in mind, we consider a square lattice
with spacing 1, $N_{x}\times N_{y}$ sites, and periodic boundary
conditions, so that $X=\mathbb{Z}_{N_{x}}\times\mathbb{Z}_{N_{y}}$
is a discretization of the flat torus $\left(\mathbb{R}/N_{x}\mathbb{Z}\right)\times\left(\mathbb{R}/N_{y}\mathbb{Z}\right)$.
Generalization to other 2-dimensional lattices is straight forward.
On this Hilbert space a gapped $r$-local Hamiltonian $H=\sum_{\mathbf{x}}H_{\mathbf{x}}$, which is in a chiral topological phase of matter,
is assumed to be given. Here, the terms $H_{\mathbf{x}}$ are supported
within a range $r$ of $\mathbf{x}$ - they are defined on $\otimes_{\left|\mathbf{y}-\mathbf{x}\right|\leq r}\mathcal{H}_{\mathbf{y}}$
and act as $0$ on all other qudits.

Fix an tensor product basis $\ket s=\otimes_{\mathbf{x}\in X}\ket{s_{\mathbf{x}}}$,
labeled by strings $s=\left(s_{\mathbf{x}}\right)_{\mathbf{x}\in X}$,
where $s_{\mathbf{x}}\in\left\{ 1,\cdots,\mathsf{d}\right\} $ labels
a basis $\ket{s_{\mathbf{x}}}$ for $\mathcal{H}_{\mathbf{x}}$. For
any vector $\mathbf{d}\in X$, the corresponding translation operator
$T^{\mathbf{d}}$ is defined in this basis, $T^{\mathbf{d}}\ket s=\ket{t^{\mathbf{d}}s}$,
with $\left(t^{\mathbf{d}}s\right)_{\mathbf{x}}=s_{\mathbf{x}+\mathbf{d}}$.
These statements assert that $\ket s$ is both an on-site and a homogeneous
basis, or \textit{on-site homogeneous }for short. Note that $T^{\mathbf{d}}$
acts as a permutation matrix on the $\ket s$s, and in particular,
has non-negative matrix elements in this basis.

In accordance with Sec.\ref{subsec:Momentum-polarization}, we  assume that the low energy description of $H$ is invariant under    $T^{\mathbf{d}}$, as defined above.   In doing so, we exclude the possibility of generic background gauge fields  for any on-site symmetry that $H$ may posses, which is beyond the scope of this work. Nevertheless, commonly used background gauge fields,  such as those  corresponding to uniform magnetic fields with rational flux per plaquette, can easily be incorporated into our analysis, by restricting to translation vectors $\mathbf{d}$ in a sub-lattice of $X$. A restriction to  sub-lattice translations can also be used to guarantee that  $T^{\mathbf{d}}$
acts purely as a translation in the low energy TQFT description. In
particular, a lattice translation may permute the anyon types $a$ \footnote{We thank Michael Levin for pointing out this phenomenon.}. Since the number of anyons is finite, restricting to large enough translations
will eliminate this effect. An example is given by Wen's plaquette
model, where different anyons are localized
on the even/odd sites of a bipartite lattice \citep{PhysRevB.87.184402}, and a restriction to translations that maps the even (odd) sites to themselves will be made.

Finally, it is  assumed that $H$ is \textit{locally stoquastic}: it is term-wise
stoquastic in a local basis. This means that a local unitary operator
$U$ exists, such that the conjugated Hamiltonian $H'=UHU^{\dagger}$
 is a sum of local terms $H_{\mathbf{x}}'=UH_{\mathbf{x}}U^{\dagger}$,
which have non-positive matrix elements in the on-site homogeneous
basis, $\bra sH_{\mathbf{x}}'\ket{\tilde{s}}\leq0$ for all basis
states $\ket s,\ket{\tilde{s}}$. Note that we include the diagonal
matrix elements in the definition, without loss of generality. 

The term \textit{local unitary} used above refers to a depth-$D$
quantum circuit, a product $U=U_{D}\cdots U_{1}$ where each $U_{i}$
is itself a product of unitary operators with non-overlapping supports\footnote{The support of a unitary $u=e^{ih}$ is the support of its Hermitian
generator $h$.} of diameter $w$. It follows that $H'$ has a range $r'=r+2r_{U}$,
where $r_{U}=Dw$ is the range of $U$. Equivalently, we may take
$U$ to be a finite-time evolution with respect to an $\tilde{r}$-local,
smoothly time-dependent, Hamiltonian $\tilde{H}\left(t\right)$, given
by the time-ordered exponential $U=\text{TO}e^{-i\int_{0}^{1}\tilde{H}\left(t\right)dt}$.
The two types of locality requirements are equivalent, as finite-time
evolutions can be efficiently approximated by finite-depth circuits,
while finite-depth circuits can be written as finite-time evolutions
over time $D$ with piecewise constant $w$-local Hamiltonians \citep{Lloyd1073,zeng2019quantum}.

\subsection{Constraining $c$ and $\left\{ \theta_{a}\right\} $}

\begin{figure}[t]
\begin{centering}
\includegraphics[width=0.45\columnwidth]{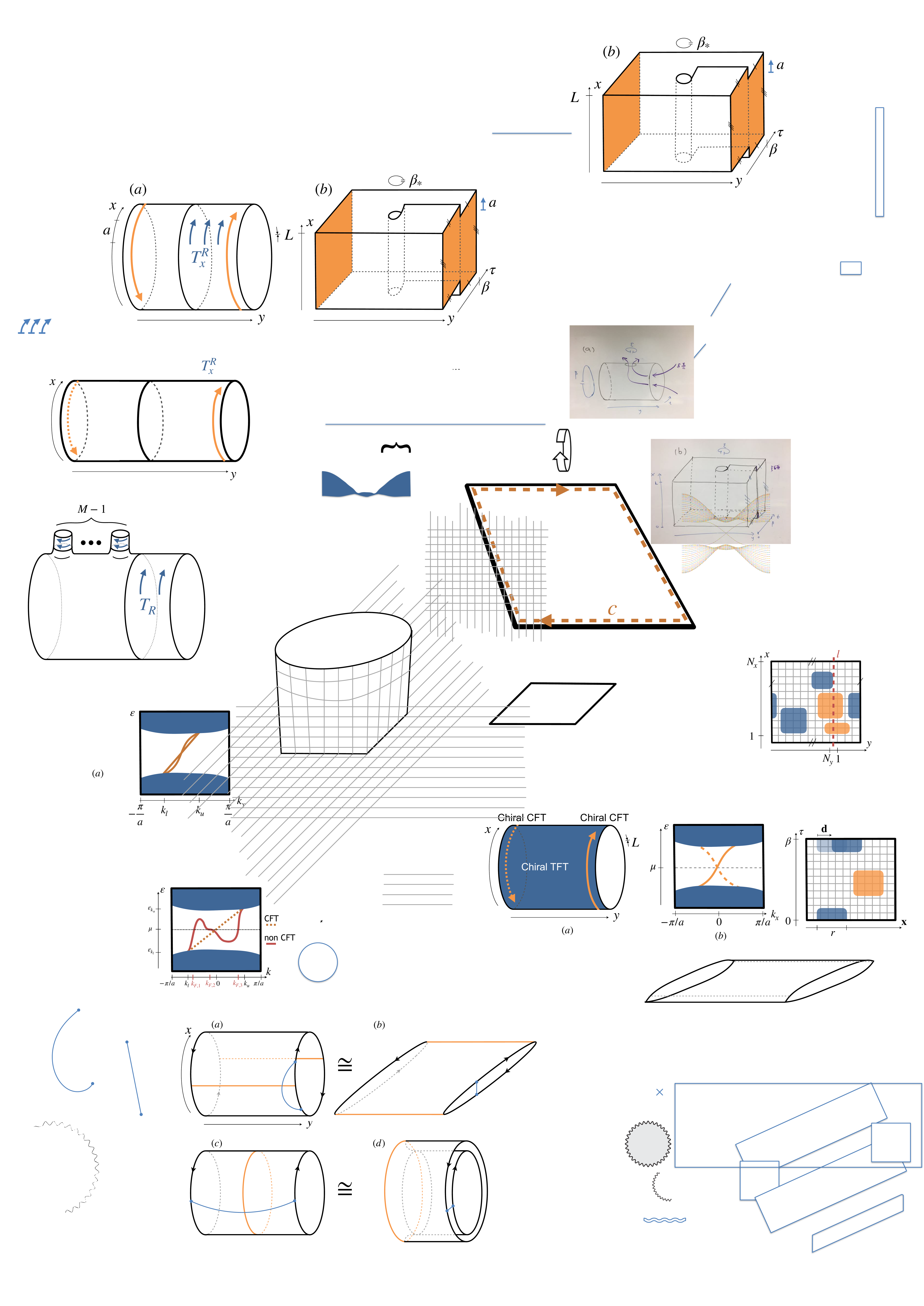}
\par\end{centering}
\caption{Cutting the torus to a cylinder along the line $l$. Orange areas
mark the supports of Hamiltonian terms $H_{\mathbf{x}}'$ which are
removed from $H'$, while blue areas mark the supports of terms which
are kept. \label{fig:cutting} }
\end{figure}

In order to discuss the momentum polarization, we need to map the
stoquastic Hamiltonian $H'$ from the torus $X$ to a cylinder $C$.
This is done by choosing a translation vector $\mathbf{d}\in X$,
and then cutting the torus $X$ along a line $l$ parallel to $\mathbf{d}$. To simplify
the presentation we restrict attention to the case $\mathbf{d}=\left(1,0\right)$.
All other cases amount to a lattice-spacing redefinition, see Appendix
\ref{subsec:Cutting-the-torus-2}. The cylinder $C=\mathbb{Z}_{N_{x}}\times\left\{ 1,\dots,N_{y}\right\} $
is then obtained from the torus $X=\mathbb{Z}_{N_{x}}\times\mathbb{Z}_{N_{y}}$
by cutting along the line $l=\left\{ \left(i,1/2\right):\;i\in\mathbb{Z}_{N_{x}}\right\} $.
A stoquastic Hamiltonian on the cylinder can be obtained from that
on the torus by removing all local terms $H'_{\mathbf{x}}$ whose
support overlaps $l$, see Fig.\ref{fig:cutting}. Note that this
procedure may render $H'$ acting as $0$ on certain qudits $\mathcal{H}_{\mathbf{x}}$
with $\mathbf{x}$ within a range $r'$ of $l$, but this does not
bother us. Since all terms $H_{\mathbf{x}}'$ are individually stoquastic,
this procedure leaves $H'$, now defined on the cylinder, stoquastic.
One can similarly map $H$ and $U$ to the cylinder $C$ such that
the relation $H'=UHU^{\dagger}$ remains valid on $C$.

Let us now make contact with the momentum polarization Eq.\eqref{eq:12-3-1}.
Having mapped $H'$ to the cylinder, we consider the 'partition function'
\begin{align}
\tilde{Z}' & :=\text{Tr}\left(e^{-\beta H'}T_{R}\right),\label{eq:4}
\end{align}
where $T_{R}=T_{R}^{\mathbf{d}}$ is defined by $T_{R}\ket s=\ket{T_{R}s}$, $\left(T_{R}s\right)_{x,y}=s_{x+\Theta\left(y\right),y}$, and $\Theta$ is a heavy side function supported on the right half
of the cylinder. Though $\tilde{Z}'$ is generally different from
$\tilde{Z}=\text{Tr}\left(e^{-\beta H}T_{R}\right)$ appearing in
Eq.\eqref{eq:12-3-1}, it satisfies two useful properties:
\begin{enumerate}
\item $\tilde{Z}'>0$. Both $-H'$ and $T_{R}$ have non-negative entries
in the on-site basis $\ket s$, and therefore so does $e^{-\beta H'}T_{R}$.
\item $\tilde{Z}'$ satisfies Eq.\eqref{eq:12-3-1}, with non-universal $\alpha'$ in place of $\alpha$, but $c'=c$, and $h_0'\in \left\{\theta_a\right\}$. This follows from the fact that $H'=UHU^{\dagger}$ is in the same phase of matter as $H$, and therefore $c'=c$, and $\left\{ \theta_{a}'\right\} =\left\{ \theta_{a}\right\} $. Indeed, treating $U$ as a finite time evolution,
we have $H\left(\lambda\right)=U\left(\lambda\right)HU\left(\lambda\right)^{\dagger}$,
where $U\left(\lambda\right):=\text{TO}e^{-i\int_{0}^{\lambda}\tilde{H}\left(t\right)dt}$,
as a deformation from $H$ to $H'$ which maintains locality and preserves
the bulk-gap. In fact, since the full spectrum on the cylinder is $\lambda$-independent, we have
 $h_{0}'=h_{0}$ for all $N_{x}$.
\end{enumerate}
Combining the
two above properties leads to

\begin{align}
1= & \tilde{Z}'/\left|\tilde{Z}'\right|\label{eq:6}\\
= & \exp2\pi i\left[\epsilon'N_{x}+\frac{1}{N_{x}}\left(h_{0}-\frac{c}{24}\right)+o\left(N_{x}^{-1}\right)\right],\nonumber 
\end{align}
where $\epsilon':=\text{Im}\left(\alpha'\right)/2\pi$. The non-universal integer part of $h_{0}$ can
then be eliminated by raising Eq.\eqref{eq:6} to the $N_{x}$th power,

\begin{align}
1 & =e^{2\pi i\epsilon'N_{x}^{2}}\theta_{0}\left(N_{x}\right)e^{-2\pi ic/24}+o\left(1\right),\label{eq:6-1}
\end{align}
where we used $\theta_{0}=e^{2\pi ih_{0}}$, and $o\left(1\right)\rightarrow0$
as $N_{x}\rightarrow\infty$. We also indicated explicitly the possible
$N_{x}$-dependence of $\theta_{0}$, as described in Sec.\ref{subsec:Boundary-finite-size}.
We proceed under the assumption that no accidental degeneracies occur
on the cylinder, so that $\theta_{0}\left(N_{x}\right)\in\left\{ \theta_{a}\right\} $
for all $N_{x}$, deferring the degenerate case to Appendix \ref{subsec:Dealing-with-accidental}.
Now, for rational $\epsilon'=n/m$, the series $e^{2\pi i\epsilon'N_{x}^{2}}$
($N_{x}\in\mathbb{N}$) covers periodically a subset $S$ of the $m$th
roots of unity, including $1\in S$. On the other hand, for irrational
$\epsilon'$ the series $e^{2\pi i\epsilon'N_{x}^{2}}$ is dense in
the unit circle. Combined with the fact that $\theta_{0}\left(N_{x}\right)$
is valued in the finite set $\left\{ \theta_{a}\right\} $, while
$c$ is $N_{x}$-independent, Equation \eqref{eq:6-1} implies that
$\epsilon'$ must be rational, and that the values attained by $\theta_{0}\left(N_{x}\right)e^{-2\pi ic/24}$
cover the set $S$ periodically, for large enough $N_{x}$. It follows
that $1\in S\subset\left\{ \theta_{a}e^{-2\pi ic/24}\right\} $. We
therefore have 

\begin{description}
\item [{Result$\;$1\label{Result 1}}] If a local bosonic Hamiltonian
$H$ is both locally stoquastic and in a chiral topological phase
of matter, then one of the corresponding topological spins satisfies
$\theta_{a}=e^{2\pi ic/24}$. Equivalently, a bosonic chiral topological
phase of matter where $e^{2\pi ic/24}$ is not the topological spin
of some anyon, i.e $e^{2\pi ic/24}\notin\left\{ \theta_{a}\right\} $,
admits no local Hamiltonians which are locally stoquastic.
\end{description}
The above result can be stated in terms of the topological $\mathbf{T}$-matrix,
which is the representation of a Dehn twist on the torus ground state
subspace, and has the spectrum $\text{Spec}\left(\mathbf{T}\right)=\left\{ \theta_{a}e^{-2\pi ic/24}\right\} _{a}$
\citep{doi:10.1142/S0129055X90000107,kitaev2006anyons,PhysRevLett.110.236801,PhysRevB.88.195412,PhysRevLett.110.067208,PhysRevB.91.125123}. 
\begin{description}
\item [{Result$\;$1'\label{Result 1'}}] If a local bosonic Hamiltonian
$H$ is is both locally stoquastic and in a chiral topological phase
of matter, then the corresponding $\mathbf{T}$-matrix satisfies $1\in\text{Spec}\left(\mathbf{T}\right)$.
Equivalently, a bosonic chiral topological phase of matter where $1\notin\text{Spec}\left(\mathbf{T}\right)$,
admits no local Hamiltonians which are locally stoquastic.
\end{description}
The above result is our main statement for bosonic phases of matter.
The logic used in its derivation will be extended in the following
sections, where we generalize Result \hyperref[Result 1]{1} to systems
which are fermionic, spontaneously-chiral, or both. 

\section{Spontaneous chirality\label{sec:Spontaneous-chirality}}

The invariants $h_{0}$ and $c$ change sign under both time reversal
$\mathcal{T}$ and parity (spatial reflection) $\mathcal{P}$, and
therefore require a breaking of $\mathcal{T}$ and $\mathcal{P}$
down to $\mathcal{PT}$ to be non-vanishing. The momentum polarization
Eq.\eqref{eq:12-3-1} is valid if this symmetry breaking is explicit,
i.e $H$ does not commute with $\mathcal{P}$ and $\mathcal{T}$ separately.
Here we consider the case where $H$ is $\mathcal{P},\mathcal{T}$-symmetric,
but these are broken down to $\mathcal{PT}$ spontaneously, as in
e.g intrinsic topological superfluids and superconductors \citep{volovik2009universe,PhysRevB.100.104512,rose2020hall}.
We first generalize Eq.\eqref{eq:12-3-1} to this setting, and then
use this generalization to obtain a spontaneously-chiral analog of
Result \hyperref[Result 1]{1}.

Note that the physical time-reversal $\mathcal{T}$ is an \textit{on-site}
anti-unitary operator acting \textit{identically} on all qudits, which
implies $\left[\mathcal{T},T_{R}\right]=0$, while $\mathcal{P}$
is a unitary operator that maps the qudit at $\mathbf{x}$ to that
at $P\mathbf{x}$, where $P$ is the nontrivial element in $O\left(2\right)/SO\left(2\right)$,
e.g $\left(x,y\right)\mapsto\left(-x,y\right)$. 

\subsection{Momentum polarization for spontaneously-chiral Hamiltonians \label{subsec:Momentum-polarization-for}}

For simplicity, we begin by assuming that $H$ is 'classically symmetry
breaking' - it has two exact ground states on the cylinder, already
at finite system sizes. We therefore have two ground states $\ket{\pm}$,
such that $\ket -$ is obtained from $\ket +$ by acting with either
$\mathcal{T}$ or $\mathcal{P}$. In particular, $\ket{\pm}$ have
opposite values of $h_{0}$ and $c$. The $\beta\rightarrow\infty$
density matrix is then $e^{-\beta H}/Z=\left(\rho_{+}+\rho_{-}\right)/2$,
where $\rho_{\pm}=\ket{\pm}\bra{\pm}$, and this modifies the right
hand side of Eq.\eqref{eq:12-3-1} to its real part,
\begin{align}
\tilde{Z}:= & \text{Tr}\left(T_{R}e^{-\beta H}\right)\label{eq:12-3-1-1}\\
= & Ze^{-\delta N_{x}}\cos2\pi\left[\epsilon N_{x}+\frac{2\pi}{N_{x}}\left(h_{0}-\frac{c}{24}\right)+o\left(N_{x}^{-1}\right)\right],\nonumber 
\end{align}
where $-\delta\pm2\pi i\epsilon$ are the values of the non-universal
$\alpha$ obtained from Eq.\eqref{eq:12-3-1}, by replacing $e^{-\beta H}$
by $\rho_{\pm}$ . Indeed, it follows from $\left[\mathcal{T},T_{R}\right]=0$
that if two density matrices are related by $\rho_{-}=\mathcal{T}\rho_{+}\mathcal{T}^{-1}$,
then $\tilde{Z}_{\pm}:=Z_{\pm}\text{Tr}\left(T_{R}\rho_{+}\right)$
are complex conjugates, $\tilde{Z}_{-}=\tilde{Z}_{+}^{*}$. 

Now, for a generic symmetry breaking Hamiltonian $H$, exact ground
state degeneracy happens only in the infinite volume limit \citep{sachdev_2011}.
At finite size, the two lowest lying eigenvalues of $H$ would be
separated by an exponentially small energy difference
$\Delta E=O\left(e^{-fN_{x}^{\lambda}}\right)$, with some $f>0,\lambda>0$.
The two corresponding eigenstates would be $\mathcal{T},\mathcal{P}$-even/odd,
of the form $W\left[\ket +\pm\ket -\right]$, where $W$ is a $\mathcal{T},\mathcal{P}$-invariant
local unitary \citep{zeng2019quantum}.  One can think of these statements as resulting from the existence
of a bulk-gap preserving and $\mathcal{T},\mathcal{P}$-symmetric
deformation of $H$ to a 'classically symmetry breaking' Hamiltonian\footnote{The canonical example is the transverse field Ising model $H\left(g\right)=-\sum_{i=1}^{N_{x}}\left(Z_{i}Z_{i+1}+gX_{i}\right)$
in 1+1d. Exact ground state degeneracy appears at finite $N_{x}$
only for $g=0$, though spontaneous symmetry breaking occurs for all
$\left|g\right|<1$, where a splitting $\sim\left|g\right|^{N_{x}}$
appears.}. 

In the generic setting, we have 
\begin{align}
e^{-\beta H}/Z & =W\left(\rho_{+}+\rho_{-}\right)W^{\dagger}/2+O\left(\beta\Delta E\right),
\end{align}
and, following our treatment of the local unitary $U$ in the previous
section, Equation \eqref{eq:12-3-1-1} remains valid, with modified
$\delta,\epsilon$, but unchanged $h_{0}-c/24$. This statement holds
for temperatures much higher than $\Delta E$ and much smaller that
the CFT energy spacing, $\Delta E\ll\beta^{-1}\ll N_{x}^{-1}$,
or more accurately $\beta^{-1}=o\left(N_{x}^{-1}\right)$ and $\beta\Delta E=o\left(N_{x}^{-1}\right)$
(cf. Sec.\ref{subsec:Momentum-polarization}).   Note that the
universal content of Eq.\eqref{eq:12-3-1-1} is the absolute value
$\left|h_{0}-c/24\right|$, since the cosine is even and $\text{sgn}\left(\epsilon\right)$
is non-universal. 

\subsection{Constraining $c$ and $\left\{ \theta_{a}\right\} $ }

Let us now assume that a gapped and local Hamiltonian $H$ is $\mathcal{T}$,$\mathcal{P}$-symmetric,
and is locally stoquastic, due to a unitary $U$. It follows that
$\tilde{Z}'=\text{Tr}\left(T_{R}e^{-\beta H'}\right)>0$, where $H'=UHU^{\dagger}$.
If $U$ happens to be $\mathcal{T}$,$\mathcal{P}$-symmetric, then
so is $H'$, and Eq.\eqref{eq:12-3-1-1} holds for $\tilde{Z}'$,
with $\delta',\epsilon'$ in place of $\delta,\epsilon$. For a general
$U$, we have 
\begin{align}
e^{-\beta H'}/Z & '=UW\left(\rho_{+}+\rho_{-}\right)W^{\dagger}U^{\dagger}/2+O\left(\beta\Delta E\right),
\end{align}
where $UW$ need not be $\mathcal{T}$,$\mathcal{P}$-symmetric. As
result, $\tilde{Z}'$ satisfies a weaker form of Eq.\eqref{eq:12-3-1-1},
\begin{align}
0<\tilde{Z}'=\left(Z'/2\right) & \sum_{\sigma=\pm}e^{-\delta_{\sigma}'N_{x}}e^{2\pi i\sigma\left[\epsilon_{\sigma}'N_{x}+\frac{1}{N_{x}}\left(h_{0}-\frac{c}{24}\right)+o\left(N_{x}^{-1}\right)\right]},\label{eq:10}
\end{align}
where $\delta_{+}',\epsilon_{+}'$ may differ from $\delta_{-}',\epsilon_{-}'$,
and we also indicated the positivity of $\tilde{Z}'$. Now, if $\delta_{+}'\neq\delta_{-}'$,
one of the chiral contributions is exponentially suppressed relative
to the other as $N_{x}\rightarrow\infty$, and we can apply the analysis
of Sec.\ref{sec:No-stoquastic-Hamiltonians}. If $\delta_{+}'=\delta_{-}'$,
we obtain
\begin{align}
0< & \sum_{\sigma=\pm}\exp2\pi i\sigma\left[\epsilon_{\sigma}'N_{x}+\frac{1}{N_{x}}\left(h_{0}-\frac{c}{24}\right)+o\left(N_{x}^{-1}\right)\right],\label{eq:11-1}
\end{align}
in analogy with Eq.\eqref{eq:6}.Unlike Eq.\eqref{eq:6}, taking
the $N_{x}$th power of this equation does not eliminate the mod 1
ambiguity in $h_{0}$. This corresponds to the fact that, as opposed
to explicitly chiral systems, stacking copies of a spontaneously chiral
system does not increase its net chirality. One can replace $T_{R}$
in $\tilde{Z}'$ with a larger half-translation $T_{R}^{m}$, which
would multiply the argument of the cosine by $m$. However, since
the largest translation on the cylinder is obtained for $m\approx N_{x}/2$,
this does not eliminate the mod 1 ambiguity in $h_{0}$. Moreover,
even if it so happens that $\epsilon'_{+}=\epsilon'_{-}=0$, Equation
\eqref{eq:11-1} does not imply $h_{0}-c/24=0$ (mod 1) since $N_{x}$
is large. 

\begin{figure}[h]
\begin{centering}
\includegraphics[width=0.5\columnwidth]{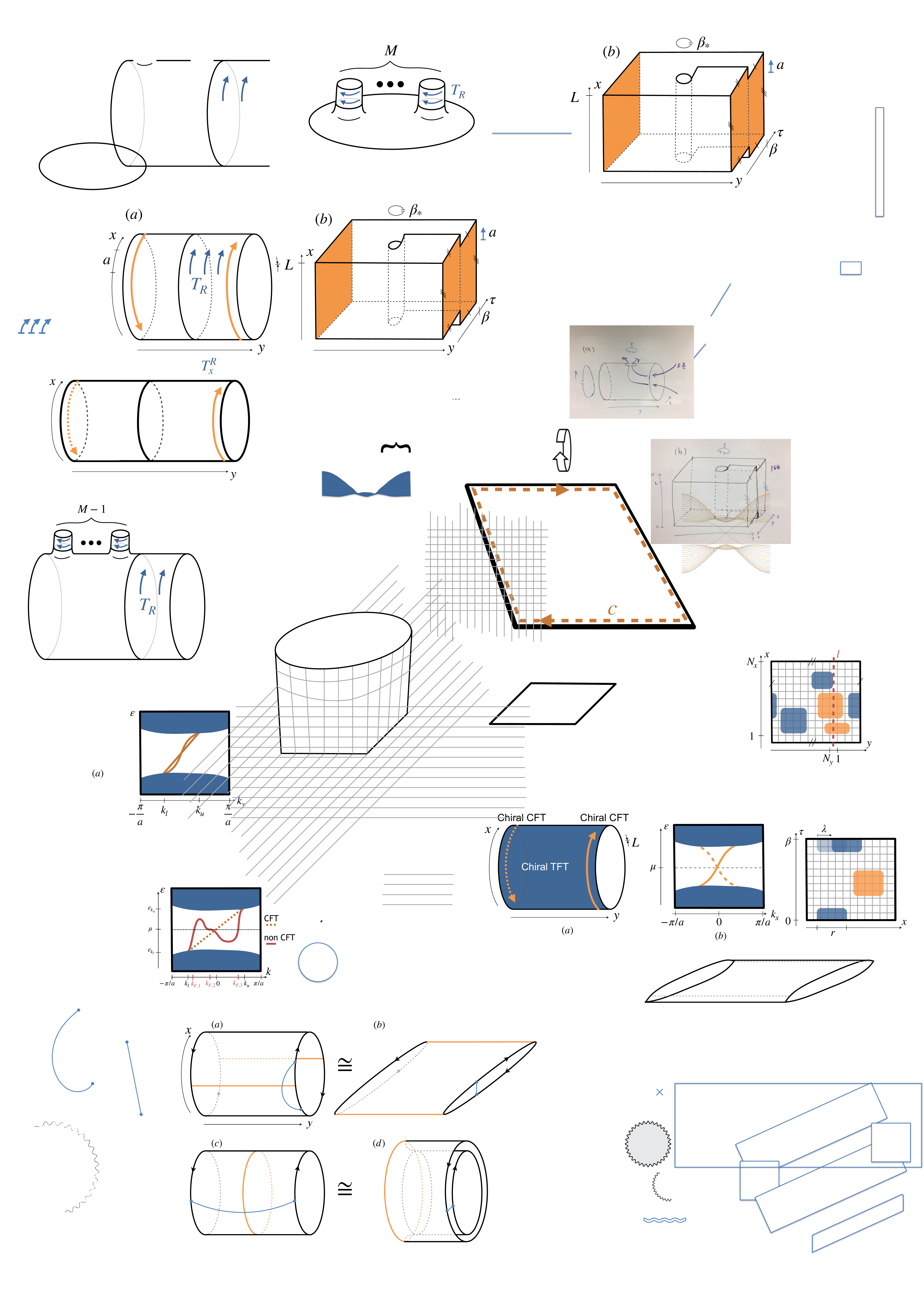}
\par\end{centering}
\caption{Bagpipes construction. We attach $M$ identical cylinders, or 'pipes',
to the given lattice, and define the half translation $T_{R}$ to
act on their top halves, as indicated by blue arrows. The contributions
of the pipes to the momentum polarization adds, producing the factor
$M$ in Eq.\eqref{eq:12-4}. \label{fig:bagpipes}}
\end{figure}

In order to make progress, we make use of the bagpipes construction
illustrated in Fig.\ref{fig:bagpipes}. We attach
$M$ identical cylinders, or 'pipes', to the given lattice, and act
with $T_{R}$ on these cylinders. The global topology of the given
lattice is unimportant - all that is needed is a large enough disk
in which the construction can be applied. The construction does require some form of homogeneity in order to have a unique extension of the Hamiltonian $H'$ to the pipes, and which will be identical for all pipes. We will assume a strict translation symmetry with respect to a sub-lattice, but we believe that this assumption can be relaxed.

The resulting surface, shown in Fig.\ref{fig:bagpipes}, has negative
curvature at the base of each pipe, which requires a finite number
of lattice disclinations in this region. In order to avoid any possible
ambiguity in the definition of $H'$ at a disclination, one can simply
remove any local term $H_{\mathbf{x}}'$ whose support contains a
disclination, which amounts to puncturing a hole around each disclination.
The resulting boundary components do not contribute to the momentum
polarization since $T_{R}$ acts on these as the identity. 

With the construction at hand, the identical contributions of all
cylinders to $\tilde{Z}'$ add, which implies
\begin{align}
0< & \sum_{\sigma=\pm}\exp2\pi i\sigma M\left[\epsilon_{\sigma}'N_{x}+\frac{1}{N_{x}}\left(h_{0}-\frac{c}{24}\right)+o\left(N_{x}^{-1}\right)\right].\label{eq:12-4}
\end{align}
Setting $M=N_{x}$ gives 
\begin{align}
0< & e^{2\pi i\epsilon_{+}'N_{x}^{2}}\theta_{0}\left(N_{x}\right)e^{-2\pi ic/24}\\
 & +e^{-2\pi i\epsilon_{-}'N_{x}^{2}}\theta_{0}^{*}\left(N_{x}\right)e^{2\pi ic/24}+o\left(1\right),\nonumber 
\end{align}
where we indicate explicitly the possible $N_{x}$-dependence of $\theta_{0}$.
This is the spontaneously chiral analog of Eq.\eqref{eq:6-1}, and
can be analyzed similarly. Since $\theta_{0}\left(N_{x}\right)$ is
valued in the finite set $\left\{ \theta_{a}\right\} $, both $\epsilon'_{\pm}$
must be rational, $\epsilon'_{\pm}=n_{\pm}/m_{\pm}$. Restricting
then to $N_{x}=n_{x}m_{+}m_{-}$, such that $e^{2\pi i\epsilon_{\pm}N_{x}^{2}}=1$,
and $\theta_{0}$ attains a constant value $\theta_{a}$ for large
enough $n_{x}$, we have 
\begin{align}
0<\text{Re}\left(\theta_{a}e^{-2\pi ic/24}\right),\label{eq:14}
\end{align}
for some anyon $a$. Repeating the analysis with $k$ times more pipes
$M=kN_{x}$, replaces $\theta_{a}e^{-2\pi ic/24}$ in Eq.\eqref{eq:14}
with its $k$th power, for all $k\in\mathbb{N}$. This infinite set
of equations then implies $\theta_{a}e^{-2\pi ic/24}=1$. To summarize,

\begin{description}
\item [{Result$\;$2\label{Result 2}}] If a local bosonic Hamiltonian
$H$ is both locally stoquastic and in a spontaneously-chiral topological
phase of matter, then one of the corresponding topological spins satisfies
$\theta_{a}=e^{2\pi ic/24}$. Equivalently, a bosonic spontaneously-chiral
topological phase of matter where $e^{2\pi ic/24}$ is not the topological
spin of some anyon, i.e $e^{2\pi ic/24}\notin\left\{ \theta_{a}\right\} $,
admits no local Hamiltonians which are locally stoquastic.
\end{description}
This extends Result \hyperref[Result 1]{1} beyond explicitly-chiral Hamiltonians,
and clarifies that the essence of the intrinsic sign problem we find
is the macroscopic, physically observable, condition $e^{2\pi ic/24}\notin\left\{ \theta_{a}\right\} $,
as opposed to the microscopic absence (or presence) of time reversal
and reflection symmetries. 

\section{DQMC: locality, homogeneity, and geometric manipulations\label{sec:Determinantal-quantum-Monte}}

In order to obtain fermionic analogs of the bosonic results of the
previous sections, we first need to establish a framework in which
such results can be obtained. In this section we develop a formalism
that unifies and generalizes the currently used DQMC algorithms and
design principles, and implement within it the geometric manipulations
used in previous sections, in a sign-free manner. Since we wish to
treat the wide range of currently known DQMC algorithms and design
principles on equal footing, the discussion will be more abstract
than the simple setting of locally stoquastic Hamiltonians used above.
In particular, Sections \ref{subsec:Local-determinantal-QMC}-\ref{subsec:Local-and--homogeneous}
lead up to the definition of \textit{locally sign-free DQMC}, which
is our fermionic analog of a locally stoquastic Hamiltonian. This
definition is used later on in Sec.\ref{sec:No-sign-free-DQMC} to
formulate Result \hyperref[Result 1F]{1F} and Result \hyperref[Result 2F]{2F},
the fermionic analogs of Results \hyperref[Result 1]{1} and \hyperref[Result 2]{2}.
The new tools needed to establish these results are the sign-free
geometric manipulations described in Sec.\ref{sec:Sign-free-geometric-manipulation}.

\subsection{Local DQMC\label{subsec:Local-determinantal-QMC}}

In the presence of bosons and fermions, the many-body Hilbert space
is given by $\mathcal{H}=\mathcal{H}_{\text{F}}\otimes\mathcal{H}_{\text{B}}$,
where $\mathcal{H}_{\text{F}}$ is a fermionic Fock space, equipped
with an on-site occupation basis $\ket{\nu}_{\text{F}}=\prod_{\mathbf{x},\alpha}\left(f_{\mathbf{x},\alpha}^{\dagger}\right)^{\nu_{\mathbf{x},\alpha}}\ket 0_{\text{F}}$,
$\nu_{\mathbf{x},\alpha}\in\left\{ 0,1\right\} $, generated by acting
with fermionic (anti-commuting) creation operators $f_{\mathbf{x},\alpha}^{\dagger}$
on the Fock vacuum $\ket 0_{\text{F}}$. The product is taken with
respect to a fixed ordering of fermion species $\alpha\in\left\{ 1,\cdots,\mathsf{d}_{\text{F}}\right\} $
and lattice sites $\mathbf{x}\in X$. We will also make use of the
single-fermion space $\mathcal{H}_{1\text{F}}\cong\mathbb{C}^{\left|X\right|}\otimes\mathbb{C}^{\mathsf{d}_{\text{F}}}$,
spanned by $\ket{\mathbf{x},\alpha}_{\text{F}}=f_{\mathbf{x},\alpha}^{\dagger}\ket 0_{\text{F}}$,
where $\left|X\right|=N_{x}N_{y}$ is the system size. As in Sec.\ref{sec:No-stoquastic-Hamiltonians},
$\mathcal{H}_{\text{B}}$ is a many-qudit Hilbert space with local
dimension $\mathsf{d}$. It can also be a bosonic Fock space where
$\mathsf{d}=\infty$.

We consider local fermion-boson Hamiltonians $H$, of the form 
\begin{align}
H=\sum_{\mathbf{x},\mathbf{y}}f_{\mathbf{x}}^{\dagger}h_{0}^{\mathbf{x},\mathbf{y}}f_{\mathbf{y}}+H_{I},\label{eq:11}
\end{align}
where the free-fermion Hermitian matrix $h_{0}^{\mathbf{x},\mathbf{y}}$
is $r_{0}$-local, it vanishes unless $\left|\mathbf{x}-\mathbf{y}\right|\leq r_{0}$,
and we suppress, here and in the following, the fermion species indices.
The Hamiltonian $H_{I}$ describes all possible $r_{0}$-local interactions
which preserve the fermion parity $\left(-1\right)^{N_{f}}$, where
$N_{f}=\sum_{\mathbf{x}}f_{\mathbf{x}}^{\dagger}f_{\mathbf{x}}$,
including fermion-independent terms $H_{\text{B}}$ as in Sec.\ref{sec:No-stoquastic-Hamiltonians}.
Thus $H_{I}$ is of the form 
\begin{align}
H_{I}= & H_{\text{B}}+\sum_{\mathbf{x},\mathbf{y}}f_{\mathbf{x}}^{\dagger}K_{\text{B}}^{\mathbf{x},\mathbf{y}}f_{\mathbf{y}}\\
 & +\sum_{\mathbf{x},\mathbf{y},\mathbf{z},\mathbf{w}}f_{\mathbf{x}}^{\dagger}f_{\mathbf{y}}^{\dagger}V_{\text{B}}^{\mathbf{x},\mathbf{y},\mathbf{z},\mathbf{w}}f_{\mathbf{z}}f_{\mathbf{w}}+\cdots,\nonumber 
\end{align}
where $K_{\text{B}}^{\mathbf{x},\mathbf{y}}$ (for all $\mathbf{x},\mathbf{y}\in X$)
is a local bosonic operator with range $r_{0}$, and vanishes unless
$\left|\mathbf{x}-\mathbf{y}\right|\leq r_{0}$, and similarly for
$V_{\text{B}}^{\mathbf{x},\mathbf{y},\mathbf{z},\mathbf{w}}$, which
vanishes unless $\mathbf{x},\mathbf{y},\mathbf{z},\mathbf{w}$ are
contained in a disk or radius $r_{0}$. In Eq.\eqref{eq:11} dots
represent additional pairing terms of the form $ff$, $f^{\dagger}f^{\dagger}$,
or $ffff$, $f^{\dagger}f^{\dagger}f^{\dagger}f^{\dagger}$, as well
as terms with a higher number of fermions, all of which are $r_{0}$-local
and preserve the fermion parity.

Since locality is defined in terms of anti-commuting Fermi operators,
a local stoquastic basis is not expected to exist, and accordingly,
the sign problem appears in any QMC method in which the Boltzmann
weights are given in terms of Hamiltonian matrix elements  in a local
basis \citep{troyer2005computational,li2019sign}. For this reason,
the methods used to perform QMC in the presence of fermions are distinct
from the ones used in their absence. These are collectively referred
to as DQMC \citep{PhysRevD.24.2278,Assaad,Santos_2003,li2019sign,berg2019monte},
and lead to the imaginary time path integral representation of the
partition function $Z=\text{Tr}\left(e^{-\beta H}\right)$,
\begin{align}
Z & =\int D\phi D\psi e^{-S_{\phi}-S_{\psi,\phi}}\label{eq:2}\\
 & =\int D\phi e^{-S_{\phi}}\text{Det}\left(D_{\phi}\right)\nonumber \\
 & =\int D\phi e^{-S_{\phi}}\text{Det}\left(I+U_{\phi}\right),\nonumber 
\end{align}
involving a bosonic field $\phi$ with an action $S_{\phi}$, and
a fermionic (grassmann valued) field $\psi$, with a quadratic action
$S_{\psi,\phi}=\sum_{\mathbf{x},\mathbf{x}'}\int\text{d}\tau\overline{\psi}_{\mathbf{x},\tau}\left[D_{\phi}\right]_{\mathbf{x},\mathbf{y}}\psi_{\mathbf{y},\tau}$
defined by the $\phi$-dependent single-fermion operator $D_{\phi}$.
In the third line of Eq.\eqref{eq:2} we assumed the Hamiltonian form
$D_{\phi}=\partial_{\tau}+h_{\phi\left(\tau\right)}$, and used a
standard identity for the determinant in terms of the single-fermion
imaginary-time evolution operator $U_{\phi}=\text{TO}e^{-\int_{0}^{\beta}h_{\phi\left(\tau\right)}\text{d}\tau}$
\citep{PhysRevD.24.2278}, where $\text{TO}$ denotes the time ordering.
The field $\phi$ ($\psi$) is defined on a continuous imaginary-time
circle $\tau\in\mathbb{R}/\beta\mathbb{Z}$, with periodic (anti-periodic)
boundary conditions, and on the spatial lattice $X$. The second
and third lines of Eq.\eqref{eq:2} define the Monte Carlo phase space
$\left\{ \phi\right\} $ and Boltzmann weight 
\begin{align}
p\left(\phi\right) & =e^{-S_{\phi}}\text{Det}\left(D_{\phi}\right)\label{eq:18-2}\\
 & =e^{-S_{\phi}}\text{Det}\left(I+U_{\phi}\right).\nonumber 
\end{align}

In applications, the DQMC representation \eqref{eq:2} may be obtained
from the Hamiltonian $H$ in a number of ways. If a Yukawa type model
is assumed as a starting point \citep{berg2019monte}, i.e $H_{I}=H_{\text{B}}+\sum_{\mathbf{x},\mathbf{y}}f_{\mathbf{x}}^{\dagger}K_{\text{B}}^{\mathbf{x},\mathbf{y}}f_{\mathbf{y}}$,
then the action $S_{\phi}$ is obtained from the Hamiltonian $H_{\text{B}}$,
and $h_{\phi\left(\tau\right)}=h_{0}+K_{\text{B}}$. Alternatively,
the representation \eqref{eq:2} may be obtained through a Hubbard\textendash Stratonovich
decoupling and/or a  series expansion of fermionic self-interactions
\citep{PhysRevLett.82.4155,chandrasekharan2013fermion,wang2015split}.
Such is the case e.g when there are no bosons $\mathcal{H}=\mathcal{H}_{\text{F}}$,
and $H_{I}=\sum f_{\mathbf{x}}^{\dagger}f_{\mathbf{y}}^{\dagger}V^{\mathbf{x},\mathbf{y},\mathbf{z},\mathbf{w}}f_{\mathbf{z}}f_{\mathbf{w}}$.

To take into account and generalize the above relations between $H$
and the corresponding DQMC representation, we will only assume (i)
that the effective single-fermion Hamiltonian $h_{\phi\left(\tau\right)}$
reduces to the free fermion matrix $h^{\left(0\right)}$ in the absence
of $\phi$, i.e $h_{\phi\left(\tau\right)=0}=h_{0}$, (ii) that the
boson field $\phi$ is itself an $r_{0}$-local object\footnote{Thus $\phi$ is a map from sets of lattice sites with diameter less
than $r_{0}$, such as links, plaquettes etc., to a fixed vector space
$\mathbb{C}^{k}$. Additionally, the $\phi$ integration in \eqref{eq:2}
runs over all such functions. As an example, restricting to constant
functions $\phi$ leads to non-local all to all interactions between
fermions.}, and (iii) that the $r_{0}$-locality of $h_{0}$ and $H_{I}$ implies
the $r$-locality of $S_{\phi}$ and $h_{\phi\left(\tau\right)}$,
where $r$ is some function of $r_{0}$, independent of system size.
The physical content of these assumptions is that the fields $\psi$
and operators $f$ correspond to the same physical fermion\footnote{Technically, via the fermionic coherent state construction of the
functional integral \citep{altland2010condensed}.}, and that the boson $\phi$ mediates \textit{all} fermionic interactions
$H_{I}$, and therefore corresponds to both the physical bosons in
$\mathcal{H}_{\text{B}}$ and to composite objects made of an even
number of fermions within a range $r_{0}$ (e.g a Cooper pair $\phi\sim ff$).

We can therefore write 
\begin{align}
S_{\phi} & =\sum_{\tau,\mathbf{x}}S_{\phi;\tau,\mathbf{x}},\label{eq:16}\\
h_{\phi\left(\tau\right)} & =\sum_{\mathbf{x}}h_{\phi\left(\tau\right);\mathbf{x}},\nonumber 
\end{align}
where each term $S_{\phi;\tau,x}$ depends only on the values of $\phi$
at points $\left(\mathbf{x}',\tau'\right)$ with $\left|\tau-\tau'\right|,\left|\mathbf{x}-\mathbf{x}'\right|\leq r$,
and similarly, each term $h_{\phi\left(\tau\right);\mathbf{x}}$ is
supported on a disk of radius $r$ around $\mathbf{x}$, and depends
on the values of $\phi\left(\tau\right)$ at points $\mathbf{x}$
within this disk.

Note that even though $H$ is Hermitian, we do not assume the same
for $h_{\phi\left(\tau\right)}$. Non-Hermitian $h_{\phi\left(\tau\right)}$s
naturally arise in Hubbard\textendash Stratonovich decouplings, see
e.g \citep{PhysRevB.71.155115,wang2015split}. Even when $h_{\phi\left(\tau\right)}$
is Hermitian for all $\phi$, its time-dependence implies that $U_{\phi}$
is non-Hermitian, and therefore $\text{Det}\left(I+U_{\phi}\right)$
in Eq.\eqref{eq:18-2} is generically complex valued \citep{PhysRevD.24.2278}.
This is the generic origin of the sign problem in DQMC. Section \ref{subsec:Local-and--homogeneous}
below describes the notion of \textit{fermionic design principles},
algebraic conditions on $U_{\phi}$ implying $\text{Det}\left(I+U_{\phi}\right)\geq0$,
and defines what it means for such design principles to be local and
homogenous. 

In the following analysis, we exclude the case of 'classically-interacting fermions',
where $\phi$ is time-independent and $h_{\phi}$ is Hermitian. In
this case the fermionic weight $\text{Det}\left(I+e^{-\beta h_{\phi}}\right)$
is trivially non-negative, and sign-free DQMC is always possible,
provided $S_{\phi}\in\mathbb{R}$. We view such models as 'exactly
solvable', on equal footing with free-fermion and commuting projector
models. Given a phase of matter, the possible existence of exactly
solvable models is independent of the possible existence of sign-free
models. Even when an exactly solvable model exists, QMC simulations
are of interest for generic questions, such as phase transitions due
to deformations of the model \citep{hofmann2019search}. In particular,
Ref.\citep{PhysRevLett.119.127204} utilized a classically-free description
of Kitaev's honeycomb model to obtain the thermal Hall conductance
and chiral central charge, which should be contrasted with the intrinsic
sign problem we find in the corresponding phase of matter, see Table \ref{tab:1} and Sec.\ref{sec:No-sign-free-DQMC}.

\subsection{Local and homogenous fermionic design principles\label{subsec:Local-and--homogeneous}}

The representation \eqref{eq:2} is sign-free if $p\left(\phi\right)=e^{-S_{\phi}}\text{Det}\left(I+U_{\phi}\right)\geq0$
for all $\phi$. A design principle then amounts to a set of polynomially
verifiable properties \footnote{That is, properties which can be verified in a polynomial-in-$\beta\left|X\right|$
time. As an example, given a local Hamiltonian, deciding whether there
exists a local basis in which it is stoquastic is NP-complete \citep{marvian2018computational,klassen2019hardness}.
In particular, one does not need to perform the exponential operation
of evaluating $p$ on every configuration $\phi$ to assure that $p\left(\phi\right)\geq0$.
Had this been possible, there would be no need for a Monte Carlo sampling
of the phase space $\left\{ \phi\right\} $.} of $S_{\phi}$ and $h_{\phi\left(\tau\right)}$ that guarantee that
the complex phase of $\text{Det}\left(I+U_{\phi}\right)$ is opposite
to that of $e^{-S_{\phi}}$. For the sake of presentation, we restrict
attention to the case where $S_{\phi}$ is manifestly real valued,
and $\text{Det}\left(I+U_{\phi}\right)\geq0$ due to an algebraic
condition on the operator $U_{\phi}$, which we write as $U_{\phi}\in\mathcal{C}_{U}$.
This is assumed to follow from an algebraic condition on $h_{\phi\left(\tau\right)}$,
written as $h_{\phi\left(\tau\right)}\in\mathcal{C}_{h}$, manifestly
satisfied for all $\phi\left(\tau\right)$. The set $\mathcal{C}_{h}$
is assumed to be closed under addition, while $\mathcal{C}_{U}$ is
closed under multiplication: $h_{1}+h_{2}\in\mathcal{C}_{h}$ for
all $h_{1},h_{2}\in\mathcal{C}_{h}$, and $U_{1}U_{2}\in\mathcal{C}_{U}$
for all $U_{1},U_{2}\in\mathcal{C}_{U}$. 

The simplest example, where $\mathcal{C}_{U}=\mathcal{C}_{h}$ is
the set of matrices obeying a fixed time reversal symmetry, is discussed
in Sec.\ref{subsec:Example:-time-reversal}. In Appendix \ref{subsec:Locality-of-known}
we review all other design principles known to us, demonstrate that
most of them are of the simplified form above, and generalize our
arguments to those that are not. Comparing with the bosonic Hamiltonians
treated in Sec.\ref{sec:No-stoquastic-Hamiltonians}, we note that
$\mathcal{C}_{h}$ is analogous to the set of stoquastic Hamiltonians
$H$ in a fixed basis, while $\mathcal{C}_{U}$ is analogous to the
resulting set of matrices $e^{-\beta H}$ with non-negative entries.

Design principles, as defined above (and in the literature), are purely
algebraic conditions, which carry no information about the underlying
geometry of space-time. However, as demonstrated in Sec.\ref{subsec:Example:-time-reversal},
in order to allow for local interactions, mediated by an $r_{0}$-local
boson $\phi$, a design principle must also be local in some sense.
We will adopt the following definitions, which are shown to be satisfied
by all physical applications of design principles that we are aware
of, in Sec.\ref{subsec:Example:-time-reversal} and Appendix \ref{subsec:Locality-of-known}.

\paragraph*{Definition (term-wise sign-free):}

We say that a DQMC representation is term-wise sign-free due to a
design principle $\mathcal{C}_{h}$, if each of the local terms $S_{\phi;\tau,\mathbf{x}},h_{\phi\left(\tau\right);\mathbf{x}}$
obey the design principle separately, rather than just they sums
$S_{\phi},h_{\phi\left(\tau\right)}$. Thus $S_{\phi;\tau,\mathbf{x}}$
is real valued, and $h_{\phi\left(\tau\right);\mathbf{x}}\in\mathcal{C}_{h}$,
for all $\tau,\mathbf{x}$.

\medskip{}

This is analogous to the requirement in Sec.\ref{subsec:Setup} that
$H'$ be term-wise stoquastic. Note that even when a DQMC representation
is term-wise sign-free, the resulting Boltzmann weights $p\left(\phi\right)$
are sign-free in a non-local manner: $\text{Det}\left(I+U_{\phi}\right)$
involves the values of $\phi$ at all space-time points, and splitting
the determinant into a product of local terms by the Leibniz formula
reintroduces signs, which capture the fermionic statistics. In this
respect, the ``classical'' Boltzmann weights $p\left(\phi\right)$
are always non-local in DQMC.

\paragraph*{Definition (on-site homogeneous design principle):\label{par:on-site-homogeneous}}

A design principle is said to be on-site homogenous if any permutation
of the lattice sites $\sigma\in S_{X}$ obeys it. That is, the operator
\begin{align}
 & O_{\left(\mathbf{x},\alpha\right),\left(\mathbf{x}',\alpha'\right)}^{\left(\sigma\right)}=\delta_{\mathbf{x},\sigma\left(\mathbf{x}'\right)}\delta_{\alpha,\alpha'},\label{eq:20}
\end{align}
viewed as a single-fermion imaginary-time evolution operator, obeys
the design principle, $O^{\left(\sigma\right)}\in\mathcal{C}_{U}$,
for all $\sigma\in S_{X}$.

\medskip{}

This amounts to the statement that the design principle treats all
lattice sites on equal footing, since it follows that $U_{\phi}\in\mathcal{C}_{U}\Rightarrow O^{\left(\sigma\right)}U_{\phi}O^{\left(\tilde{\sigma}\right)}\in\mathcal{C}_{U}$
for all permutations $\sigma,\tilde{\sigma}$. It may be that a design
principle is on-site only with respect to a sub-lattice $X'\subset X$.
In this case we simply treat $X'$ as the spatial lattice, and add
the finite set $X/X'$ to the $\mathsf{d}_{\text{F}}$ internal degrees
of freedom. Comparing with Sec.\ref{subsec:Setup}, on-site homogenous
design principles are analogous to the set of Hamiltonians $H'$ which
are stoquastic in an on-site homogenous basis - any qudit permutation
operator has non-negative entries in this basis, like the imaginary
time evolution $e^{-\beta H'}$.

\medskip{}

With these two notions of locality and homogeneity in design principles,
we now define the DQMC analog of locally stoquastic Hamiltonians (see
Sec.\ref{sec:No-stoquastic-Hamiltonians}). 

\medskip{}

\paragraph*{Definition (locally sign-free DQMC):}

Given a local fermion-boson Hamiltonian $H$, we say that $H$ allows
for a locally sign-free DQMC simulation, if there exists a local unitary
$U$, such that $H'=UHU^{\dagger}$ has a local DQMC representation
\eqref{eq:2}, which is term-wise sign-free due to an on-site homogeneous
design principle.

\medskip{}

Note that the DQMC representation \eqref{eq:2} is not of the Hamiltonian
but of the partition function, and clearly $Z'=\text{Tr}\left(e^{-\beta H'}\right)=\text{Tr}\left(e^{-\beta H}\right)=Z$.
What the above definition entails, is that it is $H'$, rather than
$H$, from which the DQMC data $S_{\phi},h_{\phi\left(\tau\right)}$
is obtained, as described in Sec.\ref{subsec:Local-determinantal-QMC}.
This data is then assumed to be term-wise sign-free due to an on-site
homogeneous design principle. The local unitary $U$ appearing in
the above definition is generally fermionic \citep{PhysRevB.91.125149}:
it can be written as a finite time evolution $U=\text{TO}e^{-i\int_{0}^{1}\tilde{H}\left(t\right)dt}$,
where $\tilde{H}$ is a local fermion-boson Hamiltonian, which is
either piecewise-constant or smooth as a function of $t$, c.f Sec.\ref{subsec:Setup}.

\medskip{}

\subsection{Example: time reversal design principle\label{subsec:Example:-time-reversal}}

To demonstrate the above definitions in a concrete setting, consider
the time-reversal design principle, defined by an anti-unitary operator
$\mathsf{T}$ acting on the single-fermion Hilbert space $\mathcal{H}_{1\text{F}}\cong\mathbb{C}^{\left|X\right|}\otimes\mathbb{C}^{\mathsf{d}_{\text{F}}}$,
such that $\mathtt{\mathsf{T}}^{2}=-I$. The set $\mathcal{C}_{h}$
contains all $\mathsf{T}$-invariant matrices, $\left[\mathsf{T},h_{\phi\left(\tau\right)}\right]=0$.
It follows that $\left[\mathsf{T},U_{\phi}\right]=0$, so that $\mathcal{C}_{U}=\mathcal{C}_{h}$
in this case, and this implies $\text{Det}\left(I+U_{\phi}\right)\geq0$
\citep{Hands_2000,PhysRevB.71.155115}. 

A sufficient condition on $\mathsf{T}$ that guarantees that the design
principle it defines is on-site homogenous is that it is of the form
$\mathsf{T}_{0}=I_{\left|X\right|}\otimes\mathsf{t}$, where $I_{\left|X\right|}$
is the identity matrix on $\mathbb{C}^{\left|X\right|}$, and $\mathsf{t}$
is an anti-unitary on $\mathbb{C}^{\mathsf{d}_{\text{F}}}$ that squares
to $-I_{\mathsf{d}_{F}}$. Equivalently, $\mathsf{T}$ is block diagonal,
with identical blocks $\mathsf{t}$ corresponding to the lattice sites
$\mathbf{x}\in X$. It is then clear that the permutation matrices
$O^{\left(\sigma\right)}$ defined in Eq.\eqref{eq:20} commute with
$\mathsf{T}$, so $O^{\left(\sigma\right)}\in\mathcal{C}_{U}$ for
all $\sigma\in S_{X}$. Note that the design principle $\mathsf{T}$
may correspond to a \textit{physical} time-reversal $\mathcal{T}$,
discussed in Sec.\ref{sec:Spontaneous-chirality}, only if it is on-site
homogenous, which is why we distinguish the two in our notation.

Additionally, if the operator $\mathsf{T}$ is $r_{\mathsf{T}}$-local
with some range $r_{\mathsf{T}}\geq0$, then any local $h_{\phi\left(\tau\right)}$
which is sign-free due to $\mathsf{T}$ can be made term-wise sign-free.
Indeed, if $\left[\mathsf{T},h_{\phi\left(\tau\right)}\right]=0$
then
\begin{align}
h_{\phi\left(\tau\right)} & =\frac{1}{2}\left(h_{\phi\left(\tau\right)}+\mathsf{T}h_{\phi\left(\tau\right)}\mathsf{T}^{-1}\right)\label{eq:21}\\
 & =\sum_{\mathbf{x}}\frac{1}{2}\left(h_{\phi\left(\tau\right);\mathbf{x}}+\mathsf{T}h_{\phi\left(\tau\right);\mathbf{x}}\mathsf{T}^{-1}\right)\nonumber \\
 & =\sum_{\mathbf{x}}\tilde{h}_{\phi\left(\tau\right);\mathbf{x}},\nonumber 
\end{align}
where $\tilde{h}{}_{\phi\left(\tau\right);\mathbf{x}}$ is now supported
on a disk of radius $r+2r_{\mathsf{T}}$ and commutes with $\mathsf{T}$,
for all $\mathbf{x}$. We see that the specific notion of $r_{\mathsf{T}}$-locality
coincides with the general notion of 'term-wise sign free'. In particular,
$\mathsf{T}=\mathsf{T}_{0}$ has a range $r_{\mathsf{T}}=0$, and
can therefore be applied term-wise. 

The above statements imply that if $\mathsf{T}=u\mathsf{T}_{0}u^{\dagger}$,
where $u$ is a single-fermion local unitary, and $H$ has a local
DQMC representation which is sign-free due to $\mathsf{T}$, then
$H$ allows for a locally sign-free DQMC simulation. Indeed, extending
$u$ to a many-body local unitary $U$, we see that $H'=UHU^{\dagger}$
admits a local DQMC representation where $\left[\mathsf{T}_{0},h_{\phi\left(\tau\right)}'\right]=0$.
Since $\mathsf{T}_{0}$ is on-site homogenous, and $h_{\phi\left(\tau\right)}'$
can be assumed term-wise sign-free (see Eq.\eqref{eq:21}), we have
the desired result. As demonstrated in Appendix \ref{subsec:Locality-of-known},
much of the above analysis carries over to other known design principles.

All realizations of $\,\mathsf{T}$ presented in Ref.\citep{PhysRevB.71.155115}
in the context of generalized Hubbard models, and in Ref.\citep{berg2019monte}
in the context of quantum critical metals, have the on-site homogeneous
form $\mathsf{T}_{0}$, and therefore correspond to locally sign-free
DQMC simulations.

We now consider a few specific time-reversal design principles $\mathsf{T}$.
The physical spin-1/2 time reversal $\mathsf{T}=\mathcal{T}^{\left(1/2\right)}$,
where $\mathcal{T}_{\left(\mathbf{x},\alpha\right),\left(\mathbf{x}',\alpha'\right)}^{\left(1/2\right)}=\delta_{\mathbf{x},\mathbf{x}'}\varepsilon_{\alpha\alpha'}\mathcal{K}$,
and $\alpha,\alpha'\in\left\{ \uparrow,\downarrow\right\} $ correspond
to up and down spin components,  is an on-site homogeneous design
principle, which accounts for the absence of signs in the attractive
Hubbard model \citep{PhysRevB.71.155115}. The composition $\mathsf{T}=\mathcal{M}\mathcal{T}^{\left(1/2\right)}$
of $\mathcal{T}^{\left(1/2\right)}$ with a modulo 2 translation,
$\mathcal{M}_{\left(\mathbf{x},\alpha\right),\left(\mathbf{x}',\alpha'\right)}=\delta_{\left(-1\right)^{x},\left(-1\right)^{x'+1}}\delta_{x_{e},x_{e}'}\delta_{y,y'}\delta_{\alpha,\alpha'}$,
where $x_{e}=2\left\lfloor x/2\right\rfloor $ is the even part of
$x$, is an on-site homogeneous design principle with respect to the
sub-lattice$X'=\left\{ \left(2x_{1},x_{2}\right):\;\mathbf{x}\in X\right\} $,
but not with respect to $X$. On the other hand, the composition $\mathsf{T}=\mathcal{P}^{\left(0\right)}\mathcal{T}^{\left(1/2\right)}$
of $\mathcal{T}^{\left(1/2\right)}$ with a spin-less reflection (or
parity) $\mathcal{P}_{\left(\mathbf{x},\alpha\right),\left(\mathbf{x}',\alpha'\right)}^{\left(0\right)}=\delta_{x,-x'}\delta_{y,y'}\delta_{\alpha\alpha'}$,
is not on-site homogeneous with respect to any sub-lattice. 

\begin{figure}[t]
\begin{centering}
\includegraphics[width=1\columnwidth]{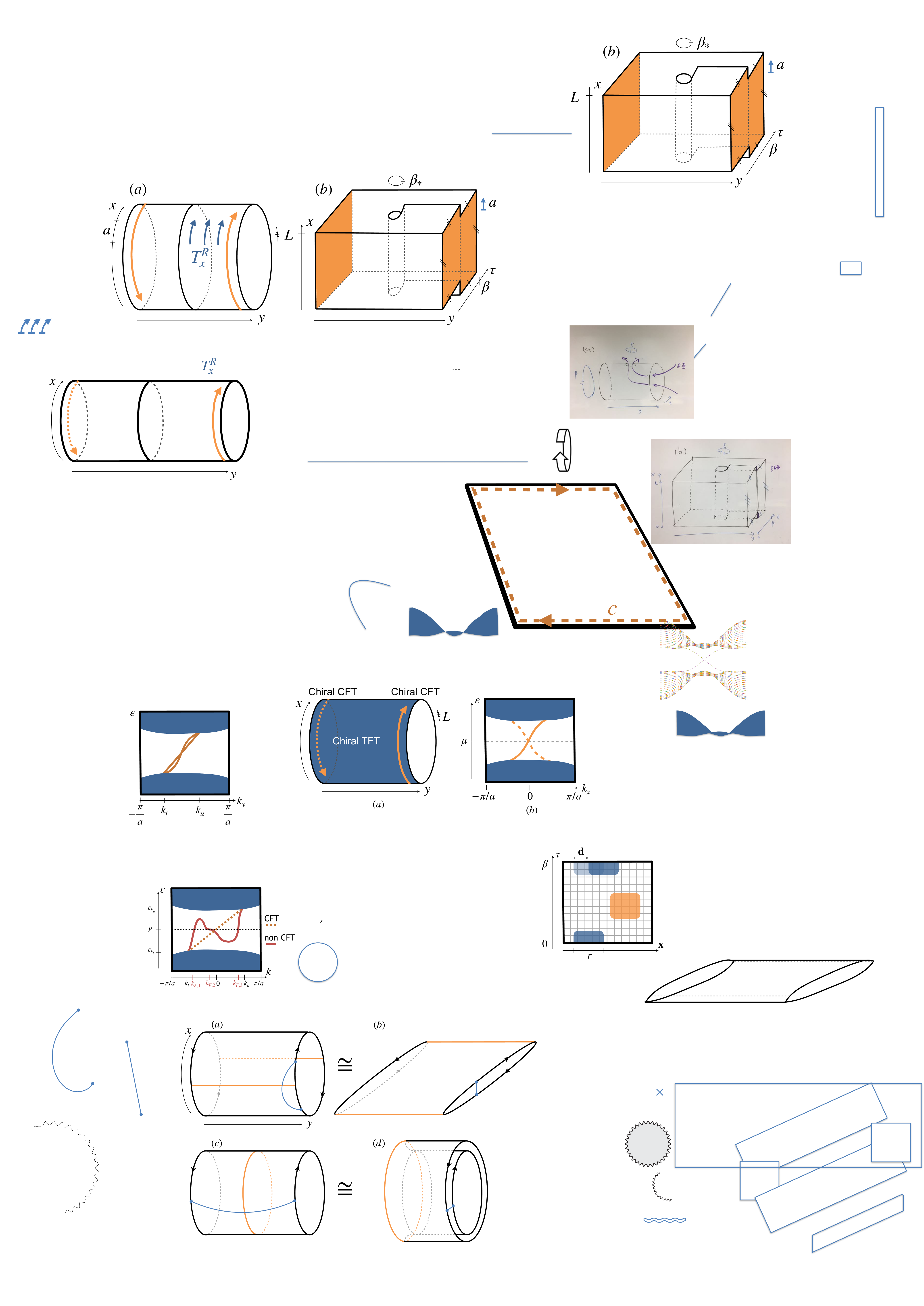}
\par\end{centering}
\caption{$\mathcal{P}\mathcal{T}$ symmetry as a 'non-local design principle'
for chiral topological matter. (a), (c): $\mathcal{P}\mathcal{T}$
symmetry, where $\mathcal{P}$ is a reflection (with respect to the
orange lines) and $\mathcal{P}\mathcal{T}$ is an on-site time-reversal,
is a natural symmetry in chiral topological phases. If $\left(\mathcal{P}\mathcal{T}\right)^{2}=-I$,
as is the case when $\mathcal{P}=\mathcal{P}^{\left(0\right)}$ is
spin-less and $\mathcal{T}=\mathcal{T}^{\left(1/2\right)}$ is spin-full,
it implies the non-negativity of fermionic determinants. Nevertheless,
as $\mathcal{P}\mathcal{T}$ is non-local, it only allows for QMC
simulations with $\mathcal{P}\mathcal{T}$ invariant bosonic fields,
which mediate non-local interactions (blue lines) between fermions.
Arrows indicate the chirality of boundary degrees of freedom. (b),
(d): Such non-local interactions effectively fold the system into
a non-chiral locally-interacting system supported on half the cylinder,
where $\mathcal{P}\mathcal{T}$ acts as an on-site time reversal.
In particular, the boundary degrees of freedom are now non-chiral.
Thus, $\mathcal{P}\mathcal{T}$ does not allow for sign-free QMC simulations
of chiral topological matter. More generally, fermionic design principles
must be local in order to allow for sign-free DQMC simulations of
local Hamiltonians. \label{fig:-symmetry-as}}
\end{figure}

The latter example is clearly non-local, and we use it to demonstrate
the necessity of locality in design principles. As discussed in Sec.\ref{sec:Spontaneous-chirality},
the breaking of $\mathcal{P}$ and $\mathcal{T}$ down to $\mathcal{P}\mathcal{T}$
actually defines the notion of chirality, and therefore $\mathcal{P}\mathcal{T}$
is a natural symmetry in chiral topological matter. Accordingly, the
design principle $\mathsf{T}=\mathcal{P}^{\left(0\right)}\mathcal{T}^{\left(1/2\right)}$
applies to a class of models for chiral topological phases, see Appendix
\ref{subsec:A-non-local-design}. This seems to allow, from the naive
algebraic perspective, for a sign-free DQMC simulation of certain
chiral topological phases. However, the weights $p\left(\phi\right)$
will only be non-negative for bosonic configurations $\phi$ which
are is invariant under $\mathsf{T}=\mathcal{P}^{\left(0\right)}\mathcal{T}^{\left(1/2\right)}$.
Restricting the $\phi$ integration in Eq.\eqref{eq:2} to such configurations
leads to non-local interactions between fermions $\psi$, coupling
the points $\left(x,y\right)$ and $\left(-x,y\right)$. These interactions
effectively fold the non-local chiral system into a local non-chiral
system of half of space, see Fig.\ref{fig:-symmetry-as}. Thus, $\mathsf{T}=\mathcal{P}^{\left(0\right)}\mathcal{T}^{\left(1/2\right)}$
does \textit{not} allow for sign-free DQMC simulations of chiral topological
matter.

\subsection{Sign-free geometric manipulations in DQMC\label{sec:Sign-free-geometric-manipulation}}

Let $Z$ be a partition function in a local DQMC form \eqref{eq:2},
on the discrete torus $X=\mathbb{Z}_{N_{x}}\times\mathbb{Z}_{N_{y}}$
and imaginary time circle $S_{\beta}^{1}=\mathbb{R}/\beta\mathbb{Z}$,
which is term-wise sign-free due to an on-site homogenous design principle.
In this section we show that it is possible to cut $X$ to the cylinder
$C$, and subsequently introduce a screw dislocation in the space-time
$C\times S_{\beta}^{1}$, which corresponds to the momentum polarization
\eqref{eq:12-3-1}, while maintaining the DQMC weights $p\left(\phi\right)$
non-negative.

\subsubsection{Introducing spatial boundaries}

Given a translation $T^{\mathbf{d}}$ ($\mathbf{d}\in X$), we can
cut the torus $X$ along a line $l$ parallel to $\mathbf{d}$, and
obtain a cylinder $C$ where $T^{\mathbf{d}}$ acts as a translation
within each boundary component, as in Sec.\ref{sec:No-stoquastic-Hamiltonians}.
Given the QMC representation \eqref{eq:2}, the corresponding representation
on $C$ is obtained by eliminating all local terms $S_{\phi;\tau,\mathbf{x}},h_{\phi\left(\tau\right);\mathbf{x}}$
whose support overlaps $l$, as in Fig.\ref{fig:cutting}. This procedure
may render $S_{\phi},h_{\phi\left(\tau\right)}$ independent of certain
degrees of freedom $\phi\left(\mathbf{x},\tau\right),\psi\left(\mathbf{x},\tau\right)$,
with $\mathbf{x}$ within a range $r$ of $l$, in which case we simply
remove such degrees of freedom from the functional integral \eqref{eq:2}\footnote{For $r_{0}$-local $\phi$, which is defined on links, plaquettes,
etc., we also remove from the functional $\phi$ integration those
links, plaquettes, etc. which overlap $l$.}. Since $S_{\phi;\tau,\mathbf{x}},h_{\phi\left(\tau\right);\mathbf{x}}$
obey the design principle for every $\mathbf{x},\tau$, the resulting
$S_{\phi},h_{\phi\left(\tau\right)}$ still obey the design principle
and the weights $p\left(\phi\right)$ remain real and non-negative.

\subsubsection{Introducing a screw dislocation in space-time}

Let us now restrict attention to $\mathbf{d}=\left(1,0\right)$, and
make contact with the momentum polarization \eqref{eq:12-3-1}. Given
a partition function on the space-time $C\times S_{\beta}^{1}$, consider
twisting the boundary conditions in the time direction,
\begin{align}
 & \phi_{\tau+\beta,x,y}=\phi_{\tau,x-\lambda\Theta\left(y\right),y},\label{eq:5-1}\\
 & \psi_{\tau+\beta,x,y}=-\psi_{\tau,x-\lambda\Theta\left(y\right),y}.\nonumber 
\end{align}
Note that $\lambda\in\mathbb{Z}_{N_{x}}$, since $x\in\mathbb{Z}_{N_{x}}$.
In particular, the full twist $\lambda=N_{x}$ is equivalent to the
untwisted case $\lambda=0$, which is equivalent to the statement
that the modular parameter of the torus is defined mod 1 (see e.g
example 8.2 of \citep{nakahara2003geometry}). The case $\lambda=0$
gives the standard boundary conditions, where the partition function
is, in Hamiltonian terms, just $Z=\text{Tr}\left(e^{-\beta H}\right)$.
In this case $Z>0$ since $H$ is Hermitian, though its QMC representation
$Z=\sum_{\phi}p\left(\phi\right)$ will generically involve complex
valued weights $p$. The twisted case $\lambda=1$ includes the insertion
of the half-translation operator 
\begin{align}
\tilde{Z} & =\text{Tr}\left(T_{R}e^{-\beta H}\right),\label{eq:7-2}
\end{align}
which appears in the momentum polarization \eqref{eq:12-3-1}. Since
$T_{R}$ is unitary rather than hermitian, $\tilde{Z}$ itself will
generically be complex. However,

\paragraph*{Claim:}

If $Z$ has a DQMC representation \eqref{eq:2}, with $p\left(\phi\right)\geq0$
term-wise due to an on-site homogeneous design principle $\mathcal{C}_{U}$,
then $\tilde{Z}$ has a QMC representation $\tilde{Z}=\sum_{\phi}\tilde{p}\left(\phi\right)$,
with $\tilde{p}\left(\phi\right)\geq0$. In particular, $\tilde{Z}\geq0$.

Proof of the claim is provided below. It revolves around two physical
points: (i) For the boson $\phi$, we only use the fact that all boundary
conditions, and those in Eq.\eqref{eq:5-1} in particular, are locally
invisible. (ii) For the fermion $\psi$, the local invisibility of
boundary conditions does not suffice, and the important point is that
translations do not act on internal degrees of freedom, and therefore
correspond to permutations of the lattice sites. The same holds for
the half translation $T_{R}$. This distinguishes translations from
internal symmetries, as well as from all other spatial symmetries,
which involve point group elements, and generically act non-trivially
on internal degrees of freedom. For example, a $C_{4}$ rotation will
act non-trivially on spin-full fermions.

\paragraph*{Proof:}

We first consider the fermionic part of the Boltzmann weight, $\text{Det}\left(I+U_{\phi}\right)$.
The Hamiltonian $h_{\phi\left(\tau\right)}$ depends on the values
of $\phi$ at a single time slice $\tau$, and is therefore unaffected
by the twist in bosonic boundary conditions. It follows that $U_{\phi}$
is independent of the twist in bosonic boundary conditions. On the
other hand, the fermionic boundary conditions in \eqref{eq:5-1} correspond
to a change of the time evolution operator $U_{\phi}\mapsto T_{R}U_{\phi}$,
in analogy with \eqref{eq:7-2}. Since the $\mathcal{C}_{U}$ is on-site,
and $T_{R}=O^{\left(\sigma\right)}$ is a permutation operator, with
$\sigma:\left(x,y\right)\mapsto\left(x+\Theta\left(y\right),y\right)$,
we have $T_{R}U_{\phi}\in\mathcal{C}_{U}$, and $\text{Det}\left(I+T_{R}U_{\phi}\right)\geq0$.

\begin{figure}[t]
\begin{centering}
\includegraphics[width=0.5\columnwidth]{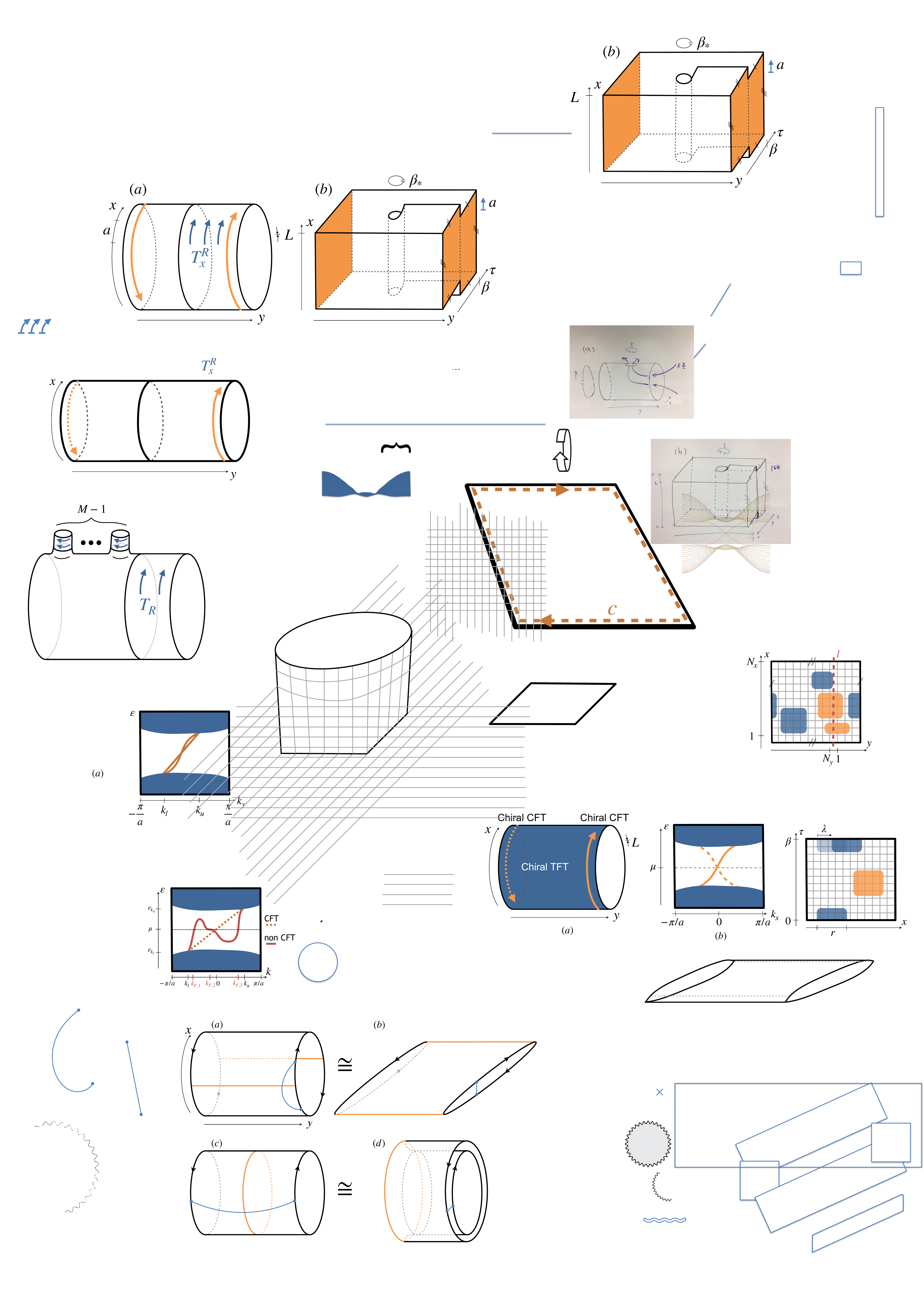}
\par\end{centering}
\caption{\label{fig:BoundaryConditions}Implementing the bosonic boundary conditions
\eqref{eq:5-1}. The lattice lies in the $x-\tau$ plane, at $y>0$
where the boundary conditions are non trivial. The orange area marks
the support, of diameter $r$, of a local term $S_{\phi;\tau,\mathbf{x}}$
which is unaffected by the boundary conditions. Blue areas correspond
to the support of a local term which is affected by the boundary conditions,
with pale blue indicating the un-twisted case $\lambda=0$.}
\end{figure}

Let us now consider the bosonic part of the Boltzmann weight $e^{-S_{\phi}}$,
where each of the local terms $S_{\phi;\tau,\mathbf{x}}$ is manifestly
real valued for all $\phi$. We  assume that the imaginary time
circle $S_{\beta}^{1}$ is discretized, such that the total number
of space-time points $\left(\tau,\mathbf{x}\right)=u\in U$ is finite.
Such a discretization is common in DQMC algorithms \citep{PhysRevD.24.2278,chandrasekharan2013fermion},
and the continuum case can be obtained by taking the appropriate limit.
The term $S_{\phi;\tau,\mathbf{x}}$ can then be written as a composition
$f\circ g_{V}$, where $f$ is a real valued function, and $g_{V}:\left(\phi_{u}\right)_{u\in U}\mapsto\left(\phi_{u}\right)_{u\in V}$
chooses the values of $\phi$ on which $S_{\phi;\tau,\mathbf{x}}$
depends, where $V\subset U$ is the support of $S_{\phi;\tau,\mathbf{x}}$.
The bosonic boundary conditions \eqref{eq:5-1} then amount to a modification
of the support $V\mapsto V_{\lambda}$, as depicted in Fig.\ref{fig:BoundaryConditions},
but not of the function $f$, which remains real valued. In particular,
for $\lambda=1$ we have $S_{\phi;\tau,\mathbf{x}}\mapsto\tilde{S}_{\phi;\tau,\mathbf{x}}=f\circ g_{V_{1}}$,
and $S_{\phi}\mapsto\tilde{S}_{\phi}=\sum_{\tau,\mathbf{x}}\tilde{S}_{\phi;\tau,\mathbf{x}}\in\mathbb{R}$.
 Combining the above conclusions for the bosonic and fermionic parts
of $\tilde{p}\left(\phi\right)=e^{-\tilde{S}_{\phi}}\text{Det}\left(I+T_{R}U_{\phi}\right)$,
we find that $\tilde{p}\left(\phi\right)\geq0$ for all $\phi$.

\section{Excluding sign-free DQMC for chiral topological matter\label{sec:No-sign-free-DQMC}}

We are now ready to demonstrate the existence of an intrinsic sign
problem in chiral topological matter comprised of bosons \textit{and}
fermions, using the machinery of Sections \ref{sec:No-stoquastic-Hamiltonians}-\ref{sec:Determinantal-quantum-Monte}. 

Let $H$ be a gapped local fermion-boson Hamiltonian on the discrete
torus, which allows for a locally sign-free DQMC simulation. Unpacking
the definition, this means that $H'=UHU^{\dagger}$ has a local DQMC
representation which is term-wise sign-free due to an on-site homogeneous
design principle. As shown in Sec.\ref{sec:Sign-free-geometric-manipulation},
this implies that $\tilde{Z}':=\text{Tr}\left(T_{R}e^{-\beta H'}\right)$,
written on the cylinder, also has a local DQMC representation, obeying
a local and on-site design principle, and as a result, $\tilde{Z}'>0$.
Now, as shown in Sec.\ref{sec:No-stoquastic-Hamiltonians}, the positivity
of $\tilde{Z}'$ implies $\theta_{a}=e^{2\pi ic/24}$ for some anyon
$a$. We therefore have the fermionic version of Result \hyperref[Result 1]{1},
\begin{description}
\item [{Result$\;$1F\label{Result 1F}}] If a local fermion-boson Hamiltonian
$H$, which is in a chiral topological phase of matter, allows for
a locally sign-free DQMC simulation, then one of the corresponding
topological spins satisfies $\theta_{a}=e^{2\pi ic/24}$. Equivalently,
a chiral topological phase of matter where $e^{2\pi ic/24}$ is not
the topological spin of some anyon, i.e $e^{2\pi ic/24}\notin\left\{ \theta_{a}\right\} $,
admits no local fermion-boson Hamiltonians for which locally sign-free
DQMC simulation is possible.
\end{description}
As shown in Sec.\ref{sec:Spontaneous-chirality}, the positivity of
$\tilde{Z}'$ implies $\theta_{a}=e^{2\pi ic/24}$ for some anyon
$a$, even if chirality appears only spontaneously. We therefore obtain
the fermionic version of Result \hyperref[Result 2]{2},
\begin{description}
\item [{Result$\;$2F\label{Result 2F}}] If a local fermion-boson Hamiltonian
$H$, which is in a spontaneously-chiral topological phase of matter,
allows for a locally sign-free DQMC simulation, then one of the corresponding
topological spins satisfies $\theta_{a}=e^{2\pi ic/24}$. Equivalently,
a spontaneously-chiral topological phase of matter where $e^{2\pi ic/24}$
is not the topological spin of some anyon, i.e $e^{2\pi ic/24}\notin\left\{ \theta_{a}\right\} $,
admits no local fermion-boson Hamiltonians which allow for a locally
sign-free DQMC simulation.\textit{}
\end{description}

In stating these results, we do not restrict to fermionic phases,
because bosonic phases may admit a fermionic description, for which
DQMC is of interest. When a bosonic phase admits a fermionic description,
the bosonic field $\phi$ in Eq.\eqref{eq:2} will contain a $\mathbb{Z}_{2}$
gauge field that couples to the fermion parity $\left(-1\right)^{N_{f}}$
of $\psi$. An important series of examples is given by the non-abelian Kitaev
spin liquids, which admit a description in terms of gapped Majorana
fermions with an odd Chern number $\nu$, coupled to a $\mathbb{Z}_{2}$
gauge field \citep{kitaev2006anyons}. As described in Table \ref{tab:1}, the criterion $e^{2\pi ic/24}\notin\left\{ \theta_{a}\right\} $ applies to the Kitaev spin liquid, for all  $\nu\in2\mathbb{Z}-1$. Result \hyperref[Result 1]{1} then excludes
the possibility of locally stoquastic Hamiltonians for the microscopic
description in terms of spins, while Result \hyperref[Result 1F]{1F}
excludes the possibility of locally sign-free DQMC simulations in
the emergent fermionic description.

\section{Conjectures: beyond chiral matter\label{sec:Generalization-and-extension} }

In Sections \ref{sec:No-stoquastic-Hamiltonians}-\ref{sec:No-sign-free-DQMC}
we established a criterion for the existence of intrinsic sign problems
in chiral topological matter: if $e^{2\pi ic/24}\notin\left\{ \theta_{a}\right\} $,
or equivalently $1\notin\text{Spec}\left(\mathbf{T}\right)$ (see
Result \hyperref[Result 1']{1'}), then an intrinsic sign problem exists.
Even if taken at face value, this criterion never applies to non-chiral bosonic topological phases, where
$c=0$, due to the vacuum topological spin $1\in\left\{ \theta_{a}\right\} $.
The same statement applies to all bosonic phases with $c\in24\mathbb{Z}$.
In this section we propose a refined criterion for intrinsic sign
problems in topological matter, which non-trivially applies to both
chiral \textit{and} non-chiral cases, and also unifies the results
of this paper with those obtained by other means in our parallel work
\citep{Paper1}.

Reference \citep{PhysRevLett.115.036802} proposed the 'universal
wave-function overlap' method for characterizing topological order
from any basis $\left\{ \ket i\right\} $ for the ground state subspace
of a local gapped Hamiltonian $H$ on the torus $X$. The method is
based on the conjecture 
\begin{align}
\bra i\mathbf{T}_{\text{m}}\ket j= & e^{-\alpha_{\mathbf{T}}A+o\left(A^{-1}\right)}\mathbf{T}_{ij},\label{eq:1-3}
\end{align}
where $A$ is the area of the torus, $\alpha_{\mathbf{T}}$ is a non-universal
complex number with non-negative real part, the microscopic Dehn-twist
operator $\mathbf{T}_{\text{m}}$ implements the Dehn twist $\left(x,y\right)\mapsto\left(x+y,y\right)$
on the Hilbert space, and $\mathbf{T}_{ij}$ are the entries of the
topological $\mathbf{T}$-matrix that characterizes the phase of $H$,
in the basis $\left\{ \ket i\right\} $. The same statement applies
to any element $\mathbf{M}$ of the mapping class group of the torus,
isomorphic to $SL\left(2,\mathbb{Z}\right)$, with $\mathbf{M}$ in
place of $\mathbf{T}$ in Eq.\eqref{eq:1-3}. The non-universal exponential
suppression of the overlap is expected because $\mathbf{M}_{\text{m}}$
will not generically map the ground-state subspace to itself, but
if $\mathbf{M}_{\text{m}}$ happens to be a symmetry of $H$, then
$\alpha_{\mathbf{M}}=0$ \citep{PhysRevB.85.235151,PhysRevLett.110.067208}.
Though we are not aware of a general analytic derivation of Eq.\eqref{eq:1-3},
it was verified analytically and numerically in a large number of
examples in Refs.\citep{PhysRevLett.115.036802,PhysRevB.91.125123,PhysRevB.91.075114,PhysRevB.90.205114,PhysRevB.95.235107},
for Hamiltonians in both chiral and non-chiral phases.  

Note the close analogy between Eq.\eqref{eq:1-3} and the momentum
polarization \eqref{eq:12-3-1}, where the microscopic Dehn-twist
$\mathbf{T}_{\text{m}}$ on the torus and the half translation $T_{R}$
on the cylinder play a similar role, and non-universal extensive contributions
are followed by sub-extensive universal data. To make this analogy
clearer, and make contact with the analysis of Sections \ref{sec:No-stoquastic-Hamiltonians}
and \ref{sec:No-sign-free-DQMC}, we consider the object $Z_{\mathbf{T}}=\text{Tr}\left(\mathbf{T}_{\text{m}}e^{-\beta H}\right)$,
which satisfies 
\begin{align}
Z_{\mathbf{T}}= & Ze^{-\alpha_{\mathbf{T}}A+o\left(A^{-1}\right)}\text{Tr}\left(\mathbf{T}\right),\label{eq:25}
\end{align}
and can be interpreted as either the (unnormalized) thermal expectation
value of $\mathbf{T}_{\text{m}}$, or the partition function on a
space-time twisted by $\mathbf{T}$, in analogy with Sec.\ref{subsec:Momentum-polarization}.
 Equation \eqref{eq:25} is valid for temperatures $\Delta E\ll1/\beta\ll E_{\text{g}}$,
much lower than the bulk gap $E_{\text{g}}$ and much higher than
any finite size splitting in the ground state-subspace, $\Delta E=o\left(A^{-1}\right)$.

Just like $T_{R}$, the operator $\mathbf{T}_{\text{m}}$ acts as
a permutation of the lattice sites. Therefore, following Sections
\ref{sec:No-stoquastic-Hamiltonians} and \ref{sec:No-sign-free-DQMC},
if $H$ is either locally stoquastic, or admits a locally sign-free
DQMC simulation, then $\text{Tr}\left(\mathbf{T}\right)\ge0$. In
terms of $c$ and $\left\{ \theta_{a}\right\} $, this implies $e^{-2\pi ic/24}\sum_{a}\theta_{a}=\text{Tr}\left(\mathbf{T}\right)\geq0$,
where the sum runs over all topological spins. 

The last statement applies to both bosonic and fermionic Hamiltonians.
For bosonic Hamiltonians, it can be strengthened by means of the Frobenius-Perron
theorem. If $H'=UHU^{\dagger}$ is stoquastic in the on-site basis
$\ket s$, Hermitian, and has a degenerate ground state subspace,
then this subspace can be spanned by an orthonormal basis $\ket{i'}$
with positive entries in the on-site basis, $\braket s{i'}\geq0$,
see e.g Ref.\citep{Paper1}. This implies that 
\begin{align}
0\leq\bra{i'}\mathbf{T}_{\text{m}}\ket{j'}= & e^{-\alpha_{\mathbf{T}}'A+o\left(A^{-1}\right)}\mathbf{T}_{i'j'},\label{eq:2-3-1}
\end{align}
where $\alpha_{\mathbf{T}}'$ is generally different from $\alpha_{\mathbf{T}}$,
but the matrix $\mathbf{T}_{i'j'}$ has the same spectrum as $\mathbf{T}_{ij}$
in Eq.\eqref{eq:1-3}. This is a stronger form of \eqref{eq:25},
which implies $\mathbf{T}_{i'j'}\geq0$. Since $\mathbf{T}_{i'j'}$
is also unitary, it is a permutation matrix, $\mathbf{T}_{i'j'}=\delta_{i',\sigma\left(j'\right)}$
for some $\sigma\in S_{N}$, where $N$ is the number of ground states.
In turn, this implies that the spectrum of $\mathbf{T}$ is a disjoint
union of complete sets of roots of unity,
\begin{align}
\left\{ \theta_{a}e^{-2\pi ic/24}\right\} _{a=1}^{N} & =\text{Spec}\left(\mathbf{T}\right)=\bigcup_{k=1}^{K}R_{n_{k}},\label{eq:27-1}
\end{align}
where $R_{n_{k}}$ is the set of $n_{k}$th roots of unity, $n_{k},K\in\mathbb{N}$,
and $\sum_{k=1}^{K}n_{k}=N$.  Therefore, 
\begin{description}
\item [{Conjecture$\;$1}] A bosonic topological phase of matter where
$\left\{ \theta_{a}e^{-2\pi ic/24}\right\} $ is not a disjoint union
of complete sets of roots of unity, admits no local Hamiltonians which
are locally stoquastic.
\end{description}
In particular, this implies an intrinsic sign problem whenever $1\notin\left\{ \theta_{a}e^{-2\pi ic/24}\right\} $,
thus generalizing Result \hyperref[Result 1]{1}. Moreover, the above
statement applies non-trivially to phases with $c\in24\mathbb{Z}$.
In particular, for non-chiral phases, where $c=0$, it reduces to
the result established in Ref.\citep{Paper1}, thus generalizing it
as well. The simplest example for a non-chiral phase
with an intrinsic sign problem is the doubled semion phase, where
$\left\{ \theta_{a}\right\} =\left\{ 1,i,-i,1\right\} $ \citep{PhysRevB.71.045110}. 

Though we are currently unaware of an analog of the Frobenius-Perron
theorem that applies to DQMC, we expect that an analogous result can
be established for fermionic Hamiltonians.
\begin{description}
\item [{Conjecture$\;$1F}] A topological phase of matter where $\left\{ \theta_{a}e^{-2\pi ic/24}\right\} $
is not a complete set of roots of unity, admits no local fermion-boson
Hamiltonians for which locally sign-free DQMC simulation is possible.
\end{description}
The above conjectures suggest a substantial improvement over the criterion
$e^{2\pi ic/24}\notin\left\{ \theta_{a}\right\} $. To demonstrate
this, we go back to the 
$1/q$ Laughlin phases and $SU\left(2\right)_{k}$
Chern-Simons theories considered in Table \ref{tab:1}. We find a conjectured
intrinsic sign problem in \textit{all} of the first one-thousand bosonic
Laughlin phases ($q$ even), fermionic Laughlin phases ($q$ odd),
and $SU\left(2\right)_{k}$ Chern-Simons theories. In particular, we note that the prototypical $1/3$ Laughlin phase is not captured by the criterion $e^{2\pi ic/24}\notin\left\{ \theta_{a}\right\}$, but is conjectured to be intrinsically sign-problematic.

\section{\label{sec:Discussion-and-outlook}Discussion and outlook }

In this paper we established the existence of intrinsic sign problems
in a broad class of chiral topological phases, namely those where
$e^{2\pi ic/24}$ does not happen to be the topological spin of an
anyon. Since these intrinsic sign problems persist even when chirality,
or time reversal symmetry breaking, appears spontaneously, they are
rooted in the macroscopic and observable data $c$, $\left\{ \theta_{a}\right\} $,
rather than the microscopic absence (or presence) of time reversal
symmetry. Going beyond the simple setting of stoquastic Hamiltonians, we provided
the first treatment of intrinsic sign problems in fermionic systems.
In particular, we constructed a general framework which describes
all DQMC algorithms and fermionic design principles that we are aware
of, including the state of art design principles \citep{wang2015split,wei2016majorana,li2016majorana,wei2017semigroup}
which are only beginning to be used by practitioners. Owing to its
generality, it is likely that our framework will apply to additional
design principles which have not yet been discovered, insofar as they
are applied locally. We also presented conjectures that strengthen our results, and unify
them with those obtained in Refs.\citep{hastings2016quantum,Paper1},
under a single criterion in terms of $c$ and $\left\{ \theta_{a}\right\} $.
These conjectures also imply intrinsic sign problems in many topological
phases not covered by existing results. 

Conceptually, our results  show that the sign problem
is not \textit{only} a statement of computational complexity: it is,
in fact, intimately connected with the physically observable properties
of quantum matter. Such a connection has long been heuristically appreciated
by QMC practitioners, and is placed on a firm and quantitative footing
by the discovery of intrinsic sign problems. 

Despite the progress made here, our understanding of intrinsic sign
problems is still in its infancy, and many open questions remain:

\paragraph*{Quantum computation and intrinsic sign problems}

Intrinsic sign-problems relate the physics of topological phases to
their computational complexity, in analogy with the classification
of topological phases which enable universal quantum computation \citep{Freedman:2002aa,nayak2008non}.
As we have seen, many phases of matter that are known to be universal
for quantum computation are also intrinsically sign-problematic, supporting
the paradigm of 'quantum advantage' or 'quantum supremacy' \citep{Arute:2019aa}.
Determining whether intrinsic sign-problems appear in \textit{all}
phases of matter which are universal for quantum computation is an
interesting open problem. Additionally, we identified intrinsic sign problems in many topological phases which are not universal for quantum computation. The intermediate complexity of such phases between classical and quantum computation is another interesting direction for future work. 

\paragraph*{Unconventional superconductivity and intrinsic sign problems}

As described in the introduction, a major motivation for the study of
intrinsic sign problems comes from long standing open problems in
fermionic many-body systems, the nature of high temperature superconductivity
 in particular. It is currently believed that many high temperature
superconductors, and the associated repulsive Hubbard models, are
 \textit{non-chiral} $d$-wave superconductors \citep{kantian2019understanding,berg2019monte},
in which we did not identify an intrinsic sign problem. The optimistic possibility that the sign problem can in fact
be cured in repulsive Hubbard models is therefore left open, though this has not yet been
accomplished in the relevant regime of parameters, away from half filling, despite intense research efforts 
\citep{PhysRevX.5.041041}. Nevertheless, the state of the art DMRG
results of Ref.\citep{kantian2019understanding} do not exclude the
possibility of a \textit{chiral} $d$-wave superconductor ($\ell=\pm 2$ in Table \ref{tab:1}). In this case we do find an intrinsic sign problem, which would account for the notorious sign problems observed in repulsive Hubbard models. More speculatively, it is possible that the mere proximity of repulsive Hubbard models to a chiral $d$-wave phase stands behind their notorious sign problems. The possible effect of an intrinsic sign problem in a given phase on the larger phase diagram was recently studied in Ref.\citep{zhang2020non}. There is also evidence
for chiral $d$-wave superconductivity in doped graphene and related
materials  \citep{PhysRevB.84.121410,Black_Schaffer_2014}, and our results therefore suggest the impossibility of sign-free QMC simulations of these. We believe that the study
of intrinsic sign problems in the context of unconventional superconductivity is a promising direction for future work.

\paragraph*{Non-locality as a possible route to sign-free QMC}

The intrinsic sign problems identified in this work add to existing
evidence for the complexity of chiral topological phases - these do
not admit local commuting projector Hamiltonians \citep{PhysRevB.89.195130,potter2015protection,PhysRevB.98.165104,kapustin2019thermal},
nor do they admit local Hamiltonians with a PEPS state as an exact
ground state \citep{PhysRevLett.111.236805,PhysRevB.90.115133,PhysRevB.92.205307,PhysRevB.98.184409}.

Nevertheless, relaxing the locality requirement does lead to positive
results for the simulation of chiral topological matter using commuting
projectors or PEPS. First, commuting projector Hamiltonians can be
obtained if the local bosonic or fermionic degrees of freedom are
replaced by anyonic (and therefore non-local) excitations of an underlying
chiral topological phase \citep{PhysRevB.97.245144}. Second, chiral
topological Hamiltonians can have a PEPS ground state if they include
interactions (or hopping amplitudes) that slowly decay as a power-law
with distance.

One may therefore hope that sign-free QMC simulations of chiral topological
matter can also be performed if the locality requirements made in
Sec.\ref{sec:Determinantal-quantum-Monte} are similarly relaxed.
Do such 'weakly-local' sign-free models exist?

\paragraph*{Easing intrinsic sign problems}

In this paper we proved the existence of an intrinsic sign problem
in chiral topological phases of matter, but we did not quantify the
\textit{severity} of this sign problem, which is an important concept
in both practical applications and theory of QMC. The severity of
a sign problem is quantified by the smallness of the average sign,
 $\left\langle \text{sgn}\right\rangle :=\sum p/\sum\left|p\right|$,
of the QMC weights $p$ with respect to the distribution $\left|p\right|$.  Since $\left\langle \text{sgn}\right\rangle $ can be viewed as the
ratio of two partition functions, it  obeys the generic scaling 
$\left\langle \text{sgn}\right\rangle \sim e^{-\Delta\beta N}$, with
$\Delta\geq0$, as $\beta N\rightarrow\infty$ \citep{troyer2005computational,hangleiter2019easing}.
A sign problem exists when $\Delta>0$, in which case QMC simulations
require exponential computational resources, and this is what the
intrinsic sign problem we identified implies for 'most' chiral topological
phases of matter. From the point of view of computational complexity,
all that matters is whether $\Delta=0$ or $\Delta>0$, but for practical
applications the value of $\Delta$ is very important, see e.g \citep{PhysRevB.84.121410}.
One may hope for a possible refinement of our results that provides
a lower bound $\Delta_{0}>0$ for $\Delta$, but since we have studied
\textit{topological} phases of matter, we view this as unlikely. It
may therefore be possible to obtain fine-tuned models and QMC methods
that lead to a $\Delta$ small enough to be \textit{practically} useful.
More generally, it may be possible to search for such models algorithmically,
thus \textit{easing} the intrinsic sign problem \citep{hangleiter2019easing,torlai2019wavefunction}. 

\paragraph*{Possible extensions}

The chiral central charge only appears modulo 24 in our results. Nevertheless,
the full value of $c$ is physically meaningful, as reviewed in the
introduction. Does an intrinsic sign problem exist in all phases with
$c\neq0$? The results of Ref.\citep{ringel2017quantized} strongly
suggest this. 

The arguments of Ref.\citep{Paper1} and Sec.\ref{sec:Generalization-and-extension}
apply equally well to any element of the modular group, rather than
just the topological $\mathbf{T}$-matrix, implying that the spectrum
of all elements decomposes into full sets of roots of unity. This may imply a more restrictive constraint on the TFT data than conjectured in
Sec.\ref{sec:Generalization-and-extension}.

It is believed that all SPT phases can be characterized by universal
complex phases acquired by their partition functions, when placed
on certain non-trivial space-times \citep{PhysRevB.90.235113,Kapustin:2015aa,RevModPhys.88.035001,PhysRevB.95.205139,PhysRevB.98.035151},
in analogy with Eq.\eqref{eq:12-3-1}. Loosely speaking, for an SPT
with on-site symmetry group $G$, the relevant space-time would be
obtained by purely geometric manipulations as performed in this paper,
along with a twisting of boundary conditions by elements $g\in G$.
Since each $G$ acts on-site,  we do not expect intrinsic sign problems
whenever a non-trivial $g$ is required to detect the SPT \citep{PhysRevB.95.174418,PhysRevB.85.045114, PhysRevB.86.045106,gazit2016bosonic}. Nevertheless,
it may be possible to obtain weaker statements, constraining the possible
bases in which a Hamiltonian in a $G$-SPT may be stoquastic. Such
constraints may be more useful for designing sign-free models than
the stronger intrinsic sign problems discussed in this paper. Similar
questions arise in the context of topologically ordered phases, enriched
by an on-site symmetry. 

Finally, going beyond gapped topological phases, are there intrinsically
sign-problematic phases which are not gapped, not topological, or
both? It is the authors' hope that answers to some of these questions
will shed new light on the formidable quantum many-body problem. 

\begin{acknowledgments}
O.G. is grateful to his wife Adi Cohen-Golan, for her invaluable support
in the completion of this work during the Coronavirus lockdown in
Israel. We thank Ryan Thorngren for participation in early stages
of this work and feedback on the manuscript, and Snir Gazit for providing
the much needed practitioner's point of view on DQMC and the sign
problem. We also benefited from discussions with Ady Stern, Ari Turner,
Ciarán Hickey, Erez Berg, Eyal Cornfeld, Eyal Leviatan, Johannes Stephan Hofmann,
Michael Levin, Paul Wiegmann, and Raquel Queiroz.  This work was
supported by the Israel Science Foundation (ISF, 2250/19), the Deutsche Forschungsgemeinschaft
(DFG,  German Research Foundation, CRC/Transregio 183, EI 519/7-1), and the European Research Council
(ERC), under Project LEGOTOP and the European Union's Horizon 2020
research and innovation program (grant agreement No. 771537).
\end{acknowledgments}


\appendix

\section{Further details regarding Eq.\eqref{eq:12-3-1}\label{subsec:Further-details-regarding}}

This appendix involves basic facts in CFT, which can be found in e.g
\citep{ginsparg1988applied,di1996conformal}.

\subsection{Definition of $h_{0}$ and ambiguities in its value \label{subsec:Definition-of-}}

A chiral topological phase of matter has a finite-dimensional ground
state subspace on the spatial torus. A basis $\left\{ \ket a\right\} _{a=1}^{N}$
for the torus ground state subspace exists, such that each state $\ket a$
corresponds to a conformal family in the boundary CFT \citep{PhysRevB.85.235151,PhysRevB.88.195412,PhysRevLett.110.236801},
constructed over a primary with right/left moving conformal weights
$h_{a}^{\left(l\right)},h_{a}^{\left(r\right)}\geq0$. The corresponding
chiral and total conformal weights are then given by $h_{a}=h_{a}^{\left(l\right)}-h_{a}^{\left(r\right)}$
and $h_{a}^{+}=h_{a}^{\left(l\right)}+h_{a}^{\left(r\right)}$, respectively.
The chiral and total central charges of the CFT are similarly defined
in terms of the left/right moving central charges, $c=c^{\left(l\right)}-c^{\left(r\right)}$
and $c^{+}=c^{\left(l\right)}+c^{\left(r\right)}$. 

When the torus is cut to a cylinder with finite circumference $L$,
the ground state degeneracy is lifted, generically leaving a unique
ground state. The lowest energy eigenstates on the cylinder can also
be labeled as $\left\{ \ket a\right\} _{a=1}^{N}$. Each $\ket a$
corresponds to a non-universal choice of state in the conformal family
labeled by $h_{a}^{\left(l\right)},h_{a}^{\left(r\right)}$ , which
need not be the primary, as demonstrated explicitly in Appendix \ref{subsec:Beyond-the-assumption}
below. 

If the boundary is described by an idealized CFT, all $\ket a$s correspond
to primaries, and the corresponding energies are given by $E_{a}=\left(4\pi v/L\right)\left(h_{a}^{+}-c^{+}/24\right)$,
relative to the ground state energy on the torus, where $v$ is the
velocity of the CFT and $L$ is the circumference of the cylinder.
These expressions receive exponentially small corrections of $O\left(Le^{-R/\xi}\right)$
and $O\left(Re^{-L/\xi}\right)$, where $\xi$ is the bulk correlation
length and $R$ is the length of the cylinder \citep{PhysRevLett.110.067208}.
The cylinder ground state then corresponds to the CFT ground state,
the primary with minimal $h_{a}^{+}$. 

More generally, each state $\ket a$ corresponds to either a primary
or a descendent, and has conformal weights $h_{a}^{\left(l\right)}+n_{a}^{\left(l\right)},h_{a}^{\left(r\right)}+n_{a}^{\left(r\right)}$,
where $n_{a}^{\left(l\right)},n_{a}^{\left(r\right)}\in\mathbb{N}_{0}$.
The corresponding energies $E_{a}$ differ from the idealized $\left(4\pi v/L\right)\left(h_{a}^{+}-c^{+}/24\right)$,
and the choice of conformal family $a_{0}$ with minimal $E_{a_{0}}$
is non-universal. In terms of $n_{a}=n_{a}^{\left(l\right)}-n_{a}^{\left(r\right)}$,
we then define $h_{0}:=h_{a_{0}}+n_{a_{0}}$, the chiral conformal
weight associated with the cylinder ground state $\ket{a_{0}}$. The
value of $h_{0}$ therefore carries two ambiguities: a choice of conformal
family $a_{0}\in\left\{ a\right\} $, and the choice of a state in
the conformal family, $n_{a_{0}}\in\mathbb{N}_{0}$. As described
in Sec.\ref{subsec:Boundary-finite-size}, the only universal statement
is $\theta_{0}=e^{2\pi ih_{0}}\in\left\{ \theta_{a}\right\} $, where
$\theta_{a}=e^{2\pi ih_{a}}$ are the topological spins of bulk anyons. 

The result of Ref.\citep{PhysRevB.88.195412} for the momentum polarization
is given terms of the low lying cylinder eigenstates $\ket a$,
\begin{align}
 & \bra aT_{R}\ket a=\exp\left[\alpha N_{x}+\frac{2\pi i}{N_{x}}\left(h_{a}-\frac{c}{24}\right)+o\left(N_{x}^{-1}\right)\right],\label{eq:17}
\end{align}
where the lattice spacing is set to 1, $N_{x}=L$.  It follows that
the thermal expectation value $\tilde{Z}/Z=\text{Tr}\left(T_{R}e^{-\beta H}\right)/Z$
is equal to $\exp\left[\alpha N_{x}+\frac{2\pi i}{N_{x}}\left(h_{0}-\frac{c}{24}\right)+o\left(N_{x}^{-1}\right)\right]$,
if the temperature $\beta^{-1}$ is much lower than the boundary energy
differences $\sim N_{x}^{-1}$, namely $\beta^{-1}=o\left(N_{x}^{-1}\right)$,
as described in Sec.\ref{sec:Signs-from-geometric}.

\subsection{The value of $h_{0}$ in fermionic phases of matter\label{subsec:The-value-of-h0}}

Fermionic topological phases are microscopically comprised of fermions
(and possibly bosons), and have the fermion parity $\left(-1\right)^{N_{f}}$
as a global symmetry \citep{Kapustin:2015aa,freed2016reflection,PhysRevB.95.235140,aasen2019fermion}.
It is therefore useful to probe such phases with a background $\mathbb{Z}_{2}$
gauge field corresponding to $\left(-1\right)^{N_{f}}$, or a spin
structure.  For our purposes, this amounts to considering both periodic
and anti-periodic boundary conditions around non-contractible cycles
in space-time. 

In the main text we were only interested in locally sign-free QMC
representations of thermal partition functions, and sign-free geometric
manipulations that can be performed to these. We therefore restricted
attention to thermal boundary conditions in the imaginary time direction
(see Sec.\ref{subsec:Local-determinantal-QMC}), and to periodic boundary
conditions around the spatial cylinder. These boundary conditions
cannot generically be modified without introducing signs into the
QMC weights.

Here we provide a fuller picture by considering the behavior of $h_{0}$
with both periodic and anti-periodic boundary conditions, in the closed
$x$ direction of the spatial cylinder. Since $h_{0}$ is a ground
state property, the time direction is open and does not play a role. 

For a fermionic chiral topological phase, the boundary CFT is also
fermionic. The primary conformal weights $\left\{ h_{a}\right\} $
then depend on the choice of boundary conditions (in the $x$ direction),
and as a result, so will the set of topological spins $\left\{ \theta_{a}\right\} $
in which $\theta_{0}=e^{2\pi ih_{0}}$ is valued. In particular, the
vacuum spin $\theta_{I}=0$ will not be included in $\left\{ \theta_{a}\right\} $
for periodic boundary conditions, while for anti-periodic boundary
conditions, both the vacuum $\theta_{I}=1$ \textit{and} the spin
$\theta_{\psi}=-1$ of the microscopic fermion will appear \citep{ginsparg1988applied,PhysRevLett.110.236801}.
Note that $\theta_{\psi}$ does not correspond to an emergent fermion,
as in e.g the toric code \citep{KITAEV20032}, and therefore does
not imply an additional ground state on the torus. 

As an example, consider the series of Laughlin phases at filling $1/q$
, with $q\in\mathbb{N}$, all of which have the chiral central charge
$c=1$. 
First, for $q\in2\mathbb{N}$ the phase is bosonic, and we consider
only periodic boundary conditions. The primary conformal weights are
given by $h_{a}=a^{2}/2q$ \citep{PhysRevB.89.125303,hu2020microscopic},
with $a\in\mathbb{N}_{0}$. The topological spins $\theta_{a}=e^{2\pi ih_{a}}$
depend only on $a\mod q$, and the $q$ spins $\left\{ \theta_{a}\right\} _{a=0}^{q-1}$ (appearing in Table \ref{tab:1})
correspond to the $q$ degenerate ground states on the
torus. In particular, the vacuum spin $\theta_{I}=1$ is obtained
for $a=0$.

For $q\in2\mathbb{N}-1$ the phase is fermionic, and we consider both
periodic and anti-periodic boundary conditions. For periodic boundary
conditions the weights are given by $h_{a}=\left(a+1/2\right)^{2}/2q$
\citep{hu2020microscopic}. As in the bosonic case, $\theta_{a}=e^{2\pi ih_{a}}$
depend only on $a\mod q$, with $\left\{ \theta_{a}\right\} _{a=0}^{q-1}$ (appearing in Table \ref{tab:1})
corresponding to the $q$ degenerate ground states on the torus. Unlike
the bosonic case, the vacuum spin $\theta_{I}=1$ is not included
in $\left\{ \theta_{a}\right\} _{a=0}^{q-1}$. For anti-periodic boundary
conditions, the weights are given by $h_{a}=a^{2}/2q$ as in the bosonic
case \citep{PhysRevB.89.125303}. The set $\left\{ \theta_{a}\right\} _{a=0}^{q-1}$
again corresponds to the $q$ torus ground states, but now $\theta_{\psi}=\theta_{a=q}=-1$
is an additional topological spin that corresponds to the physical
Fermion $\psi$ \citep{bonderson2007non}.

The simplest fermionic Laughlin phase is given by $q=1$, and corresponds
to a Chern insulator with Chern number $\nu=1$ \citep{PhysRevLett.61.2015,qi2008topological},
which is studied in detail in Appendix \ref{subsec:Beyond-the-assumption}
below. The Chern insulator has a unique ground state on the torus,
and accordingly, there is a unique topological spin $\theta_{\sigma}=e^{2\pi i\left(1/8\right)}$
for periodic boundary conditions on the cylinder, and two topological
spins $\theta_{I}=1,\theta_{\psi}=-1$ for anti-periodic boundary
conditions. Here $\psi$ corresponds the physical fermions from which
the Chern insulator is comprised. The object carrying the spin $\theta_{\sigma}$
is the complex analog of the celebrated Majorana zero mode supported
on vortices in the bulk of a $p+ip$ superconductor \citep{read2000paired,kitaev2006anyons}.

\section{Momentum polarization with non CFT boundaries\label{subsec:Beyond-the-assumption}}

As reviewed in Sec.\ref{sec:Signs-from-geometric}, the existing analytic
derivation of Eq.\eqref{eq:12-3-1} relies on the CFT description
of the physical boundaries of the cylinder, and of the line $y=0$
where $T_{R}$ is discontinuous \citep{PhysRevB.88.195412}. In this
appendix we perform an analytic and numerical study that shows that,
at least for free fermions, the relevant CFT expressions and the resulting
Eq.\eqref{eq:12-3-1}, hold even if the boundary is not described
by an idealized CFT. We will however, find a number of subtleties
which have not been demonstrated in the literature, as already described
below Eq.\eqref{eq:2-2} and in Appendix \ref{subsec:Further-details-regarding}. 

\subsection{CFT finite-size correction in non CFT boundaries\label{subsec:CFT-finite-size-correction}}

The main ingredient in the analytic derivation of Eq.\eqref{eq:12-3-1}
is the expression \eqref{eq:2-2} for the finite size correction to
the momentum density in CFT \citep{PhysRevB.88.195412}. In this appendix
we show that, at least in the non-interacting case, Equation \eqref{eq:2-2}
remains valid, with $\theta_{0}=e^{2\pi ih_{0}}\in\left\{ \theta_{a}\right\} $,
even when the boundary cannot be described by a CFT. 

We will consider a Chern insulator, such as the prototypical Haldane
model \citep{PhysRevLett.61.2015}. When the boundary degrees of freedom
can be described by a CFT, they correspond to the Weyl fermion CFT,
where $c=\pm1$ and the primary conformal weights are $h_{\sigma}=\pm1/8$
($h_{I}=0,h_{\psi}=\pm1/2$) for periodic (anti-periodic) boundary
conditions, as described in Appendix \ref{subsec:The-value-of-h0}.
The sign corresponds to the two possible chiralities. More generally,
on a lattice with spacing $1$, the boundary supports a complex fermion
with an energy dispersion $\varepsilon_{k}$, where $k=k_{x}$ takes
values in the Brillouin zone $\mathbb{R}/2\pi\mathbb{Z}$ for an infinite
circumference $L=\infty$, or its discretization $\left(2\pi/L\right)\mathbb{Z}_{L}$
($\left(2\pi/L\right)\left(\mathbb{Z}_{L}+1/2\right)$), for $L<\infty$
and periodic (anti-periodic) boundary conditions. The only requirement
on $\varepsilon_{k}$ is that it be \textit{chiral}, in the sense
that it connects the two separated bulk energy bands. If the intersections
$k_{l}$ and $k_{u}$ with the lower and upper bulk bands, respectively,
satisfy $k_{l}<k_{u}$ ($k_{l}>k_{u}$), we say that the boundary
is right (left) moving, or has a positive (negative) chirality, see
Fig.\ref{fig:Schematic-band-structure,}.
\begin{figure}[t]
\begin{centering}
\includegraphics[width=0.6\columnwidth]{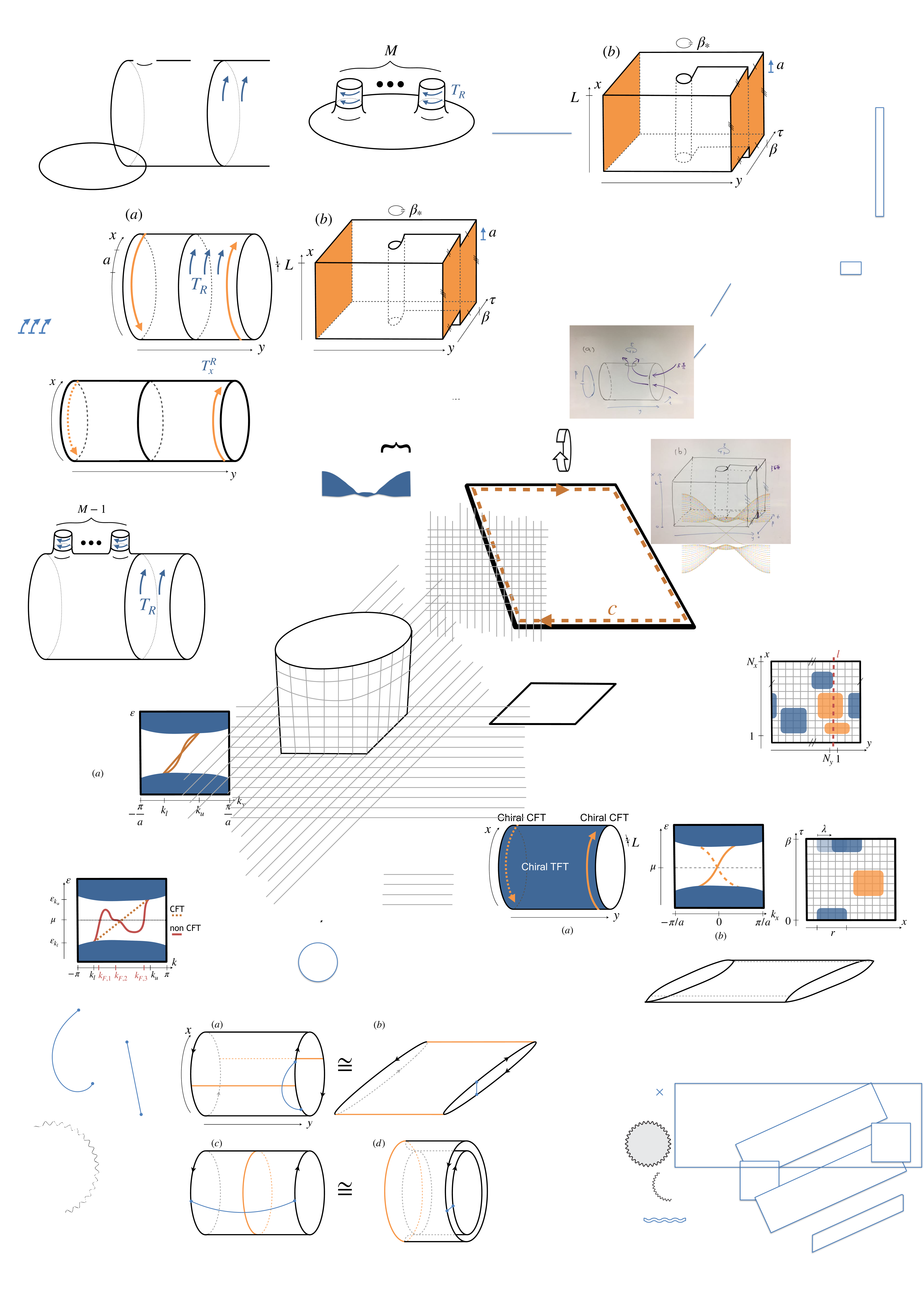}
\par\end{centering}
\caption{Schematic band structure, energy $\varepsilon$ as a function of momentum
$k=k_{x}$ in the periodic $x$ direction, of a Chern insulator on
the cylinder. The figure shows the bulk energy bands (blue) and the
chiral boundary dispersion, with two dispersion branches, on a single
boundary component (orange and red curves). The opposite chirality
branches on the second boundary component are not drawn. The momenta
$k_{l}$ and $k_{u}$ correspond to the intersections of the boundary
dispersion with the lower and upper bulk bands, respectively. Since
$k_{u}>k_{l}$, both dispersion branches have a positive chirality.
The orange line indicates the idealized linear dispersion with a single
Fermi momentum $k_{F}=0$, which corresponds to the Weyl fermion CFT.
The solid red curve corresponds to a more general chiral branch, with
three Fermi momenta $k_{F,1},k_{F,2},k_{F,3}$, where the dispersion
around $k_{F,2}$ takes a (non-generic) non-linear form. With periodic
boundary conditions around the cylinder, both dispersion branches
produce the same $L^{-2}$ correction to the momentum density in Eq.\eqref{eq:18},
with a positive chirality $+$, up to a mod 1 ambiguity: $1/12\protect\mapsto1/12+n,\;n\in\mathbb{N}$.\label{fig:Schematic-band-structure,}}
\end{figure}

 More generally, the dispersion will contain several dispersion branches
$\left\{ \varepsilon_{j,k}\right\} _{j=1}^{J}$, but since the momentum
density is additive in $j$ we restrict attention to a single branch.
Without loss of generality, we also fix the chemical potential $\mu=0$,
in which case the Fermi momentum $k_{F}$ satisfies $\varepsilon_{k_{F}}=0$.
The value of $k_{F}$ plays an important role is the subsequent analysis. 

The simplest dispersion that satisfies the above requirements is the
linear one $\varepsilon_{k}=v\left(k-k_{F}\right)$. For $k_{F}=0$
this corresponds to the Weyl fermion CFT. The presence of $k_{F}\neq0$
corresponds to the addition of a chemical potential $vk_{F}$, which
breaks the conformal symmetry. The generic form is $\varepsilon_{k}=v\left(k-k_{F}\right)+O\left(k-k_{F}\right)^{2}$.
A non-generic dispersion can take the form $\varepsilon_{k}=v_{3}\left(k-k_{F}\right)^{3}+O\left(k-k_{F}\right)^{4}$,
and there may be several Fermi momenta if the dispersion is non monotonic,
see Fig.\ref{fig:Schematic-band-structure,}. 

In all cases the many-body ground state momentum is given by summing
the momenta of all filled single Fermion states $p\left(L\right)=\frac{1}{L}\sum_{\varepsilon{}_{k}<0}k$,
where the sum runs over $k\in\left(2\pi/L\right)\mathbb{Z}_{L}$ such
that $\varepsilon_{k}$ is negative and in the bulk energy gap. In
order to obtain $p$ as a continuous function of $L$, we treat the
bulk energy gap as a smooth cutoff $p\left(L\right)=\frac{1}{L}\sum_{\varepsilon_{k}<0}kC\left(\varepsilon_{k}\right)$,
where the function $C\left(\varepsilon\right)$ goes to $1$ ($0$)
fast enough as $\varepsilon$ goes to $0$ ($\varepsilon_{k_{l}}$
or $\varepsilon_{k_{u}}$)\footnote{It suffices that $C'\left(\varepsilon\right)$ vanish at $\varepsilon=0,\varepsilon_{k_{l}},\varepsilon_{k_{u}}$.}.
The cutoff $C$ represents the smooth delocalization of boundary eigenstates
as their energy nears the bulk energy bands. 

To obtain the $L$ dependence of $p\left(L\right)$, we will use the
Euler-Maclaurin formula
\begin{align}
\sum_{n=n_{1}}^{n_{2}}f\left(n\right) & =\int_{n_{1}}^{n_{2}}f\left(x\right)dx+\frac{f\left(n_{2}\right)+f\left(n_{1}\right)}{2}\label{eq:19}\\
+ & \frac{1}{6}\frac{f'\left(n_{2}\right)-f'\left(n_{1}\right)}{2!}-\frac{1}{30}\frac{f'''\left(n_{2}\right)-f'''\left(n_{1}\right)}{4!}+\mathsf{R},\nonumber 
\end{align}
where the remainder satisfies $\left|\mathsf{R}\right|\leq\frac{2\zeta\left(5\right)}{\left(2\pi\right)^{5}}\int_{n_{1}}^{n_{2}}\left|f^{\left(5\right)}\left(x\right)\right|dx$.
We begin by considering periodic boundary conditions, where we set
$f\left(n\right)=\left(2\pi n/L^{2}\right)C\left(\varepsilon_{2\pi n/L}\right)$.
Assuming a single, vanishing, Fermi momentum $k_{F}=0$, we set $\left(n_{1},n_{2}\right)=\left(-\infty,0\right)$
for positive chirality, and $\left(n_{1},n_{2}\right)=\left(0,\infty\right)$
for negative chirality. Equation \eqref{eq:19} then gives 

\begin{align}
p\left(L\right)= & p\left(\infty\right)\pm\frac{2\pi}{L^{2}}\frac{1}{12}+O\left(\frac{1}{L^{4}}\right),\;\text{as }L\rightarrow\infty,\label{eq:18}
\end{align}
where $p\left(\infty\right)=\int_{\varepsilon_{k}<0}kC\left(\varepsilon_{k}\right)dk/2\pi$
and $\pm=\text{sgn}\left(k_{u}-k_{l}\right)$ is the chirality. The
$1/L^{2}$ correction in \eqref{eq:18} comes from $f'\left(0\right)=2\pi/L^{2}$
in \eqref{eq:19}. We see that the leading finite size correction
$h_{0}-c/24$ is unchanged from its CFT value $h_{\sigma}-c/24=\pm1/12$,
even when a CFT description does not apply. 

The case of a single non-zero Fermi momentum $k_{F}\neq0$ is more
interesting, as it demonstrates that the integer part of $h_{0}$
can change as a function of $L$ and $k_{F}$. The direct derivation
of the end result from the Euler-Maclaurin formula is surprisingly
lengthy, so we omit it and present a more direct route to the end
result. To be concrete, assume a positive chirality and $k_{F}>0$.
The Euler-Maclaurin formula leads to cutoff independent results, so
we can restrict attention to cutoff functions $C\left(\varepsilon_{k}\right)$
which are identically 1 for $0<k<k_{F}$. Since these can serve as
cutoff functions for the case $k_{F}=0$ as well, we can deduce the
$k_{F}\neq0$ momentum density $p\left(L,k_{F}\right)$ from the $k_{F}=0$
momentum density $p\left(L\right)$, 
\begin{align}
p\left(L,k_{F}\right) & =\frac{1}{L}\sum_{k<k_{F}}kC\left(\varepsilon_{k}\right)\\
 & =\frac{1}{L}\sum_{k<0}kC\left(\varepsilon_{k}\right)+\frac{1}{L}\sum_{0<k<k_{F}}k\nonumber \\
 & =p\left(L\right)+\frac{2\pi}{L^{2}}\sum_{l=1}^{n}l,\nonumber 
\end{align}
where $n=\left\lfloor k_{F}L/2\pi\right\rfloor $. Using Eq.\eqref{eq:18},
we then have 
\begin{align}
p\left(L,k_{F}\right)= & p\left(\infty\right)+\frac{2\pi}{L^{2}}\left[\frac{1}{12}+\sum_{l=1}^{n}l\right]+O\left(\frac{1}{L^{4}}\right),\label{eq:32}
\end{align}
where $p\left(\infty\right)$ is the momentum density at $L=\infty$
and $k_{F}=0$. We see that the value of $h_{0}-c/24$ is only equal
to the idealized CFT result $h_{\sigma}-c/24=1/12$ modulo 1, while
the integer part jumps periodically as a function of $k_{F}$ at fixed
number of sites $L$, or as the number of sites $L$ at fixed $k_{F}$.
Treating $k_{F}$ as fixed and valued in $(-\pi,\pi]$, the period
in $L$ is given by $q=\left|2\pi/k_{F}\right|\geq2$, which need
not be an integer. As described in Appendix \ref{subsec:Further-details-regarding},
the mod 1 ambiguity is attributed to $h_{0}$ rather than $c$, which
corresponds to the topological spin $\theta_{0}=\theta_{\sigma}=e^{2\pi i\left(1/8\right)}$. 

The interpretation of Eq.\eqref{eq:32} is straight forward. As the
number of sites $L$ increases, the single particle momenta $\left(2\pi/L\right)\mathbb{Z}_{L}$
become denser in the Brillouin zone $\mathbb{R}/2\pi\mathbb{Z}$.
The $n$th jump in $h_{0}$ correspond to the motion of a single particle
state with momentum $2\pi n/L$ through $k_{F}$ and into the Fermi
sea, adding a momentum density $2\pi n/L^{2}$ to the ground state. 

\begin{figure}[t]
\begin{raggedright}
\includegraphics[width=1\columnwidth]{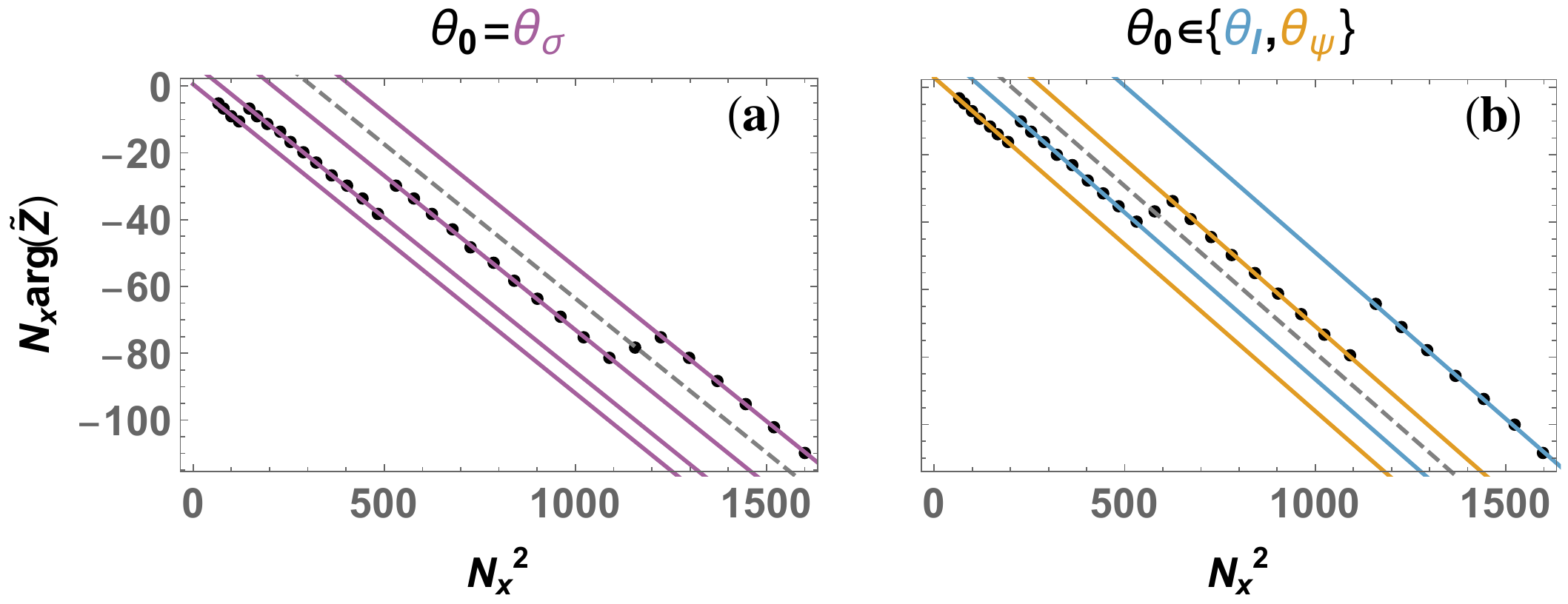}
\par\end{raggedright}
\caption{Numerical results for the momentum polarization Eq.\eqref{eq:12-3-1},
in a Chern insulator with $k_{F}\protect\neq0$. Black dots mark numerically
obtained values of $N_{x}\arg\tilde{Z}$ as a function of $N_{x}^{2}$.
(a) Periodic boundary conditions. Purple lines indicate linear fits,
with approximately the same slope and intercepts $2\pi\left(1/12+\sum_{l=1}^{n}l\right)$
with the $y$ axis, with $n=0,1,2,3$, in accordance with Eq.\eqref{eq:32}.
This allows for the extraction of the topological spin $\theta_{\sigma}=e^{2\pi i\left(1/8\right)}$.
To illustrate the possibility of accidental degeneracies, we choose
$k_{F}=3/34$, where a degeneracy occurs for $N_{x}=34$, and the
average value of $N_{x}\arg\tilde{Z}$ between the two ground states
is obtained. A grey dotted line indicates the average of the two neighboring
purple lines. (b) Anti-periodic boundary conditions. Colored lines
indicate linear fits, with approximately the same slope, and intercepts
$2\pi\left[-1/24+\sum_{l=1}^{n}\left(l-1/2\right)\right]$, with $n=1,2,3,4$,
in accordance with Eq.\eqref{eq:35-2}. The value $n=0$ is not obtained
as it occurs only for small circumferences $N_{x}<6$ where $N_{x}\arg\tilde{Z}$
is not computed. Orange lines correspond to the fermion spin $\theta_{\psi}=-1$,
while blue lines correspond to the vacuum spin $\theta_{I}=0$. To
illustrate the possibility of accidental degeneracies, we choose $k_{F}=\left(5/2\right)/24$,
where an accidental degeneracy occurs for $N_{x}=24$. \label{fig:Numerical-results-for}}
\end{figure}

Figure \ref{fig:Numerical-results-for} presents the results of numerical
computations of the momentum polarization Eq.\eqref{eq:12-3-1} in
a Chern insulator with $k_{F}\neq0$ on a square lattice. Details of the model and computations
can be found in the accompanying Mathematica notebook. In particular,
Fig.\ref{fig:Numerical-results-for}(a) verifies Eq.\eqref{eq:32}.

A subtle point, not mentioned above, is that when $k_{F}L/2\pi\in\mathbb{Z}$,
which happens only when $k_{F}/2\pi=a/b$ is rational and $L\in b\mathbb{N}$,
a single particle state with momentum exactly $k_{F}$  exists, leading
to an accidental degeneracy on the cylinder, between two many-body
ground states with momentum densities given by Eq.\eqref{eq:32} with
$n$ and $n+1$. The momentum polarization \eqref{eq:12-3-1} then
gives the average momentum density in the two ground-states, as visualized
by the grey dotted line in Fig.\ref{fig:Numerical-results-for}(a).
For such system sizes the value $\theta_{0}=-\theta_{\sigma}$ may
be obtained rather than the generic $\theta_{0}=\theta_{\sigma}$. 

The same analysis can be performed for anti-periodic boundary conditions,
where we sum over single particle momenta $k\in\frac{2\pi}{L}\left(\mathbb{Z}_{L}+\frac{1}{2}\right)$.
Equation \eqref{eq:32} is then modified to 
\begin{align}
p\left(L,k_{F}\right)= & p\left(\infty\right)\label{eq:35-2}\\
 & +\frac{2\pi}{L^{2}}\left[-\frac{1}{24}+\sum_{l=1}^{n}\left(l-\frac{1}{2}\right)\right]+O\left(\frac{1}{L^{4}}\right),\nonumber 
\end{align}
where $n=\left\lfloor \frac{k_{F}L}{2\pi}-\frac{1}{2}\right\rfloor $.
As a function of $L$, jumps in $h_{0}-c/24$ occur with the same
period $q=\left|2\pi/k_{F}\right|\geq2$, but are shifted by $q/2$.
Moreover, $h_{0}-c/24$ now attains two values modulo 1, namely $h_{I}-c/24=-1/24$
and $h_{\psi}-c/24=1/2-1/24$. 

Equation \eqref{eq:35-2} therefore demonstrates explicitly the statements
made in Appendix \ref{subsec:Definition-of-}. For $k_{F}=0$, the
cylinder ground state of the Chern insulator corresponds to the idealized
Weyl fermion CFT. A single value $h_{0}=0$ is attained, which is
the conformal weight $h_{I}$ of the CFT vacuum. A non-vanishing $k_{F}$
corresponds to the addition of a chemical potential to the CFT, which
changes the energies of the CFT states, favoring a CFT exited state
over the CFT vacuum. The cylinder ground state of the Chern insulator
may then correspond to any CFT state in the conformal family of either
the vacuum $I$ or fermion $\psi$, which need not be primary. From
the bulk TFT perspective, we see that $\theta_{0}=e^{2\pi ih_{0}}$
may be equal to either of the topological spins $\theta_{I}=1,\theta_{\psi}=-1$
as a function of $L$. 

As in the case of periodic boundary conditions, accidental degeneracies
on the cylinder occur when $k_{F}L/2\pi\in\mathbb{Z}+1/2$, changing
the value of $h_{0}$ attained from the momentum polarization to its
average over the degenerate states. For such system sizes, the value
$\theta_{0}=\pm\sqrt{\theta_{I}\theta_{\psi}}$ is obtained than
the generic $\theta_{0}\in\left\{ \theta_{I},\theta_{\psi}\right\} $. 

Equation \eqref{eq:35-2} is verified numerically in Fig.\ref{fig:Numerical-results-for}(b),
which demonstrates that the value of $\theta_{0}=e^{2\pi ih_{0}}$,
obtained from the momentum polarization \eqref{eq:12-3-1}, takes
different values in the set $\left\{ \theta_{a}\right\} $ of topological
spins as a function of system size $L$, apart from accidental degeneracies.

\subsection{No finite-size correction at finite temperature}

The line $y=0$ where $T_{R}$ jumps can be interpreted as an additional
boundary component at the 'entanglement temperature' $\beta_{*}^{-1}$.
Reference \citep{PhysRevB.88.195412} used the modular transformations
of CFT partition functions to demonstrate that when $\beta_{*}\ll L/v$,
this boundary component does not contribute to the $1/L$ correction
to $\log\tilde{Z}$. Here, we note that the same result holds for
free fermions with a general dispersion $\varepsilon_{k}$. The contribution
of the additional boundary component to $\log\tilde{Z}$ is given
by $\log\tilde{Z}{}_{*}\left(L\right)=Lf_{*}\left(L\right)$, with
the free energy density 
\begin{align}
f_{*}\left(L\right) & =\frac{1}{L}\sum_{k}\log\left(1+e^{iak}e^{-\beta_{*}\varepsilon_{k}}\right)C\left(\varepsilon_{k}\right).
\end{align}
Using Eq.\eqref{eq:19} one finds $f_{*}\left(L\right)=f_{*}\left(\infty\right)+O\left(L^{-4}\right)$
for both periodic and anti-periodic boundary conditions, which implies
$\log\tilde{Z}{}_{*}=Lf_{*}\left(\infty\right)+O\left(L^{-3}\right)$,
with no $1/L$ contribution. The complex number $f_{*}\left(\infty\right)=\int\log\left(1+e^{iak}e^{-\beta_{*}\varepsilon_{k}}\right)C\left(\varepsilon_{k}\right)\text{d}k/2\pi$
contributes to the non-universal $\alpha$ in Eq.\eqref{eq:12-3-1}.

\section{Cutting the torus along an arbitrary vector\label{subsec:Cutting-the-torus-2}}

For $\mathbf{d}=\left(d_{x},0\right)$ we restrict to $N_{x}=n_{x}d_{x},\;n\in\mathbb{N}$,
viewing $d_{x}$ as an enlarged lattice spacing, and treating $n$
as a reduced number of sites along the circumference in place of $N_{x}$.
The same logic applies to $\mathbf{d}=\left(0,d_{y}\right)$. For
$\mathbf{d}=\left(d_{x},d_{y}\right)$ with both $d_{x},d_{y}\neq0$,
we restrict to system sizes $\left(N_{x},N_{y}\right)=n\mathbf{d}$,
such that the line $l=\text{span}_{\mathbb{R}}\mathbf{d}$ is a diagonal
of the rectangle $nN_{x}\times nN_{y}$ and corresponds to a circle
on the torus $\left(\mathbb{R}/N_{x}\mathbb{Z}\right)\times\left(\mathbb{R}/N_{y}\mathbb{Z}\right)$.
Cutting $X$ along this line produces a cylinder $C$ of circumference
$L=n\left|\mathbf{d}\right|$. We then view $\left|\mathbf{d}\right|$
as a lattice spacing and $n$ as a the number of sites along the circumference.
Note that the distance between the boundary components of the resulting
cylinder is $R=nd_{x}d_{x}/\left|\mathbf{d}\right|$, and the thermodynamic
limit is indeed obtained as $n\rightarrow\infty$. With these identifications
the momentum polarization \eqref{eq:12-3-1} remains unchanged, apart
from a modification of the non-universal $\alpha$ to $\alpha\left|\mathbf{d}\right|^{2}$.\label{fn:For--we}

\section{Dealing with accidental degeneracies on the cylinder\label{subsec:Dealing-with-accidental}}

\textbf{}

As demonstrated in Appendix \ref{subsec:Beyond-the-assumption}, for
certain system sizes $N_{x}$ accidental degeneracies occur on the
cylinder, and the function $\theta_{0}\left(N_{x}\right)=e^{2\pi ih_{0}\left(N_{x}\right)}$
obtained from Eq.\eqref{eq:12-3-1} may take values outside the set
$\left\{ \theta_{a}\right\} $, namely $\theta_{0}=\pm\sqrt{\theta_{a}\theta_{b}}$
for a two-fold degeneracy. In this appendix we complete the derivation
of Result \hyperref[Result 1]{1} and \hyperref[Result 1F]{1F} by considering
the possibility of such degeneracies. 

First, even in the presence of degeneracies, $\theta_{0}$ is valued
in a finite set. Therefore, Eq.\eqref{eq:6-1} still implies that
$\epsilon'=n/m$ is rational, and that $\theta_{0}\left(N_{x}\right)e^{-2\pi ic/24}=e^{-2\pi i\epsilon'N_{x}^{2}}$
periodically covers a subset $S\ni1$ of $m$th roots of unity, for
all large enough $N_{x}$. We denote by $\mathcal{N}\subset\mathbb{N}$
the set of circumferences $N_{x}$ for which a degeneracy appears
and $\theta_{0}\left(N_{x}\right)\notin\left\{ \theta_{a}\right\} $.
 If circumferences $N_{x}\in m\mathbb{N}$, where $e^{2\pi i\epsilon'N_{x}^{2}}=1$,
are not all contained in $\mathcal{N}$, then $1=\theta_{0}\left(N_{x}\right)e^{-2\pi ic/24}\in\left\{ \theta_{a}e^{-2\pi ic/24}\right\} $,
as stated in Results \hyperref[Result 1]{1} and \hyperref[Result 1F]{1F}. 

We are left with the complementary case, where degeneracies occur
for all $N_{x}\in m\mathbb{N}$, i.e $m\mathbb{N}\subset\mathcal{N}$.
Note that this case is highly fine tuned, as it ties together the
non-universal $\epsilon'=n/m$ and the set $\mathcal{N}$ of $N_{x}$s
where accidental degeneracies appear. In order to deal with this case,
we make use of the Frobenius-Perron theorem to resolve the degenerate
ground-state subspace, without introducing signs. The analysis applies
only to the bosonic setting of Result \hyperref[Result 1]{1}, and is
similar to that made in Sec.\ref{sec:Generalization-and-extension}.
We now have a Hamiltonian $H'$ on the cylinder, which has an exactly
degenerate ground-state subspace for all $N_{x}\in m\mathbb{N}$,
and has non-positive matrix elements in the on-site homogenous basis
$\ket s$. The Frobenius-Perron theorem implies that an orthonormal
basis $\ket i$ with non-negative entries may be chosen for the ground
state subspace, $\braket si\geq0$ for all $s,i$. It follows that
the matrix elements of $T_{R}$ in the basis $\ket i$ are non-negative,
$M_{ij}:=\bra iT_{R}\ket j\geq0$. Taking the $N_{x}$th matrix power
of $M$ we have $\left(M^{N_{x}}\right)_{ij}\geq0$. Equation \eqref{eq:12-3-1}
implies that the eigenvalues of $M^{N_{x}}$ are of the form $e^{-\delta'N_{x}^{2}}e^{-2\pi i\epsilon'N_{x}^{2}+o\left(1\right)}\theta_{a}e^{-2\pi ic/24}$,
 so we can write $\left(M^{N_{x}}\right)_{ij}=e^{-\delta'N_{x}^{2}}e^{-2\pi i\epsilon'N_{x}^{2}+o\left(1\right)}T_{ij}$,
where $T_{ij}$ has eigenvalues $\left\{ \lambda\right\} \subset\left\{ \theta_{a}e^{-2\pi ic/24}\right\} $.
In particular, $T_{ij}$ is unitary. Since $e^{2\pi i\epsilon'N_{x}^{2}}=1$
for all $N_{x}\in m\mathbb{N}$, we see that $T_{ij}$ also has non-negative
entires, and is therefore a permutation matrix, containing $1$ in
its spectrum (see Sec.\ref{sec:Generalization-and-extension}). It
follows that $1\in\left\{ \lambda\right\} \subset\left\{ \theta_{a}e^{-2\pi ic/24}\right\} $,
asserting Result \hyperref[Result 1]{1}. 

We are currently unaware of an analog of the Frobenius-Perron theorem
in the context of DQMC, that may be used to resolve the degenerate
ground-state subspace without introducing signs. Instead, we will
make a physical assumption under which Result \hyperref[Result 1F]{1F}
holds. Namely, we will assume that the fine tuned constraint $\epsilon'=n/m$\textit{
and} $m\mathbb{N}\subset\mathcal{N}$ may be lifted by a sign-free
perturbation. This includes (i) perturbations to the effective single-fermion
Hamiltonian $h_{\phi\left(\tau\right)}$ that do not violate the algebraic
condition $h_{\phi\left(\tau\right)}\in\mathcal{C}_{h}$, (ii) perturbations
to the bosonic action $S_{\phi}$ that maintain its reality, and (iii)
changes of the vector $\mathbf{d}$ along which the torus is cut to
a cylinder, as described in Appendix \ref{subsec:Cutting-the-torus-2},
which will generically change the details of the boundary spectrum,
including the non-universal number $\epsilon'$ and the set $\mathcal{N}$
of $N_{x}$s where accidental degeneracies appear. A robustness of
$\epsilon'$ and the set $\mathcal{N}$, both non-universal, under
all three of the above deformations certainly goes beyond the low
energy description of a chiral TFT in the bulk and a chiral CFT on
the boundary. We also adopt this assumption in the fermionic spontaneously-chiral
setting of Result \hyperref[Result 2F]{2F}. 

A stoquastic variant of the above assumption may also be adopted to
establish the bosonic spontaneously-chiral Result \hyperref[Result 2]{2},
but a stronger statement can in fact be made, by again making use
of the Frobenius-Perron theorem to resolve the accidentally degenerate
ground states. The Frobenius-Perron theorem does not immediately complete
the derivation of Result \hyperref[Result 2]{2}, since the former is
a ground state statement, while the latter made use of the finite
temperature $\Delta E\ll\beta^{-1}\ll N_{x}^{-1}$, where $\Delta E$
is the exponentially small finite-size splitting between low lying
symmetry breaking eigenstates. This difficulty does not arise in the
'classical symmetry breaking' scenario, where $\Delta E=0$. In the
generic case $\Delta E\neq0$, we can make progress under the assumption
that the $\mathcal{T},\mathcal{P}$-even state $W\left[\ket ++\ket -\right]$
has lower energy than the $\mathcal{T},\mathcal{P}$-odd state $W\left[\ket +-\ket -\right]$,
rather than the opposite possibility. The derivation of Result \hyperref[Result 2]{2}
in Sec.\ref{subsec:Momentum-polarization-for} can then be repeated
at zero temperature. In particular, Eq.\eqref{eq:10} and its analysis
are unchanged. 

\section{A 'non-local design principle' for chiral topological matter\label{subsec:A-non-local-design}}

As stated in Sec.\ref{subsec:Example:-time-reversal}, the composition
$\mathsf{T}=\mathcal{P}^{\left(0\right)}\mathcal{T}^{\left(1/2\right)}$
of the spin-less reflection with the spin-1/2 time-reversal, naturally
provides a design-principle for a class of models for chiral topological
matter. Here we describe such $\mathsf{T}$-invariant models for chiral
topological superconductors.

The simplest model is comprised of two copies, labeled by $\sigma=\uparrow,\downarrow$,
of a spin-less $p+ip$ superconductor,

\begin{align}
H & =\sum_{\mathbf{x},\mathbf{x}',\sigma,\sigma'}\left[\psi_{\sigma,\mathbf{x}}^{\dagger}h_{\mathbf{x},\mathbf{x}',\sigma,\sigma'}\psi_{\sigma,\mathbf{x}'}+\psi_{\sigma,\mathbf{x}}^{\dagger}\Delta_{\mathbf{x}\mathbf{x}',\sigma,\sigma'}\psi_{\sigma,\mathbf{x}'}^{\dagger}+h.c\right].
\end{align}
Here 
\begin{align}
\Delta= & \begin{pmatrix}\Delta_{0}\left(d^{x}+id^{y}\right) & 0\\
0 & \Delta_{0}\left(d^{x}+id^{y}\right)
\end{pmatrix},
\end{align}
where $d_{\mathbf{x}\mathbf{x}'}^{x}$ ($d_{\mathbf{x}\mathbf{x}'}^{y}$)
is the anti-symmetric $x$ ($y$) difference operator, 
\begin{align}
d_{\mathbf{x}\mathbf{x}'}^{x} & =\left(\delta_{x,x'+1}-\delta_{x+1,x}\right)\delta_{y,y'}/2,\\
d_{\mathbf{x}\mathbf{x}'}^{y} & =\delta_{x,x'}\left(\delta_{y,y'+1}-\delta_{y+1,y'}\right)/2,\nonumber 
\end{align}
and $\Delta_{0}\in\mathbb{R}-\left\{ 0\right\} $. Additionally,
\begin{align}
h= & \begin{pmatrix}t & 0\\
0 & t
\end{pmatrix},
\end{align}
and the hopping $t$ is real and reflection symmetric, e.g 
\begin{align}
t_{\mathbf{x},\mathbf{x}'} & =\frac{1}{2}t_{0}\left(\delta_{x,x'+1}+\delta_{x+1,x'}\right)\delta_{y,y'}+\left(x\leftrightarrow y\right)-\mu,
\end{align}
with $t_0>0,\;\mu\in\mathbb{R}$. It is well known that the chemical potential
$\mu$ can be used to tune the model between gapped SPT phases with
$c=0,-1,1,$ for $\left|\mu\right|>2t_{0},-2t_{0}<\mu<0,0<\mu<2t_{0}$, respectively,
see e.g \citep{PhysRevB.98.064503}. Additionally, the Hamiltonian
is invariant under the combination of the unitary spin-less reflection
$\mathcal{P}^{\left(0\right)}:\psi_{\sigma,\left(x,y\right)}\mapsto\psi_{\sigma,\left(x,-y\right)}$
and the anti-unitary spin-full time-reversal $\mathcal{T}^{\left(1/2\right)}:\psi_{\uparrow,\mathbf{x}}\mapsto\psi_{\downarrow,\mathbf{x}},\;\psi_{\downarrow,\mathbf{x}}\mapsto-\psi_{\uparrow,\mathbf{x}}$.

The model can be written in the BdG form 
\begin{align}
H & =\sum_{\mathbf{x},\mathbf{x}'}\Psi_{\mathbf{x}}^{\dagger}h_{\text{BdG}}^{\mathbf{x},\mathbf{x}'}\Psi_{\mathbf{x}'},
\end{align}
where $\Psi_{\mathbf{x}}^{T}=\left(\psi_{\uparrow\mathbf{x}},\psi_{\downarrow\mathbf{x}},\psi_{\uparrow\mathbf{x}}^{\dagger},\psi_{\downarrow\mathbf{x}}^{\dagger}\right)$
is the Nambu spinor (a Majorana spinor), and 
\begin{align}
h_{\text{BdG}} & =\begin{pmatrix}h & \Delta\\
-\Delta^{*} & -h^{*}
\end{pmatrix}.\label{eq:48}
\end{align}
The ``single-fermion'' space on which $h_{\text{BdG}}$ acts
is $\mathcal{H}_{1\text{F}}=\mathcal{H}_{X}\otimes\mathcal{H}_{\text{spin}}\otimes\mathcal{H}_{\text{Nambu}}\cong\mathbb{C}^{\left|X\right|}\otimes\mathbb{C}^{2}\otimes\mathbb{C}^{2}$.
The spin-less reflection acts on $\mathcal{H}_{1\text{F}}$ as $\mathcal{P}^{\left(0\right)}=\mathcal{P}_{X}^{\left(0\right)}\otimes I_{2}\otimes I_{2}$,
where $\mathcal{P}_{X}^{\left(0\right)}=\delta_{x,x'}\delta_{y,-y'}$.
The spin-full time-reversal acts by $\mathcal{T}^{\left(1/2\right)}=I_{\left|X\right|}\otimes iY\otimes I_{2}\mathcal{K}$,
where $Y$ is the Pauli matrix and $\mathcal{K}$ is the complex conjugation.
The operator $\mathsf{T}=\mathcal{P}^{\left(0\right)}\mathcal{T}^{\left(1/2\right)}$
satisfies $\mathsf{T}^{2}=-I$ and $\left[\mathsf{T},h_{\text{BdG}}\right]=0$,
and is therefore a time-reversal design principle which applies to
$h_{\text{BdG}}$, implying $\text{det}\left(\partial_{\tau}+h_{\text{BdG}}\right)\ge0$.
 Since $h_{\text{BdG}}$ acts on the Majorana spinor $\Psi$, the
relevant quantity is actually the Pfaffian $\text{Pf}\left(\partial_{\tau}+h_{\text{BdG}}\right)=\sqrt{\text{det}\left(\partial_{\tau}+h_{\text{BdG}}\right)}\geq0$,
where the principle branch of the square root is chosen.

The Hamiltonian $h_{\text{BdG}}$ can be considerably generalized
while maintaining $\left[\mathsf{T},h_{\text{BdG}}\right]=0$, by
taking \begin{widetext}
\begin{align}
h= & \begin{pmatrix}t & r\\
-r^{*} & t^{*}
\end{pmatrix},\\
\Delta= & \begin{pmatrix}e^{i\alpha}\left(\left|\Delta_{x}\right|d^{x}+i\left|\Delta_{y}\right|d^{y}\right)^{\left|\ell\right|} & e^{i\tilde{\alpha}}\left(\left|\tilde{\Delta}_{x}\right|d^{x}+i\left|\tilde{\Delta}_{y}\right|d^{y}\right)^{\left|\tilde{\ell}\right|}\\
-e^{-i\tilde{\alpha}}\left(\left|\tilde{\Delta}_{x}\right|d^{x}+i\left|\tilde{\Delta}_{y}\right|d^{y}\right)^{\left|\tilde{\ell}\right|} & e^{-i\alpha}\left(\left|\Delta_{x}\right|d^{x}+i\left|\Delta_{y}\right|d^{y}\right)^{\left|\ell\right|}
\end{pmatrix},\nonumber 
\end{align}
\end{widetext}where $t_{\mathbf{x},\mathbf{x}'},r_{\mathbf{x},\mathbf{x}'}$
are general matrices, and $\ell\in2\mathbb{Z}+1$ ($\tilde{\ell}\in2\mathbb{Z}$)
is the angular momentum channel of the triplet (singlet) pairing.
This can be further generalized to a sum over all angular momentum
channels $\sum_{\ell\in2\mathbb{Z}+1}e^{i\alpha_{\ell}}\left(\left|\Delta_{\ell,x}\right|d^{x}+i\left|\Delta_{\ell,y}\right|d^{y}\right)^{\left|\ell\right|}$
and similarly for $\tilde{\ell}$. The model is Hermitian for $t=t^{\dagger}$,
$r=-r^{T}$, but this is not required to avoid the sign problem.

In order to obtain an interacting model, the parameters $\phi=\left\{ t,r,\alpha_{\ell},\tilde{\alpha}_{\tilde{\ell}},\left|\Delta_{\ell,x}\right|,\left|\Delta_{\ell,y}\right|,\left|\tilde{\Delta}_{\tilde{\ell},x}\right|,\left|\tilde{\Delta}_{\tilde{\ell},y}\right|\right\} $
can now be promoted to space-time dependent bosonic fields, with any
action $S_{\phi}\in\mathbb{R}$. The model will be sign-free as long
as $h_{\text{BdG}}$ remains $\mathsf{T}$-invariant for all configurations
$\phi$, which requires that only reflection-even configurations $\phi\left(\tau,x,y\right)=\phi\left(\tau,x,-y\right)$
are summed over. As discussed in Sec.\ref{subsec:Example:-time-reversal},
this implies non-local interactions, which effectively fold the chiral
system into a non-chiral system of half of space.

\section{Locality and homogeneity of known design principles \label{subsec:Locality-of-known}}

In this appendix we review all fermionic design principles known to
us, clarify their common features, and describe the conditions under
which they are \textit{on-site homogeneous}, imply a \textit{term-wise
sign-free} DQMC representation, and allow a \textit{locally sign-free
DQMC} simulation, as defined in Sec.\ref{subsec:Local-and--homogeneous}.
The design principles are stated as algebraic conditions satisfied
by the effective single-fermion Hamiltonian $h_{\phi}=h_{\phi\left(\tau\right)}$
and the corresponding imaginary-time evolution $U_{\phi}=\text{TO}e^{-\int_{0}^{\beta}h_{\phi\left(\tau\right)}\text{d}\tau}$,
or in terms of the operator $D_{\phi}=\partial_{\tau}+h_{\phi}$,
see Sec.\ref{subsec:Local-determinantal-QMC}.

\paragraph*{Contraction semi-groups and Majorana time reversals}

The time-reversal design principle covered in Sec.\ref{subsec:Example:-time-reversal}
is a special case of a broad class of design principles that were
recently discovered and unified \citep{li2016majorana,wei2016majorana,wei2017semigroup}.
These are stated in terms of Majorana fermions, where $\psi$ is real
and $\overline{\psi}=\psi^{T}$, in which case $h_{\phi}$ is anti-symmetric
and the determinants in \eqref{eq:2} are replaced by their square
roots. Reference \citep{wei2017semigroup} shows that if 
\begin{align}
\mathsf{J}_{1}h_{\phi}-h_{\phi}^{*}\mathsf{J}_{1} & =0,\label{eq:J1}\\
i\left(\mathsf{J}_{2}h_{\phi}-h_{\phi}^{*}\mathsf{J}_{2}\right) & \geq0,\label{eq:J2}
\end{align}
where the matrices $\mathsf{J}_{1},\mathsf{J}_{2}$ are real and orthogonal,
and obey $\mathsf{J}_{1}^{T}=\pm\mathsf{J}_{1}$, $\mathsf{J}_{2}^{T}=-\mathsf{J}_{2}$,
$\left\{ \mathsf{J}_{1},\mathsf{J}_{2}\right\} =0$, then $\text{Det}\left(I+U_{\phi}\right)\geq0$.
The equality \eqref{eq:J1} corresponds to an anti-unitary symmetry
$\mathsf{T}_{1}=\mathsf{J}_{1}\mathcal{K},\;\mathsf{T}_{1}^{2}=\pm I$,
where $\mathcal{K}$ is the complex conjugation. If the inequality
\eqref{eq:J2} is replaced by an equality, it corresponds to an additional
anti-unitary symmetry, $\mathsf{T}_{2}=\mathsf{J}_{2}\mathcal{K}$,
$\mathsf{T}_{2}^{2}=-I$. The case $\mathsf{T}_{1}^{2}=-I$ then reduces
to the standard time-reversal $\mathsf{T}$ described in Sec.\ref{subsec:Example:-time-reversal},
while $\mathsf{T}_{1}^{2}=I$ corresponds to the 'Majorana class'
of Ref.\citep{li2016majorana}. More generally, the inequality \eqref{eq:J2}
states that the left hand side is a positive semi-definite matrix,
and implies that $h_{\phi}$ is a generator of the contraction semi-group
defined by the Hermitian metric $\eta_{2}=i\mathsf{J}_{2},\;\eta_{2}^{2}=I$,
$\left[\mathsf{T}_{1},\eta_{2}\right]=0$. Explicitly, Eq.\eqref{eq:J1}-\eqref{eq:J2}
can be written as  
\begin{align}
\left[\mathsf{T}_{1},h_{\phi}\right]=0,\;\;\eta_{2}h_{\phi}+h_{\phi}^{\dagger}\eta_{2} & \geq0,\label{eq:eta2h}
\end{align}
and imply 
\begin{align}
\left[\mathsf{T}_{1},U_{\phi}\right]=0,\;\;\eta_{2}-U_{\phi}^{\dagger}\eta_{2}U_{\phi} & \geq0.\label{eq:eta2}
\end{align}

In the language of Sec.\ref{subsec:Local-and--homogeneous}, for fixed
$\mathsf{T}_{1},\eta_{2}$, the set $\mathcal{C}_{h}$ contains all
matrices $h_{\phi}$ satisfying \eqref{eq:eta2h}. It is clear that
this set is additive: $h_{1}+h_{2}\in\mathcal{C}_{h}$ for all $h_{1},h_{2}\in\mathcal{C}_{h}$.
The set $\mathcal{C}_{U}$ contains all matrices $U_{\phi}$ satisfying
Eq.\eqref{eq:eta2}, and is multiplicative: $U_{1}U_{2}\in\mathcal{C}_{U}$
for all $U_{1},U_{2}\in\mathcal{C}_{U}$. 

A sufficient condition on $\mathsf{T}_{1},\eta_{2}$ that guarantees
that the design principle they define is on-site homogenous is that
they are of the form $\mathsf{T}_{1}=I_{\left|X\right|}\otimes\mathsf{t}_{1},\eta_{2}=I_{\left|X\right|}\otimes e_{2}$,
written in terms of the decomposition $\mathcal{H}_{1\text{F}}\cong\mathbb{C}^{\left|X\right|}\otimes\mathbb{C}^{\mathsf{d}_{\text{F}}}$
of the single-fermion space.  The permutation matrices $O^{\left(\sigma\right)}$
defined in Eq.\eqref{eq:20} then commute with $\eta_{2}$ and $\mathsf{T}_{1}$.
Since $O^{\left(\sigma\right)}$ are also unitary, we have $O^{\left(\sigma\right)}\in\mathcal{C}_{U}$
for all $\sigma\in S_{X}$. All examples described in Refs.\citep{li2016majorana,wei2016majorana,wei2017semigroup}
are of the on-site homogenous form $\mathsf{T}_{1}=I_{\left|X\right|}\otimes\mathsf{t}_{1},\eta_{2}=I_{\left|X\right|}\otimes e_{2}$.

As in our discussion of $\mathsf{T}$ in Sec.\ref{subsec:Example:-time-reversal},
the locality of $\mathsf{T}_{1}=I_{\left|X\right|}\otimes\mathsf{t}_{1}$
means that it can be applied term-wise, by symmetrizing the local
terms $h_{\phi;\mathbf{x}}\mapsto\frac{1}{2}\left(h_{\phi;\mathbf{x}}+\mathsf{T}_{1}h_{\phi;\mathbf{x}}\mathsf{T}_{1}^{-1}\right)$.
A similar procedure for $\eta_{2}$ is only possible if the inequality
in Eq.\eqref{eq:eta2h} holds as an equality (as in Ref.\citep{li2016majorana}).
The contraction semi-group defined by $\eta_{2}$ then reduces to
an orthogonal group, and one can enforce the term-wise relations $\eta_{2}h_{\phi;\mathbf{x}}+h_{\phi;\mathbf{x}}^{\dagger}\eta_{2}=0$
by $h_{\phi;\mathbf{x}}\mapsto\frac{1}{2}\left(h_{\phi;x}-\eta_{2}h_{\phi;x}^{\dagger}\eta_{2}\right)$.

Collecting the above, we see that if $\mathsf{T}_{1},\eta_{2}$ can
be brought to the form $\mathsf{T}_{1}=I_{\left|X\right|}\otimes\mathsf{t}_{1},\eta_{2}=I_{\left|X\right|}\otimes e_{2}$
by the same single-fermion local unitary $u$, then a DQMC representation
which is $\mathsf{T}_{1}$-symmetric, and respects Eq.\eqref{eq:eta2h}
terms-wise, leads to a locally-sign free DQMC simulation.

\paragraph*{Split orthogonal group}

Another recently discovered design principle is defined in terms of
the split orthogonal group $O\left(n,n\right)$ \citep{wang2015split}:
if $U_{\phi}\in O\left(n,n\right)$, then the sign of $\text{Det}\left(I+U_{\phi}\right)$
depends only on the connected component of $O\left(n,n\right)$ to
which $U_{\phi}$ belongs. If the sign of $e^{-S_{\phi}}$ is manifestly
compatible with the connected component of $U_{\phi}$ in $O\left(n,n\right)$,
one has $p\left(\phi\right)=e^{-S_{\phi}}\text{Det}\left(I+U_{\phi}\right)\geq0$.
More explicitly, the statement $U_{\phi}\in O\left(n,n\right)$ implies
that $U_{\phi}$ is a real matrix and $\eta-U_{\phi}^{T}\eta U_{\phi}=0$,
where $\eta=\text{diag}\left(I_{n},-I_{n}\right)$. Restricting to
the identity component $O_{0}\left(n,n\right)$, this amounts to the
statements that $h_{\phi}$ is in the Lie algebra $o\left(n,n\right)$:
it is real and satisfies $\eta h_{\phi}+h_{\phi}^{T}\eta=0$.

In a basis independent formulation, the data that defines the design
principle is an anti-unitary $\tilde{\mathsf{T}}$, such that $\tilde{\mathsf{T}}^{2}=I$,
and a Hermitian metric $\tilde{\eta}$ with canonical form $\eta$,
such that $\left[\tilde{\mathsf{T}},\tilde{\eta}\right]=0$. The
set $\mathcal{C}_{h}$ is then given by matrices $h_{\phi}$ satisfying
\begin{align}
\left[\tilde{\mathsf{T}},h_{\phi}\right]=0,\;\; & \tilde{\eta}h_{\phi}+h_{\phi}^{\dagger}\tilde{\eta}=0,
\end{align}
while $\mathcal{C}_{U}$ is defined by 
\begin{align}
\left[\tilde{\mathsf{T}},U_{\phi}\right]=0,\;\; & \tilde{\eta}-U_{\phi}^{\dagger}\tilde{\eta}U_{\phi}=0.
\end{align}
The analogy with \eqref{eq:eta2h}-\eqref{eq:eta2} is now manifest,
with the inequalities strengthened to equalities. Accordingly, the
$O\left(n,n\right)$ design-principle is on-site homogeneous if $\tilde{\mathsf{T}}=I_{\left|X\right|}\otimes\tilde{\mathsf{t}}$
and $\tilde{\eta}=I_{\left|X\right|}\otimes\tilde{e}$. If these forms
can be obtained by conjugation of $\tilde{\mathsf{T}},\tilde{\eta}$
with the same single-fermion local unitary $u$, then a DQMC representation
which is sign-free due to $\tilde{\mathsf{T}},\tilde{\eta}$, leads
to a locally-sign free DQMC simulation. 

The above statements hold for $U_{\phi}$ in the identity component
$O_{0}\left(n,n\right)$, which is always the case when $h_{\phi}\in o\left(n,n\right)$
and $U_{\phi}=\text{TO}e^{-\int_{0}^{\beta}h_{\phi\left(\tau\right)}\text{d}\tau}$.
Time evolutions in the additional three connected components of $O\left(n,n\right)$
can be obtained by operator insertions generalizing $U_{\phi}=U_{k}\cdots U_{2}U_{1}$
to $U_{k}\cdots O_{2}U_{2}O_{1}U_{1}$, where $O\in O\left(n,n\right)/O_{0}\left(n,n\right)$
\citep{wang2015split}. These can be incorporated into the frame-work
of Sec.\ref{sec:Determinantal-quantum-Monte}, if each $O_{k}$ is
supported on a disk of radius $w$ around a site $\mathbf{x}_{k}$,
i.e $\left(O_{k}\right)_{\mathbf{x},\mathbf{y}}=\delta_{\mathbf{x},\mathbf{y}}$
if $\left|\mathbf{x}-\mathbf{x}_{k}\right|>w$ or $\left|\mathbf{y}-\mathbf{x}_{k}\right|>w$.
With this generalization, all sign-free examples described in Ref.\citep{wang2015split}
amount to locally sign-free DQMC.

\paragraph*{Solvable fermionic and bosonic actions}

Reference \citep{chandrasekharan2013fermion} described a design principle
that nontrivially relates the fermionic action $S_{\psi,\phi}=\overline{\psi}D_{\phi}\psi$
and bosonic action $S_{\phi}$. A fermionic action was termed 'solvable'
if $D_{\phi}$ has the form
\begin{align}
 & D_{\phi}=\begin{pmatrix}0 & M_{\phi}\\
-M_{\phi}^{\dagger} & 0
\end{pmatrix},\label{eq:58}
\end{align}
which clearly implies $\text{Det}\left(D_{\phi}\right)=\left|\text{Det}\left(M_{\phi}\right)\right|^{2}\geq0$.
Here the imaginary time circle $\mathbb{R}/\beta\mathbb{Z}$ is discretized
to $\mathbb{Z}_{\beta}=\mathbb{Z}/\beta\mathbb{Z}$, and $D_{\phi}$
is treated as a matrix on $\mathbb{C}^{\beta}\times\mathcal{H}_{1\text{F}}=\mathbb{C}^{\beta}\times\mathbb{C}^{\left|X\right|}\times\mathbb{C}^{\mathsf{d}_{F}}$,
with indices $\left(\tau,\mathbf{x},\alpha\right),\left(\tau',\mathbf{x}',\alpha'\right)$
for time, space, and internal degrees of freedom. For example, the
Hamiltonian form $D_{\phi}=\partial_{\tau}+h_{\phi\left(\tau\right)}$
is discretized to

\begin{widetext}
\begin{align}
\left[D_{\phi}\right]_{\left(\tau,\mathbf{x},\alpha\right),\left(\tau',\mathbf{x}',\alpha'\right)}= & \left(\delta_{\tau,\tau'}-\delta_{\tau-1,\tau'}\right)\delta_{\mathbf{x},\mathbf{x}'}\delta_{\alpha,\alpha'}+\delta_{\tau-1,\tau'}\left[h_{\phi\left(\tau\right)}\right]_{\left(\mathbf{x},\alpha\right),\left(\mathbf{x}',\alpha'\right)}.\label{eq:52}
\end{align}
\end{widetext} 

In a basis independent language, Eq.\eqref{eq:58} corresponds to
\begin{align}
 & \left\{ \Gamma,D_{\phi}\right\} =0,\;\;D_{\phi}^{\dagger}=-D_{\phi},\label{eq:53}
\end{align}
where $\Gamma$ is a 'chiral symmetry', $\Gamma^{2}=I,\;\Gamma=\Gamma^{\dagger}$.
Eq.\eqref{eq:16} is then obtained in a basis where is diagonal, $\Gamma=\text{diag}\left(I,-I\right)$.
Note however that $\Gamma$ acts on $D_{\phi}$ rather than $h_{\phi}$,
and that the form \eqref{eq:58} requires a non-canonical transformation
away from the Hamiltonian form \eqref{eq:52}. We refer to $\Gamma$
as on-site homogeneous if it is of the form $\Gamma=I_{\beta}\otimes I_{\left|X\right|}\otimes\gamma$
, and to $D_{\phi}$ as local if $D_{\phi}=\sum_{\tau,\mathbf{x}}D_{\phi;\tau,\mathbf{x}}$
where each term $D_{\phi;\tau,\mathbf{x}}$ is supported on a disk
of radius $r$ around $\left(\tau,\mathbf{x}\right)$, and depends
on the values of $\phi$ at points within this disk. The action $D_{\phi}$
is 'term-wise solvable' if each $D_{\phi;\tau,\mathbf{x}}$ satisfies
\eqref{eq:53}. Any local $D_{\phi}$ obeying \eqref{eq:53} with
$\Gamma=I_{\beta}\otimes I_{\left|X\right|}\otimes\gamma$ can be
made term-wise solvable by replacing $D_{\phi;\tau,\mathbf{x}}\mapsto\frac{1}{2}\left(D_{\phi;\tau,\mathbf{x}}-\Gamma D_{\phi;\tau,\mathbf{x}}\Gamma\right)$
and then $D_{\phi;\tau,\mathbf{x}}\mapsto\frac{1}{2}\left(D_{\phi;\tau,\mathbf{x}}-D_{\phi;\tau,\mathbf{x}}^{\dagger}\right)$.
The twisted fermionic boundary conditions in \eqref{eq:5-1} are implemented
by declaring that the index '$\left(\tau=0,\mathbf{x},\alpha\right)$'
that appears in Eq.\eqref{eq:52} corresponds to $\left(\tau=\beta,x+\lambda\Theta\left(y\right),y,\alpha\right)$
with $\lambda\neq0$. Equation \eqref{eq:53} then holds for all $\lambda$
if $\Gamma=I_{\beta}\otimes I_{\left|X\right|}\otimes\gamma$. Under
these conditions, solvable fermionic actions can then be incorporated
into the definition of locally sign-free DQMC given in Sec.\ref{sec:Determinantal-quantum-Monte}.

All examples given in Ref.\citep{chandrasekharan2013fermion} have
an on-site $\Gamma$ and local $D_{\phi}$, and it follows from the
above discussion that, under these conditions, solvable fermionic
actions can then be incorporated into the definition of locally sign-free
DQMC given in Sec.\ref{sec:Determinantal-quantum-Monte}. 

A bosonic action $S_{\phi}$ for a complex valued field $\phi=\left|\phi\right|e^{i\theta}$
was termed 'solvable' in Ref.\citep{chandrasekharan2013fermion} if
\begin{align}
 & S_{\phi}=S_{\left|\phi\right|}-\sum_{u,u'}\beta_{u,u'}\left|\phi_{u}\right|\left|\phi_{u'}\right|\cos\left(\varepsilon_{u}\theta_{u}+\varepsilon_{u'}\theta_{u'}\right),
\end{align}
where $u=\left(\mathbf{x},\tau\right)$, $u=\left(\mathbf{x}',\tau'\right)$,
and $\varepsilon_{u},\varepsilon_{u'}\in\left\{ \pm1\right\} $, and
$\beta_{u,u'}\geq0$. For such actions, it was shown that all correlators
$\int D\phi e^{-S_{\phi}}\phi_{u_{1}}\cdots\phi_{u_{k}}$ are non-negative,
and therefore $\phi$ can be added to the diagonal in \eqref{eq:58}
with a positive coupling constant $g>0$, $D_{\phi}=\begin{pmatrix}g\phi & M\\
-M^{\dagger} & g\phi
\end{pmatrix}$, without introducing signs, though $D_{\phi}$ is no longer solvable.
Solvable bosonic actions are easily incorporated into the framework
of Sec.\ref{sec:Determinantal-quantum-Monte}, as long as they are
local in the sense of Eq.\eqref{eq:16}, and in particular, $\beta_{u,u'}=0$
unless the points $u$ and $u'$ are close.

%

\end{document}